\documentclass[fleqn,usenatbib]{mnras}
\usepackage{amsmath}
\usepackage{graphicx}
\usepackage{epstopdf}
\usepackage{amssymb,txfonts}
\usepackage{xspace}
\usepackage{lineno}

\newcommand{\xte}{{\textit{RXTE}}}

\newcommand{\fermi}{{\textit{Fermi}}}
\newcommand{\agile}{{\textit{AGILE}}}
\newcommand{\swift}{{\textit{Swift}}}
\newcommand{\msun}{{\rm M}_{\sun}}

\newcommand{\g}{$\gamma$}

\def\sym{$S_YZ_6R_{30}T_{150}C_2$\xspace}
\def\slm{$S_LZ_6R_{20}T_\infty C_5$\xspace}
\def\ssm{$S_SZ_4R_{20}T_{150}C_5$\xspace}

\let\oldhat\hat
\renewcommand{\vec}[1]{\boldmath{#1}}
\renewcommand{\hat}[1]{\oldhat{\mathbf{#1}}}

\newbox\grsign \setbox\grsign=\hbox{$>$} \newdimen\grdimen \grdimen=\ht\grsign
\newbox\simpropbox
\setbox\simpropbox=\hbox{\raise.5ex\hbox{$\propto$}\llap
     {\lower.5ex\hbox{$\sim$}}}\ht2=\grdimen\dp2=0pt
\def\simprop{\mathrel{\copy\simpropbox}}

\title[Gamma-ray and radio emission from Cyg X-3]{A comprehensive study of high-energy gamma-ray and radio emission from Cyg X-3}
\author[A. A. Zdziarski et al.]
{Andrzej A. Zdziarski,$^{1}$\thanks{E-mail: aaz@camk.edu.pl (AAZ), ddenys.malyshev@astro.uni-tuebingen.de (DM), Guillaume.Dubus@univ-grenoble-alpes.fr (GD)} Denys Malyshev,$^{2}$\footnotemark[1] Guillaume Dubus,$^3$ Guy G. Pooley,$^4$\newauthor Tyrel Johnson,$^{5}$\thanks{Present address: Naval Research Laboratory, Washington, DC 20375, USA} Adam Frankowski,$^1$ Barbara de Marco,$^1$ Maria Chernyakova$^{6,7}$\newauthor and A. R. Rao$^8$\\
$^1$Nicolaus Copernicus Astronomical Center, Polish Academy of Sciences, Bartycka 18, PL-00-716 Warszawa, Poland\\
$^2$Institut f{\"u}r Astronomie und Astrophysik T{\"u}bingen, Universit{\"a}t T{\"u}bingen, Sand 1, D-72076 T{\"u}bingen, Germany \\
$^3$UJF-Grenoble 1/CNRS-INSU, Institut de Plan{\'e}tologie et d'Astrophysique de Grenoble, UMR 5274, 38041 Grenoble, France\\
$^4$Cavendish Laboratory, J. J. Thomson Avenue, Cambridge CB3 0HE, UK\\
$^5$College of Science, George Mason University, Fairfax, VA 22030, USA\\
$^6$School of Physical Sciences, Dublin City University, Glasnevin, Dublin 9, Ireland\\ 
$^7$DIAS, Fitzwiliam Place 31, Dublin 2, Ireland\\
$^8$Tata Institute of Fundamental Research, Mumbai 400005, India
}

\date{Accepted 2018 June 14. Received 2018 June 12; in original form 2018 April 20}

\pagerange{\pageref{firstpage}--\pageref{lastpage}}
\pubyear{2018}

\begin{document}

\maketitle

\label{firstpage}

\begin{abstract}
We study high-energy $\gamma$-rays observed from Cyg X-3 by the \textit{Fermi}\/ Large Area Telescope and the 15-GHz emission observed by the Ryle Telescope and the Arcminute Microkelvin Imager. We measure the $\gamma$-ray spectrum averaged over strong flares much more accurately than before, and find it well modelled by Compton scattering of stellar radiation by relativistic electrons with the power law index of $\simeq$3.5 and a low-energy cutoff at the Lorentz factor of $\sim\!10^3$. We find a weaker spectrum in the soft spectral state, but only upper limits in the hard and intermediate states. We measure strong orbital modulation during the flaring state, well modelled by anisotropic Compton scattering of blackbody photons from the donor by jet relativistic electrons. We discover a weaker orbital modulation of the 15 GHz radio emission, which is well modelled by free-free absorption by the stellar wind. We then study cross-correlations between radio, $\gamma$-ray and X-ray emissions. We find the cross-correlation between the radio and $\gamma$-ray emissions peaks at a lag less than 1 d, while we detect a distinct radio lag of $\sim$50 d with respect to the soft X-rays in the soft spectral state.  
\end{abstract}

\section{Introduction}
\label{intro}

Cyg X-3, one of the first discovered X-ray binaries \citep{giacconi67}, is a unique and puzzling system. The nature of its compact object remains uncertain; \citet*{zmb13} considered the radial velocity measurements of \citet*{hanson00} and \citet{vilhu09} as well as constraints from the donor mass-loss rate and the orbital-period change and obtained the compact-object mass range of $M_{\rm c}\simeq 2.4_{-1.1}^{+2.1} \msun$. \citet{koljonen17} did not confirm the measurements of \citet{hanson00} in their IR spectroscopic measurements, and noted it was possible that the velocity amplitude of \citet{hanson00} traced the motion of the stellar wind rather than of the star. Nevertheless, \citet{koljonen17} found the most likely mass range of $M_{\rm c}\lesssim 5\msun$. Thus, the current constraints allow either a neutron star or a low-mass black hole (BH). The presence of a BH appears to be favoured by considering various aspects of the X-ray and radio emission \citep*{h08,h09,sz08,szm08,koljonen10,koljonen18}. Also, \citet*{zmg10} have shown that the differences between the shapes of the X-ray spectra of Cyg X-3 in its hard spectral state from those of confirmed accreting BH binaries can be accounted for by Compton scattering in the strong stellar wind from the donor, which also would account for the lack of high frequencies in its power spectra \citep*{alh09}. On the other hand, \citet*{burke17} showed that weakly-magnetized neutron-star X-ray binaries in the hard state have similar spectral properties to their BH counterparts but lower electron temperatures and softer spectra (which confirms some previous studies). This may be compatible with Cyg X-3 hosting a neutron star. 

\begin{figure*}
\centerline{\includegraphics[width=18cm]{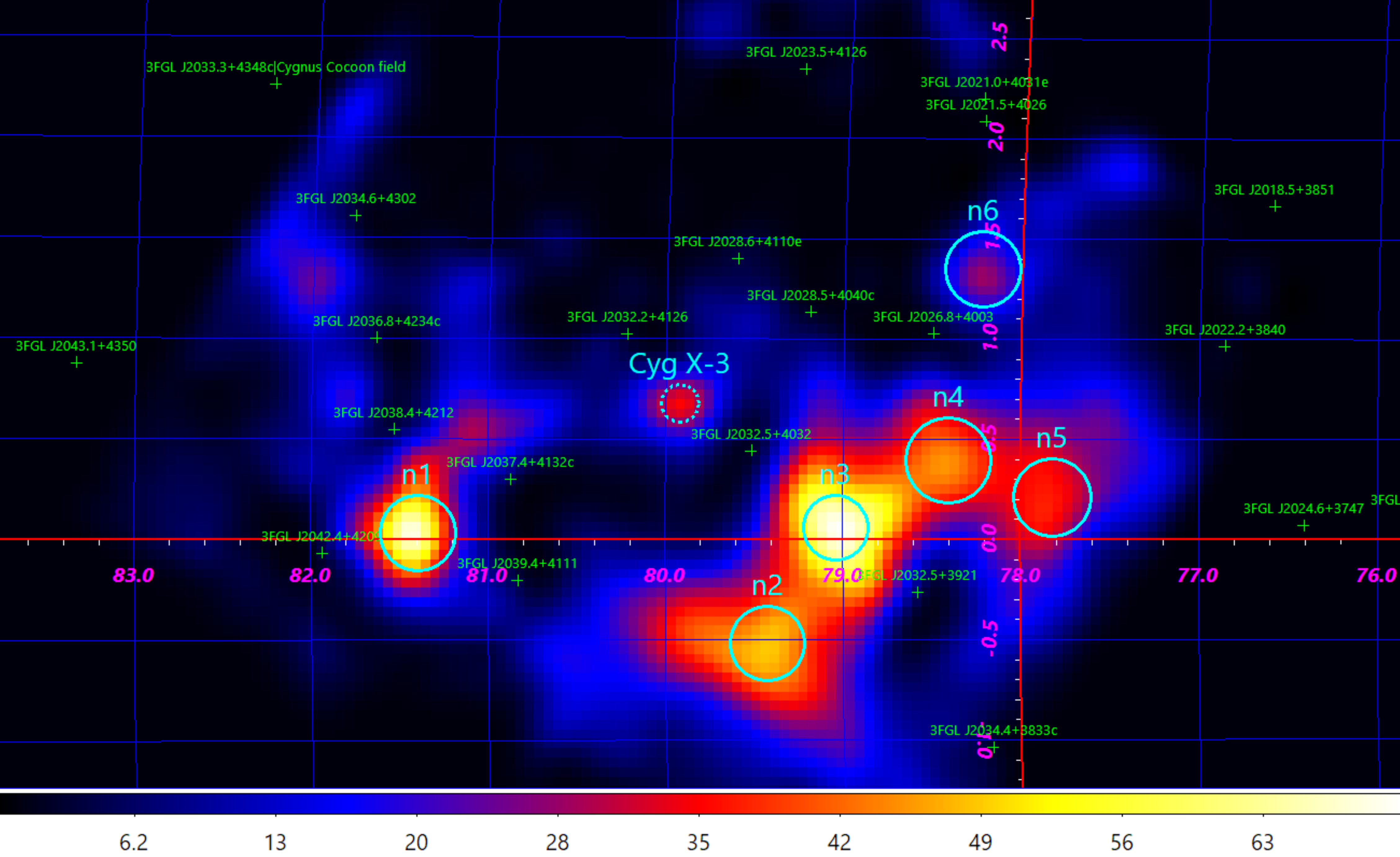}} 
\caption{The TS map (Galactic coordinates) at energies $\geq$1 GeV for the data within a $5\degr\times 5\degr$ square around the position of Cyg X-3 with the 3FGL sources subtracted, with their positions marked by the green crosses. The apparently detected sources not present in 3FGL are marked with cyan circles and denoted {\tt n1}--{\tt n6} (see Section \ref{fermi}). Cyg X-3 is marked by the dotted cyan circle in the centre.
} \label{map}
\end{figure*}

Cyg X-3 is the only known binary in the Galaxy containing both a compact object and a Wolf-Rayet star \citep*{v92,v96,v93,fender99}. Given its very short period of $P\simeq 0.2$ d (unusual for a high-mass binary), it is a likely progenitor of a close double degenerate system, after the donor explodes as a supernova \citep{belczynski13}. The merger will then be associated with emission of gravitational waves, which has important implications for the detectability of similar extragalactic systems by LIGO and VIRGO \citep{belczynski13}. 

The most recent distance estimate is the geometric one from dust scattering halos of $D\simeq 7.4\pm 1.1$ kpc \citep*{McCollough16}, which agrees well with the estimate of $7.2^{+0.3}_{-0.5}$ kpc of \citet*{lzt09} obtained with the same method. A similar distance range is also preferred based on considering constraints on the donor mass \citep{koljonen17}. At this $D$, its absorption-corrected bolometric X-ray luminosity reaches several times $10^{38}$ erg s$^{-1}$ in its brightest (soft) state, i.e., it reaches the Eddington limit for a $4\msun$ BH and exceeds it for a neutron star (\citealt*{z16}, hereafter ZSP16). Among X-ray binaries, Cyg X-3 is the brightest and most highly variable radio source \citep{mccollough99}, also showing resolved jets \citep{m01,miller_jones04,tudose07,egron17} on the size scale from a few up to several tens of mas (at 7 kpc, 25 mas corresponds to the projected distance of 1 light day). However, larger radio structures are also observed \citep*{marti01} on the scale of 1 arcsec (corresponding to 40 light days). 
 
Its high-energy (HE) \g-ray emission has been discovered by the \fermi\/ Large Area Telescope (LAT; \citealt{atwood09}) and \agile\/ \citep{tavani09} in the soft spectral state (\citealt{fermi09}, hereafter FLC09; \citealt{agile}). The GeV power-law emission and its orbital modulation appear to be due to Compton up-scattering of the stellar emission from the companion WR star by relativistic electrons in the jet \citep*{dch10b}. Cyg X-3 is one of only two X-ray binaries that are certainly powered by accretion for which HE \g-ray emission has been detected at a high statistical significance; the other one being Cyg X-1 (also a high-mass X-ray binary, hereafter HMXB), where the \g-ray emission is, however, much weaker \citep{zanin16,z17}. Among low-mass X-ray binaries, a \g-ray flare from V404 Cyg was detected at a $\sim\! 4\sigma$ significance during the 2015 outburst \citep{loh16}. The relatively strong radio and HE \g-ray emissions in Cyg X-3 may be due to interaction of the jet with the stellar wind, which is very dense near the compact object in this close Wolf-Rayet system, and subsequent formation of recollimation shocks (e.g., \citealt*{yoon16} and references therein).

The presence of a powerful jet in soft states in Cyg X-3 is significantly different from the behaviour of accreting BH low-mass X-ray binaries (LMXBs), where a short-duration transient jet can appear during hard-to-soft transitions \citep*{fender04}, and the radio emission is strongly quenched in the soft state \citep{corbel00}. As shown by \citet{koljonen10}, strong radio flares in Cyg X-3 occur during the transition from the softest (hypersoft) states to harder ones, in the opposite direction and at much lower hardness ratios than those in LMXBs. On the other hand, accreting BH binaries in the hard state feature a steady compact jet, emitting partially self-absorbed synchrotron radio-mm-IR emission. This is then similar to Cyg X-3, which has a hard state with relatively strong radio emission correlated with soft X-rays, which correlation is similar to that in BH binaries (\citealt{corbel13a}; ZSP16). Therefore, we can expect some HE \g-ray emission of Cyg X-3 in its hard state, similar to the case of Cyg X-1.

In this work, we obtain HE \g-ray spectra in the flaring, hard, intermediate and soft states. Furthermore, we present and study 15 GHz monitoring data from the Ryle Telescope and the Arcminute Microkelvin Imager (AMI). The latter data cover the entire duration of the LAT observations analysed here. We measure and model the orbital modulation of both radio and HE \g-ray emission. We then study correlations between the \g-rays, radio and X-ray emission.

\section{Data}
\label{data}

\subsection{\textit{FERMI}\/ LAT data reduction}
\label{fermi}

\begin{figure*}
\centerline{\includegraphics[width=16.5cm]{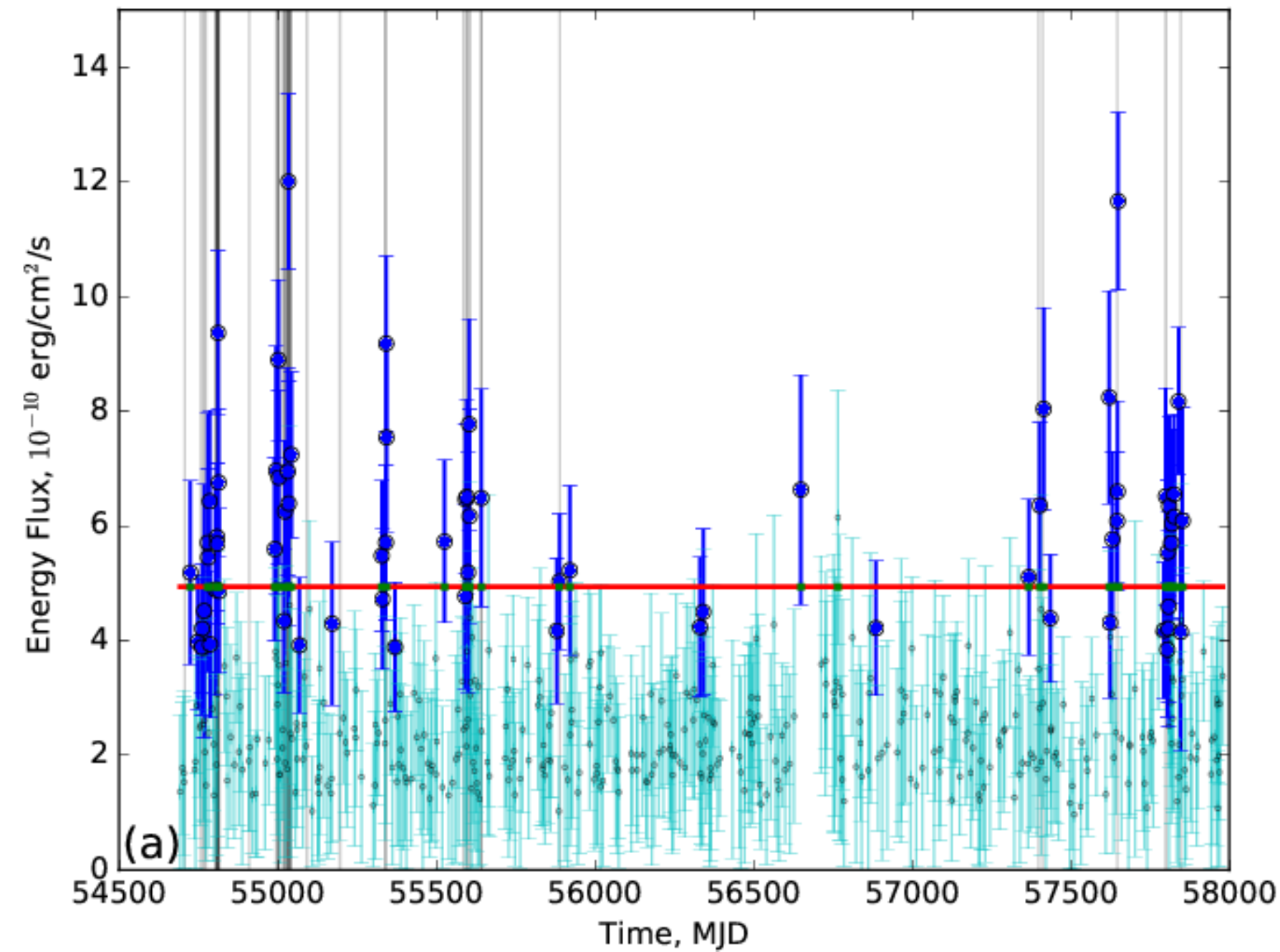}} 
\centerline{
\includegraphics[height=8.5cm]{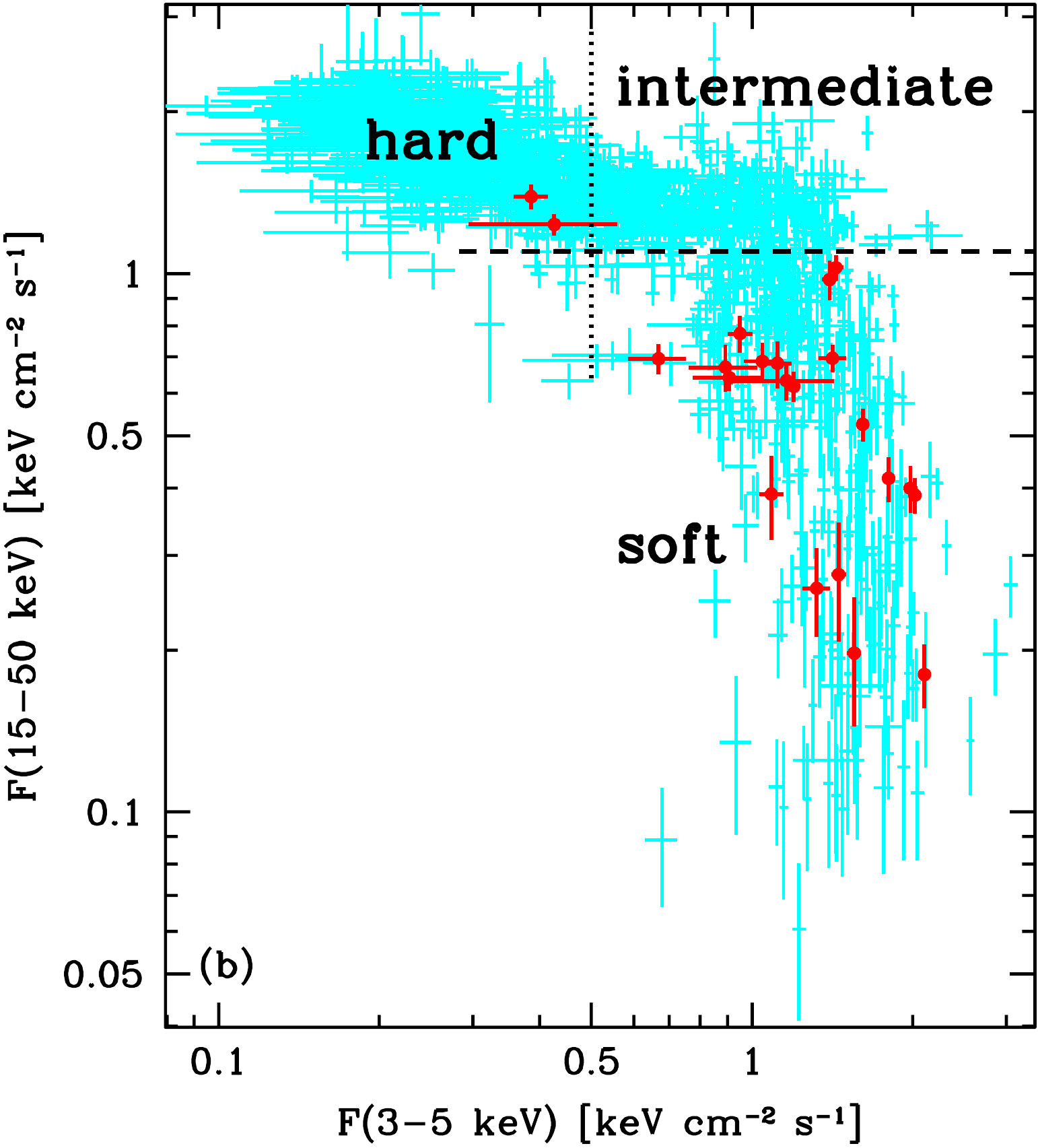}{\hskip 0.3cm}
\includegraphics[height=8.5cm]{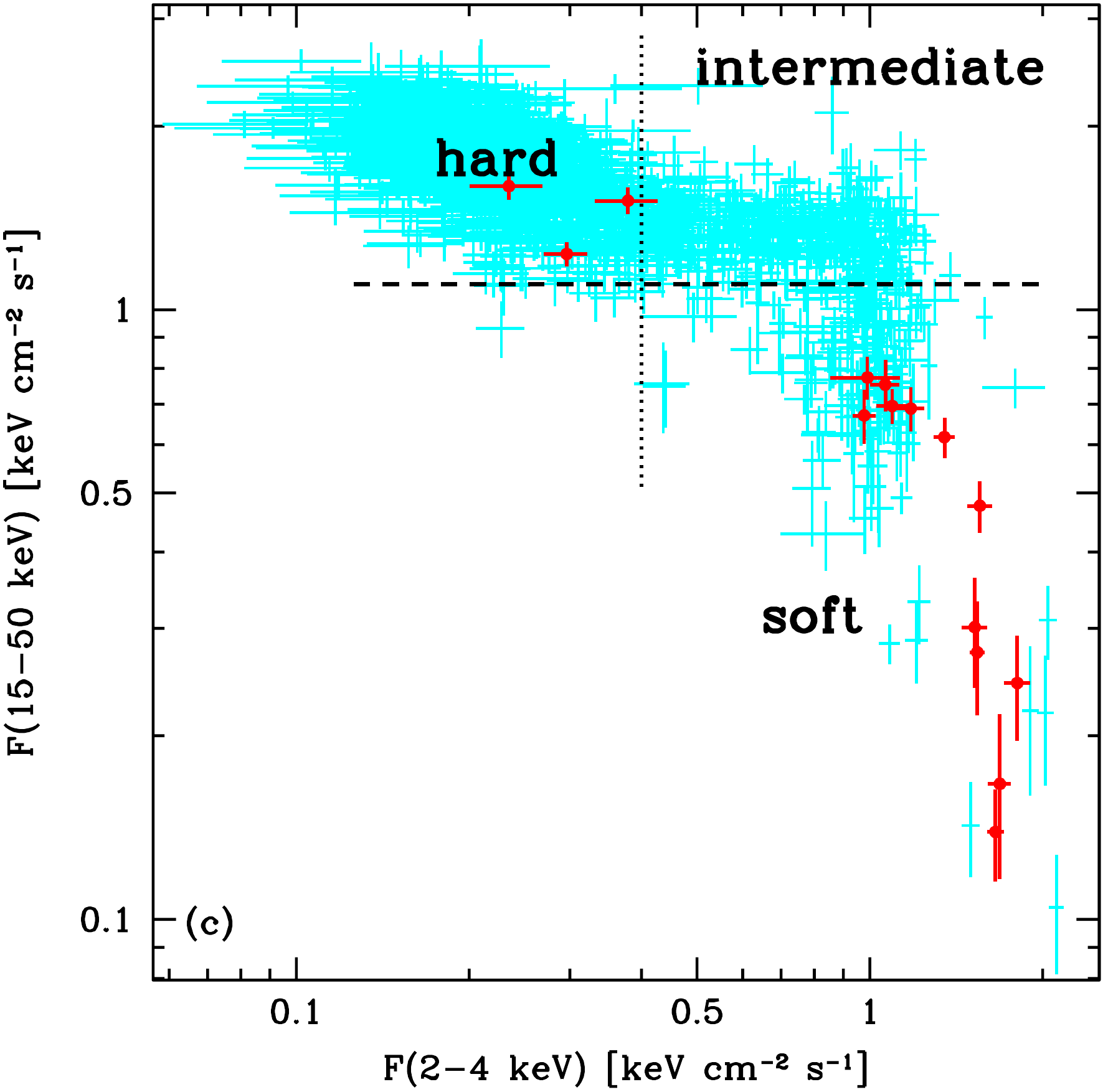}}
\caption{(a) The 1-d bin \g-ray light curve with detections by the LAT. Our criterion defining the flaring state for 1-d integration, $F(0.1$--100\,GeV$) > 4.94\times 10^{-10}$ erg cm$^{-2}$ s$^{-1}$, is shown by the red line. The grey vertical lines correspond to the previously published detections. The large blue circles and small cyan circles with flux error bars correspond to the days with TS $\geq$16 and $<$16, respectively. (b) The relationship between the daily-averaged energy fluxes in the 3--5 keV (ASM) and 15--50 keV (BAT) ranges. The hard and soft states are defined here by $F(3$--5 keV$)< 0.5$ keV cm$^{-2}$ s$^{-1}$ (corresponding to the count rate $\lesssim 2.7$ s$^{-1}$) and $F(15$--50 keV$)< 1.1$ keV cm$^{-2}$ s$^{-1}$ (corresponding to the count rate $\lesssim 0.028$ cm$^{-2}$ s$^{-1}$), respectively. These boundaries are marked by the dotted and dashed lines, respectively. The intermediate state corresponds to both the 3--5 keV and 15--50 keV fluxes above the respective boundaries. Only points with statistical significance $>3\sigma$ (required for each of the fluxes) are shown. The thick red circles with error bars correspond to MJDs with LAT flares, defined as above. (c) The same except that the 2--4 keV flux from MAXI is used. The adopted hard-state condition (the dotted line) is here $F(2$--4 keV$)< 0.4$ keV cm$^{-2}$ s$^{-1}$ (corresponding to the photon flux $\lesssim 0.14$ cm$^{-2}$ s$^{-1}$). 
}
\label{states}
\end{figure*}

We have analysed the available \fermi/LAT data (MJD 54682--57982) coming from the direction of Cyg X-3 using the latest version of the Fermi Science Tools (v10r0p5) with the P8R2\_CLEAN\_V6 instrument response functions. We have used the standard value of the zenith angle cut of $z_{\rm max}=90\degr$. 

Similarly to FLC09, we have considered the presence of the nearby \g-ray pulsar PSR J2032+4127, located about $30'$ away from Cyg X-3. That pulsar appears to be a member of a highly eccentric ($e\sim 0.95$), long-period ($\sim\! 25$--50 yr), binary with a massive Be star \citep{lyne15,takata17}, and the orbital motion causes strong variations of the pulse period. The spin ephemeris is given by \citet{lyne15}, and according to it the pulse curve of PSR J2032+4127 is dominated by two strong peaks at phases 0.13--0.19 and 0.62--0.70. However, PSR J2032+4127 is now approaching the periastron, which resulted in strong departures of the ephemeris from that of \citet{lyne15} after 2015. Based on the LAT data, we have been unable to update the ephemeris. Therefore, we have been unable to use the approach of FLC09 of using only the off-pulse intervals. Instead, we have relied on taking into account the emission of the pulsar (along with other sources in the region) in our fitting of the LAT data (see below). We also note that the pulsar flux contribution at the position of Cyg X-3 is much lower than that of Cyg X-3 when it is in the flaring state, and also no GeV flux enhancement from PSR J2032+4127 was observed while approaching the periastron \citep{takata17}. Still, we have compared the results for the Cyg X-3 spectra in different states obtained using all the data and those subtracting the pulsar peaks according to \citet{lyne15}, and found virtually no differences. Therefore, we have used the entire data in this work.

In order to take account of the broad \fermi/LAT point-spread function (PSF) at energies studied by us (80~MeV--300~GeV), we consider a large, $25\degr \times 25\degr$, region of interest (ROI) around the Cyg X-3 position. We include in the modelling all sources within the ROI from the 4-year \fermi\/ catalogue (\citealt{3rdcat}; 3FGL). We have used the standard templates for the Galactic (\texttt{gll\_iem\_v06.fits}) and extragalactic (\texttt{iso\_P8R2\_CLEAN\_V6\_v06.txt}) diffuse backgrounds. The catalogue sources were assumed to be described by the 3FGL spectral models with all parameters except the normalizations frozen to their catalogue values. In order to avoid possible systematic effects, we have also included into the model the 3FGL catalogue sources located up to $10\degr$ beyond the ROI with all parameters fixed to their catalogue values. For Cyg X-3, we adopt the power law model with the slope 2.7 reported previously in FLC09. The spectral analysis has been performed with the python tools\footnote{\url{fermi.gsfc.nasa.gov/ssc/data/analysis/scitools/python\_tutorial.html}.}. The upper limits are calculated with the \texttt{IntegralUpperLimits} python module for detection significances of TS (test statistic; see \citealt{mattox96}) $<4$ , which correspond to a 95 per cent ($\simeq 2\sigma$) probability for the energy flux to be lower than the limit.

We then built the TS map in a $5\degr \times 5\degr$ region around the position of Cyg X-3 in the 1--300~GeV energy band, see Fig.\ \ref{map}. We see a number of residuals along the Galactic plane, which we mark as {\tt n1}--{\tt n6}. Since almost all of them are at very low Galactic latitudes, where we expect the highest uncertainties, most of them seem to be diffuse residuals unaccounted for in the diffuse background model. The only residual that can be identified in any catalogue is {\tt n1}, which was present in the first \fermi\/ catalogue \citep{fermi1}, but it disappeared in 3FGL, and which appears to be associated with an H{\sc ii} region \citep*{munar11}. The map also reveals a weak point-like source at the catalogue position of Cyg X-3 with TS $\simeq 45$, corresponding to a $\gtrsim\! 6\sigma$ detection significance. Hereafter, we use the HEASARC catalogue position of Cyg X-3, (RA, Dec) = (308.107420 ; 40.957750). We also note that the Cyg X-3 position is actually consistent with that of J2032+4050 as given in the preliminary 8-yr \fermi/LAT source list\footnote{\url{https://fermi.gsfc.nasa.gov/ssc/data/access/lat/fl8y/}} (which appeared when the paper was in final stages of preparation). 

The spectral analysis in 0.08--300~GeV was performed in a set of narrow energy bins. For each energy bin, we have iteratively removed weak (TS $<1$) sources other than Cyg X-3, and then have redone the fit until no weak sources remain.

In addition, we also consider a lower energy band available to the LAT of 40--80~MeV (30--104~MeV accounting for the energy dispersion) for our brightest (flaring) spectrum, where we find an upper limit. For that energy range, we employ a method similar to that used in \citet{z17}. This range is not covered by the standard templates for Galactic and isotropic diffuse emission. Therefore, we base our analysis for the Galactic background on three different templates, \slm, \ssm and \sym, produced with the GALPROP code \citep{galprop}. Those templates are known to describe \fermi/LAT data at higher energies reasonably well \citep{ackermann12}. The spectrum of the standard isotropic background model is available down to 34~MeV, which almost covers the analysed energies, and we employ a power law continuation of that spectrum down to 30~MeV. We also use a low value zenith angle cut of $z_{\rm max}=70\degr$, instead of the standard value of $90\degr$ (which we adopt at higher energies). 

We have then performed timing analysis for Cyg X-3 in 1-d bins in a broad energy range of 0.1--10~GeV in a way similar to the above described binned spectral analysis with iterative elimination of weak sources. Fig.~\ref{states}(a) shows the 1-d bin light curve of the LAT detections. We find 486 days with the signal-to-noise ratios (SNR) $>1$ and 174 days with SNR $>2$. The detections with TS $\geq 16$ and $<16$ are plotted in blue and cyan, respectively. We confirm most of the previous detections by the LAT and \agile\/ (FLC09; \citealt{williams11,corbel12,fermi16,cheung16,fermi17a,fermi17b,piano12,agile, bulgarelli12,piano16b,piano17a,piano17b,bodaghee13,williams11}), which we show by the grey vertical lines. We also find a number of new detections. We then split the days with detections into two energy flux regions, the high-flux region which we call the flaring state and the other, with detections at lower fluxes. The boundary between the two regions, $F_{\rm b}$, is selected with the iterative procedure defined as following. At each iteration, we define the mean level  $F_{\rm low}$, and the standard deviation, $\sigma_{\rm low}$ of all detections defined as non-flaring at the previous iteration. If any, we mark all detections with a flux higher than $F_{\rm b}=F_{\rm low}+3\sigma_{\rm low}$ as flares and continue to the next iteration. The iterations stop when no new detections are attributed to the flaring state. Using this procedure, $F_{\rm b}$ is found to equal $4.94\times 10^{-10}$ erg cm$^{-2}$ s$^{-1}$, which is shown by the red horizontal line in Fig.\ \ref{states}(a). We find 49 days with the energy flux $\geq$ this limit, and we list them in Table \ref{flares}.

\begin{table}
\centering
\caption{The MJD days corresponding to the flaring state defined in Section \ref{fermi} and Fig.\ \ref{states}(a), with $F_\gamma(0.1$--$100\,{\rm GeV})>4.94\times 10^{-10}$ erg cm$^{-2}$ s$^{-1}$. The \g-ray flux and the TS are averages over a given MJD, and the letters S, I and H correspond to the soft, intermediate and hard state, respectively. \label{flares}
}
\begin{tabular}{cccc}
\hline
MJD & $F_\gamma$  [$10^{-10}$ erg cm$^{-2}$ s$^{-1}$] & TS & State \\ 
\hline 
54725 & $5.2 \pm 1.6$ & 18.5 & H \\ 
54780 & $5.7 \pm 2.3$ & 22.7 & S \\ 
54781 & $5.5 \pm 1.5$ & 23.3 & S \\ 
54786 & $6.4 \pm 1.6$ & 35.3 & S \\ 
54809 & $5.8 \pm 1.3$ & 36.8 & S \\ 
54810 & $5.7 \pm 1.4$ & 32.3 & S \\ 
54812 & $9.4 \pm 1.4$ & 83.0 & S \\ 
54814 & $6.8 \pm 1.3$ & 50.4 & S \\ 
54991 & $5.6 \pm 1.6$ & 18.5 & S \\ 
54995 & $7.0 \pm 2.2$ & 29.0 & S \\ 
55002 & $8.9 \pm 1.4$ & 67.0 & S \\ 
55003 & $6.8 \pm 1.5$ & 43.8 & S \\ 
55023 & $6.3 \pm 1.2$ & 39.9 & S \\ 
55032 & $7.0 \pm 1.8$ & 28.2 & S \\ 
55034 & $12.0\pm 1.5$ & 119.2 & S \\ 
55035 & $6.4 \pm 2.1$ & 38.7 & S \\ 
55043 & $7.2 \pm 1.5$ & 40.4 & S \\ 
55328 & $5.5 \pm 1.3$ & 29.5 & S \\ 
55341 & $5.7 \pm 1.3$ & 28.4 & S \\ 
55342 & $9.2 \pm 1.5$ & 61.5 & S \\ 
55343 & $7.5 \pm 1.6$ & 38.7 & S \\ 
55526 & $5.7 \pm 1.4$ & 24.8 & H \\ 
55592 & $6.5 \pm 1.3$ & 40.0 & S \\ 
55596 & $6.5 \pm 1.7$ & 31.1 & S \\ 
55600 & $5.2 \pm 2.1$ & 17.6 & S \\ 
55604 & $7.8 \pm 1.8$ & 37.2 & S \\ 
55605 & $6.2 \pm 1.9$ & 23.9 & S \\ 
55642 & $6.5 \pm 1.9$ & 40.5 & S \\ 
55888 & $5.0 \pm 1.2$ & 29.6 & I \\ 
55921 & $5.2 \pm 1.5$ & 19.0 & H \\ 
56649 & $6.6 \pm 2.0$ & 19.3 & H \\ 
56766 & $6.1 \pm 2.2$ & 11.5 & I \\ 
57367 & $5.1 \pm 1.4$ & 19.7 & S \\ 
57402 & $6.4 \pm 1.4$ & 34.0 & S \\ 
57414 & $8.0 \pm 1.8$ & 37.3 & S \\ 
57621 & $8.2 \pm 1.9$ & 35.9 & S \\ 
57631 & $5.8 \pm 1.5$ & 24.4 & S \\ 
57646 & $6.1 \pm 1.2$ & 39.0 & S \\ 
57647 & $6.6 \pm 1.6$ & 41.0 & S \\ 
57649 & $11.7\pm 1.5$ & 108.6 & S \\ 
57799 & $6.5 \pm 1.9$ & 42.3 & S \\ 
57805 & $5.5 \pm 1.3$ & 35.0 & S \\ 
57810 & $6.3 \pm 1.6$ & 35.7 & S \\ 
57816 & $5.7 \pm 1.4$ & 32.0 & S \\ 
57818 & $6.0 \pm 1.7$ & 42.9 & S \\ 
57825 & $6.6 \pm 1.4$ & 46.3 & S \\ 
57826 & $6.2 \pm 1.5$ & 31.7 & S \\ 
57839 & $8.2 \pm 1.3$ & 63.1 & S \\ 
57852 & $6.1 \pm 2.0$ & 15.9 & S \\ 
\hline
\end{tabular}
\end{table}

\begin{figure*}
\centerline{\includegraphics[height=7.7cm]{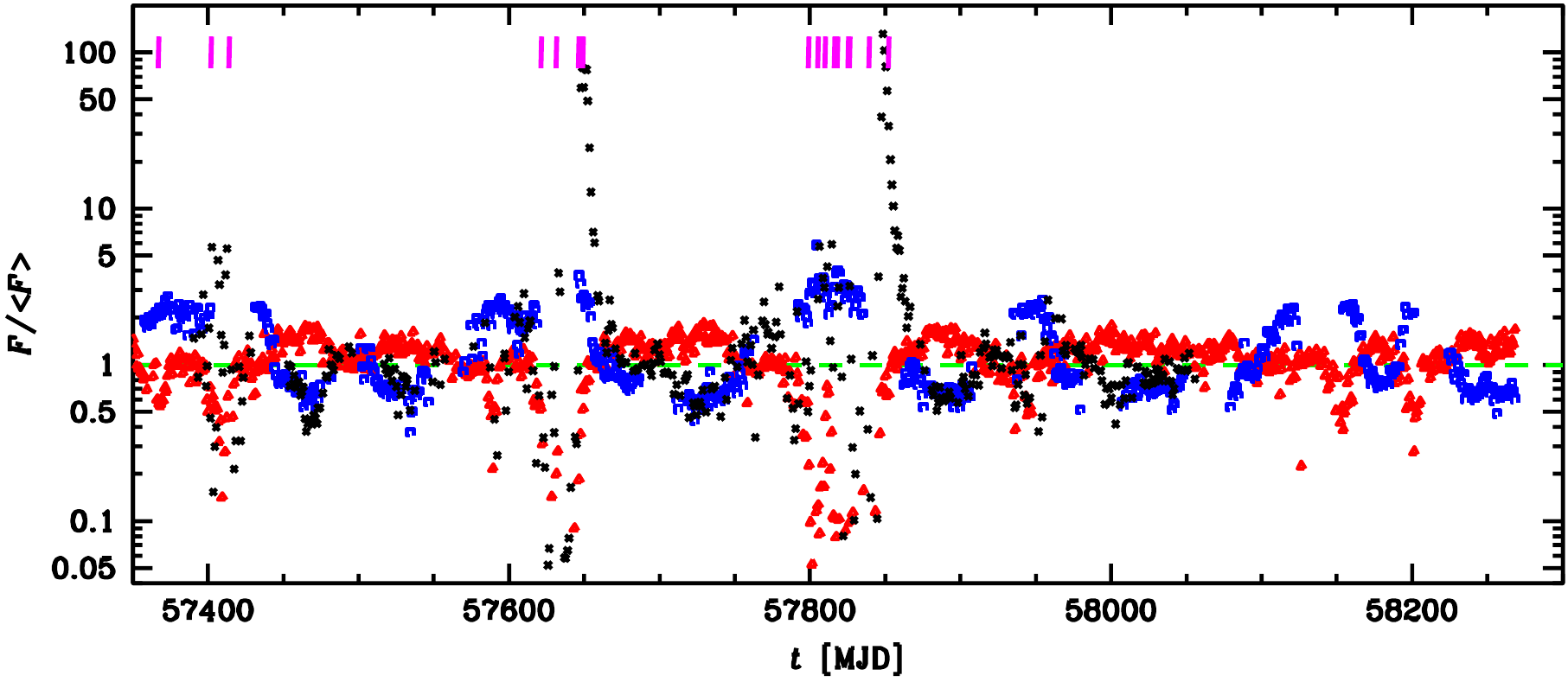}}
\caption{The recent light curves of Cyg X-3 normalized to the respective average over the total observation length. The blue squares, red triangles and black crosses show the rates normalized to the averages over all available data for the MAXI (2--10 keV), BAT (15--50 keV) and AMI (15 GHz), equal to 0.37 cm$^{-2}$ s$^{-1}$, 0.0338 cm$^{-2}$ s$^{-1}$, 0.104 Jy, respectively. Only points with the significance $\geq\! 2\sigma$ are shown; for clarity of display, the error bars are not plotted. The dashed green line corresponds to the averages. The heavy magenta vertical lines correspond to the flares of HE \g-ray emission, as defined in Section \ref{fermi}. We note that the last days of the HE \g-ray and 15-GHz data analyzed in this work are MJD 57982 and 58055, respectively.
}
\label{lcX}
\end{figure*}

We then divide the available LAT observations into the hard, intermediate and soft states based on the daily-averaged data from All-Sky Monitor (ASM; \citealt*{brs93,levine96}) on board {\it Rossi X-ray Timing Explorer\/}, the Monitor of All-sky X-ray Image (MAXI; \citealt{matsuoka09}) on board {\it International Space Station}, and the Burst Alert Telescope (BAT; \citealt{barthelmy05,m05,krimm13}) on board \swift. The long-term light curves from those detectors are given in ZSP16, and they are updated in Fig.\ \ref{lcX}. In order to determine the states, we use a method similar to that in \citet{z12a} and ZSP16, except that we use here the public 15--50 keV BAT data\footnote{\url{http://swift.gsfc.nasa.gov/results/transients/CygX-3/}} instead of custom data used in those papers. We convert the ASM and BAT count rates and the MAXI photon fluxes into energy fluxes using the scaling to the Crab, assuming its spectrum as given in ZSP16. Figs.\ \ref{states}(b--c) shows the BAT flux vs.\ the 3--5 keV ASM and 2--4 keV MAXI fluxes. The flux regions delineating the states are defined by the dashed lines. The 3--5 keV boundary of the hard state corresponds to the maximum flux with a positive correlation with the radio emission (ZSP16) and the hard/soft X-rays anti-correlation, and the 15--50 keV boundary approximately corresponds to the lowest fluxes of the hard state. For days without both soft and hard X-ray data, we interpolate the available X-ray data to infer the spectral state, as well as use the 15 GHz data and the radio/X-ray correlations as given in ZSP16. This yields 3074, 736 and 446 days with LAT coverage in the hard, intermediate and soft state, respectively.

Then, we find 43, 2 and 4 days with \g-ray flares (defined as above) in the soft, intermediate and hard state, respectively. The points corresponding to the flares for the days with both soft and hard X-ray coverage are shown in red in Figs.\ \ref{states}(b--c). They show the four flaring days in the hard state, where one point appears on both panels, i.e., it has both simultaneous ASM/BAT and MAXI/BAT coverages. The occurrence of the intermediate state for two flares was determined by interpolating the X-ray data (see above); thus, those days do not appear in Figs.\ \ref{states}(b--c). We see that while most of the days with strong \g-ray detections correspond to the soft state, there are still several detections during the intermediate and hard states, i.e., with low soft X-ray energy fluxes and high hard X-ray ones.

\subsection{The radio data}
\label{radio}

We study here radio monitoring data at 15 GHz from the Ryle Telescope, which cover MJD 49231--53905 (74181 measurements), and the Arcminute Microkelvin Imager (AMI), MJD 54612--58055 (5125 measurements). The combined data set contains 79306 measurements. The AMI Large Array is the re-built and reconfigured Ryle Telescope. \citet{pf97} describe the normal operating mode for the Ryle telescope in the monitoring observations; the observing scheme for the AMI Large Array is very similar. The new correlator has a useful bandwidth of about 4 GHz (compared with 0.35 GHz for the Ryle), but the effective centre frequency is similar. 

In order to establish the calibration parameters of the array, we use observations of a bright, nearby unresolved source interleaved with those of the main target source. Our primary calibrators were 3C 48 and 3C 286. This procedure resulted in variations in the flux calibration limited to $\la 10$ per cent from one day to another.

Throughout this work, we consider the radio emission as coming from the source associated with the accretion/outflow associated with the compact object, presumably the jet. Still, some contribution to that emission comes from the stellar wind. However, it has to be minor, given the very low radio flux levels the source achieves. Another possible contribution is from the stellar wind interacting with disc winds \citep{koljonen18}. In this work, we will not distinguish this contribution from that from the jet, given that we have only 15 GHz radio fluxes at our disposal.

\section{LAT Spectra in different states}
\label{spectra}

\begin{figure}
\centerline{\includegraphics[width=\columnwidth]{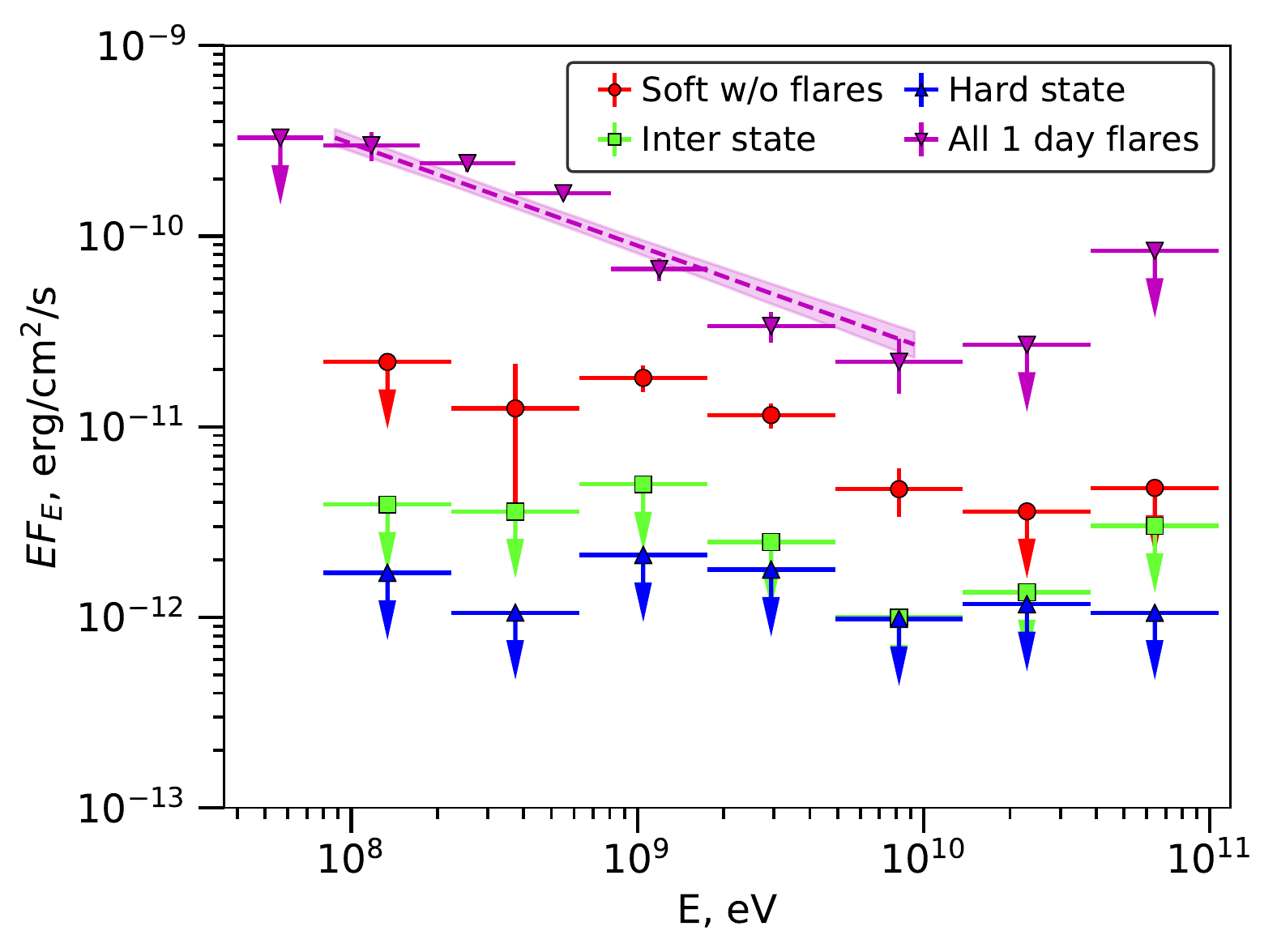}} 
\caption{The \fermi\/ LAT spectra and upper limits for the sum of all single MJDs with strong \g-ray detections (the flaring state, as defined in Fig.\ \ref{states}a), and the soft (excluding the flaring days), intermediate and hard states, shown by the magenta inverted triangles, red circles, green squares and blue triangles, respectively, with associated error bars, and arrows indicating upper limits. The dashed line and the shaded region show the best power-law fit to the flaring-state spectrum and its uncertainties.
} \label{spectra_states}
\end{figure}

\begin{figure}
\centerline{\includegraphics[width=\columnwidth]{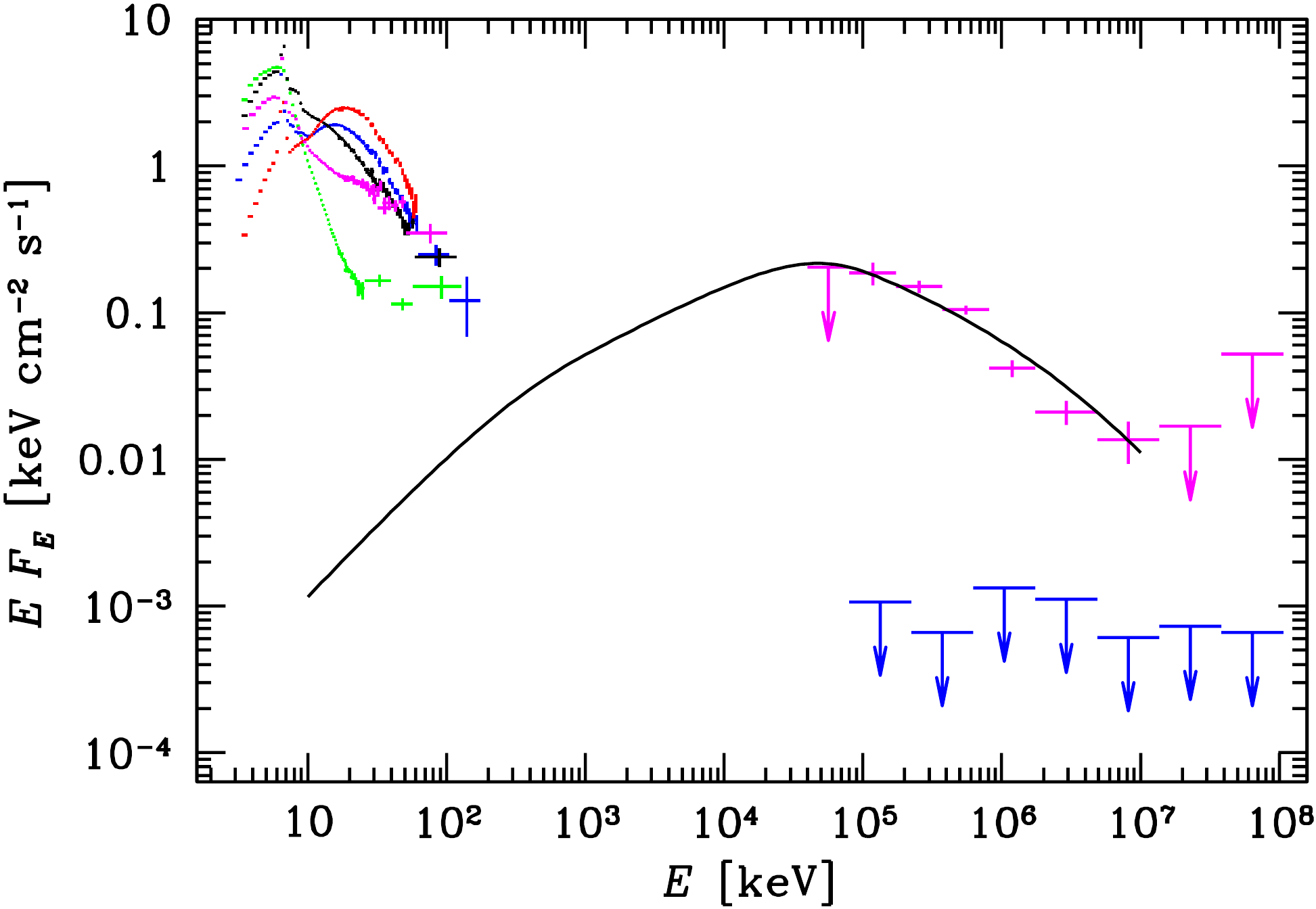}} 
\caption{The \fermi\/ LAT measurements and upper limits in the flaring state (upper \g-ray error bars) and the upper limits in the hard state (lower \g-ray error bars) compared to the average \xte\/ spectra of Cyg X-3 in the five spectral states of \citet{szm08}. The X-ray hard-state spectra are the red and blue ones, i.e, with the two top fluxes at 20 keV and bottom at 4 keV, and the remaining spectra correspond to our intermediate and soft states. The solid curve shows the Compton scattering model spectrum from \citet{z12b} with the electrons with a power-law distribution with the index of $p=3.5$ above the low-energy break at $\gamma_1=1300$ and extending up to $\gamma_2\rightarrow \infty$. The electrons scatter stellar blackbody photons with the temperature of $10^5$ K. The fitted normalization is larger by a factor 2.1 than that of \citet{z12b}.}
\label{spectrum_model}
\end{figure}

We have calculated the average LAT spectra in the hard, intermediate, soft and flaring states, as defined in Section \ref{fermi}, except that we have excluded the flaring days from the soft-state spectrum. Our results are shown in Fig.\ \ref{spectra_states}. 

We have detected the source with the significance of $\sim\!40\sigma$ during the flaring state in the 0.08--15 GeV range. A power-law fit of the 0.08--10 GeV range gives a photon spectral index of $\Gamma=2.55\pm 0.05$ at the normalization at 100 MeV of $1.9 \pm 0.2\times 10^{-8}$ MeV$^{-1}$ cm$^{-2}$ s$^{-1}$, which corresponds to the energy flux above 100 MeV of $5.5\times 10^{-10}$ erg cm$^{-2}$ s$^{-1}$. This power-law spectrum is significantly more accurately determined and slightly harder than the $\Gamma=2.70\pm 0.25$ of FLC09 (measured during MJD 54750--54820 and 54990--55045), and the integrated energy flux is slightly larger than their value of $4.0\pm 1.6\times 10^{-10}$ erg cm$^{-2}$ s$^{-1}$ at their best fits. Nevertheless, both $\Gamma$ and the flux are consistent with the FLC09 values within the uncertainties.

However, there is a visible curvature in the present flaring-state spectrum, and we have also fitted it by a log-normal distribution in the form of ${\rm d}N/{\rm d}E=N_0 (E/ E_{\rm b})^{-2} 10^{-\beta \log^2_{10}(E/E_{\rm b})}$, where $E_{\rm b}$ is the peak of it in $E F_E$. (Note that this form is equivalent to a parabola in logarithmic coordinates, see \citealt{z16a}.) We obtain $E_{\rm b}=384\pm 15$ MeV, $\beta=0.351 \pm 0.005$ and $N_0=1.36\pm 0.09 \times 10^{-9}$ cm$^{-2}$ s$^{-1}$. We find the log-normal/log-parabola fit is strongly preferable to the power-law fit, with $\Delta\chi^2\simeq -59$ for adding one free parameter, which corresponds to a significance of $\sigma\simeq 7$--8 of the presence of a curvature. We note that the log-normal model is sharply cut off at $\ga$2 GeV and it is thus much below the last detected spectral point.

We still detect Cyg X-3 in the soft state outside the flaring days in the 0.2--15 GeV range at a lower flux than that in the flaring state. This spectrum is parallel to the flaring-state one above $\sim$1 GeV, but we see a hardening at lower energies. We do not detect the source in the intermediate state. Contrary to our original expectations (see Section \ref{intro}), we have not detected Cyg X-3 in the hard state, obtaining stringent upper limits. 

Fig.\ \ref{spectrum_model} shows the flaring-state spectrum together with the average X-ray spectra of Cyg X-3 from \xte\/ \citep{szm08}. In their classification, the average spectra correspond to five spectral states, from the hardest to the softest. We have compared our flaring-state spectrum to the models of \citet{z12b}, in which relativistic electrons Compton-upscatter blackbody photons emitted by the donor, and which take into account the full Klein-Nishina cross section. Among those, the model with the steady-state electron power-law index of $p=3.5$, the minimum electron Lorentz factor of $\gamma_1=1300$, the maximum one of $\gamma_2\rightarrow \infty$, and scattering stellar blackbody photons at the temperature of $10^5$ K, fits well the current spectrum with $\Delta \chi^2\simeq -23$ with respect to the above power-law fit. Only the normalization has been fitted, yielding the flux at 1 GeV of $E F_E\simeq 0.0638$ keV cm$^{-2}$ s$^{-1}$. The Lorentz factor of $\gamma_1$ corresponds to the minimum above which the electrons are accelerated with an index of $p_{\rm acc}\simeq 2.5$. Below $\gamma_1$, the electrons are from cooling by the Compton and adiabatic losses and have the distribution given by equation (21) of \citet{z12b}. The electron spectral index is harder than that corresponding to Compton scattering in the Thomson limit, $p=2\Gamma-1=4.1$ because of the Klein-Nishina decline of the Compton cross section, which softens the spectrum. The model satisfies the constraint obtained by \citet{z12a} that the contribution of the jet emission at $\sim$100 keV is minor (based on the pattern of the orbital modulation found at 50--100 keV). We confirm the result of \citet{z12b} that the magnetic field strength in the \g-ray emitting region is relatively weak, $B\lesssim 100$ G. 

Fig.\ \ref{spectrum_model} also shows the hard-state spectrum upper limits. We can see they are quite stringent, implying any jet emission in that state to have an $EF_E$ spectrum at a level a few thousand times below the peak of the hard-state X-ray spectrum. We note that the hard-state HE \g-ray spectrum of Cyg X-1 is actually about four orders of magnitude below the peak of the hard-state spectrum \citep{zanin16,z17}. So a \g-ray spectrum at a similar relative level could still be emitted by Cyg X-3 and remain undetectable. 

\section{Orbital modulation}
\label{modulation}

\subsection{The ephemeris}
\label{ephemeris}

The period of Cyg X-3 is increasing. We take it into account by using a quadratic form of the ephemeris, 
\begin{equation}
T_n=T_0+P_0 n +c_0 n^2,\quad c_0=P_0 \dot P/2,\quad P_n=P_0 + 2c_0 n,
\label{eph}
\end{equation}
where $T_n$ is the time of an $n$-th occurrence of a zero orbital phase, approximately\footnote{We note that the template of \citet{vb89} has the minimum slightly below zero phase, which is also the case for the X-ray light curves phase-folded based on a previous ephemeris in \citet{z12a}. Furthermore, the strongest X-ray absorption may not exactly correspond to the conjunction due to a likely asymmetry of the stellar wind in this short-period binary.} related to the superior conjunction, and measured from the reference time, $T_0$, $P_0$ is the period at $T_0$, $\dot P$ is the period derivative, and $P_n$ is the period at $T_n$. The most recent ephemeris is that of \citet{bhargava17},
\begin{equation}
T_0=40949.384, \, P_0=0.19968476(3)\,{\rm d},\, c_0=5.41(2)\times 10^{-11}\,{\rm d},
\label{bhargava}
\end{equation}
where hereafter the numbers in parentheses give the uncertainty of the last digit. 

The above ephemeris is given in the Terrestrial Time MJD and is based on X-ray light curves taking into account the barycentric correction (Y. Bhargava, private communication). Thus, we consider the same time format and apply the barycentric correction to the light curves used for the orbital modulation. In order to determine the orbital phase of a measurement at a time $T$, we solve equation (\ref{eph}) for $n$ treating it as a real number, and then subtract its integer part. 

\subsection{Modulation of HE \boldmath{$\gamma$}-rays}
\label{orbital_gamma}

\begin{figure}
\centerline{\includegraphics[width=8cm]{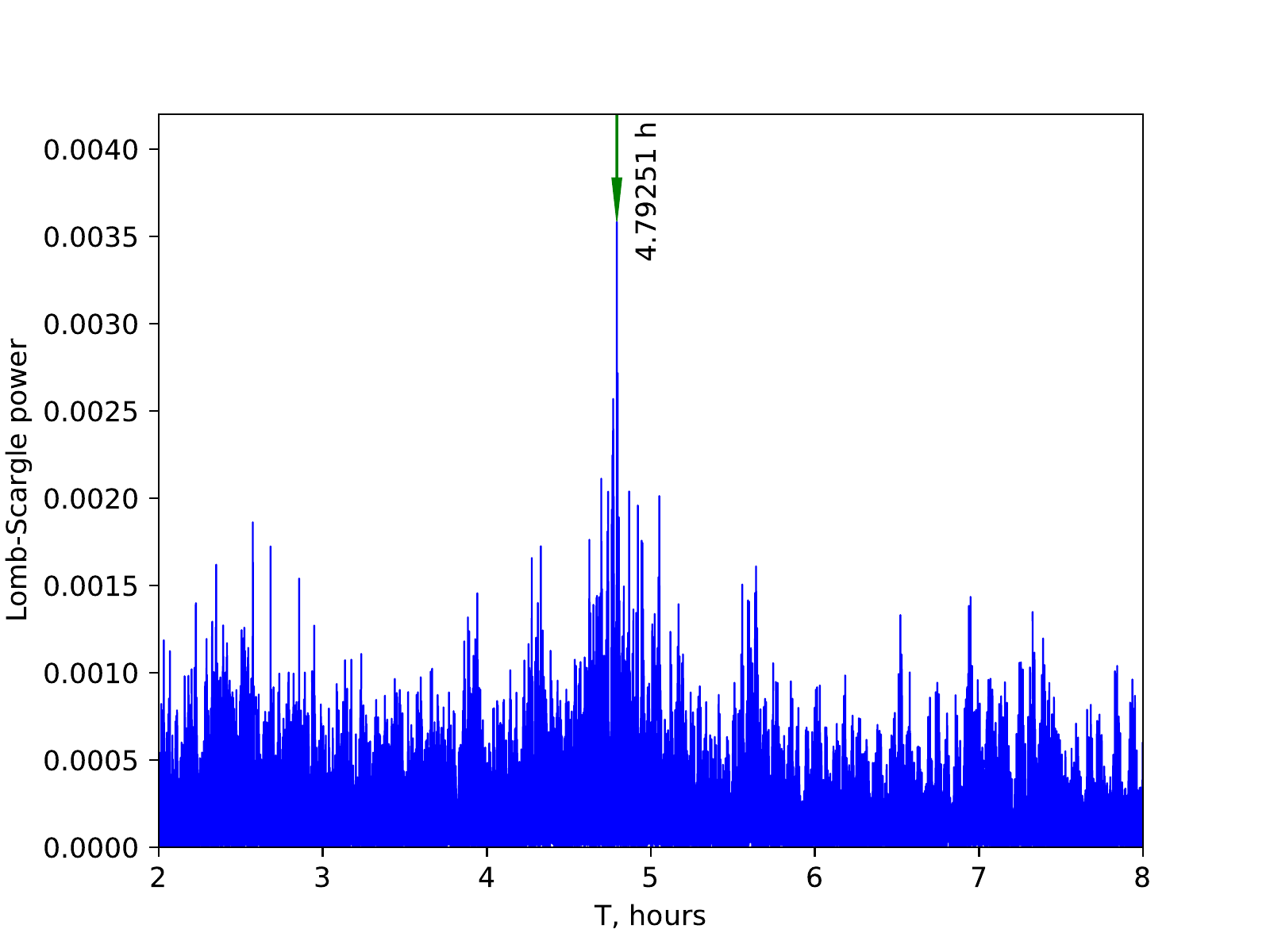}} 
\caption{The Lomb-Scargle periodogram for Cyg X-3 in the flaring state in the 0.1--100 GeV range, calculated accounting for the orbital period increase, taking into account the measurement uncertainties, and normalized with the $\chi^2$ of a constant model (which results in the power within the 0--1 range). The highest peak corresponds to the orbital period of Cyg X-3. 
} \label{periodogram}
\end{figure}

We clearly detect the period of Cyg X-3 in HE \g-rays in the flaring state. We note that the orbital period of Cyg X-3 is increasing, which could shift and smear out the peak due to the periodicity. To account for that, we convert the observation time to that of a constant period by calculating $n(T)$ for an observation time, $T$, by treating $n$ as a real number, and solving the binomial in equation (\ref{eph}). We then subtract $c_0 n^2$ from the time of an observation. The results of our Lomb-Scargle analysis for the light curve corrected in this way are shown in Fig.\ \ref{periodogram}. We find the period of 0.199688(4) d, which, given its standard deviation, agrees very well with $P_0$ of Cyg X-3 of equation (\ref{bhargava}). We have also found an analogous result on the direct flaring-state light curve, with the peak corresponding to the range of the orbital period during the epoch of the LAT observations, though with a lower peak power, reflecting a period change during that epoch.

\begin{figure}
\centerline{\includegraphics[width=7.cm]{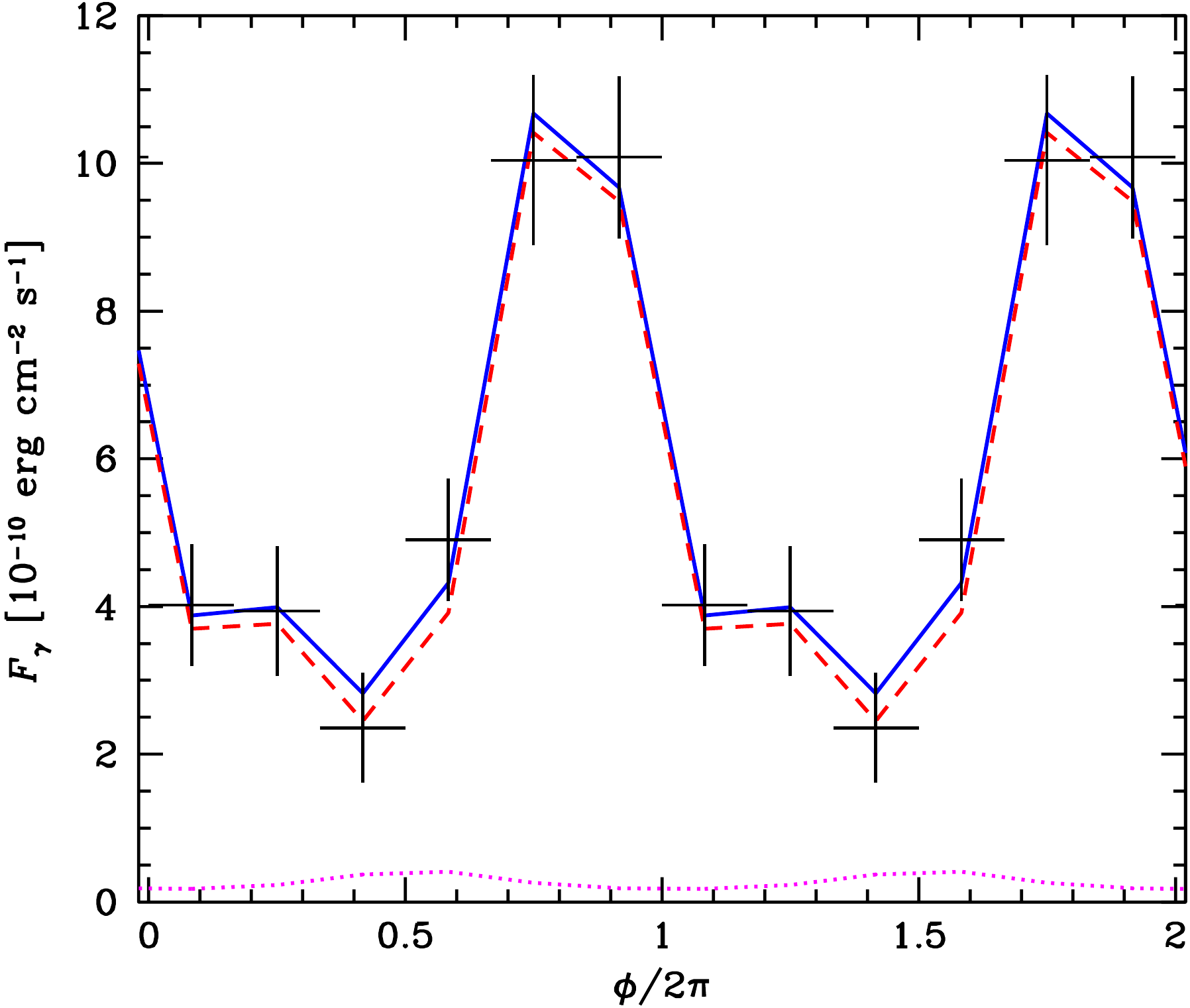}} 
\caption{The observed orbital modulation in the flaring state in the 0.1--100 GeV range shown by the error bars. The zero phase is defined from X-rays \citep{bhargava17} and it approximately corresponds to the superior conjunction. For clarity, hereafter two full phase ranges are shown. The blue solid line shows the best fit of the Compton-anisotropy model, and the red dashed and magenta dotted lines show the contributions of the jet and counterjet, respectively. 
} \label{orbital_g}
\end{figure}

We then use the ephemeris of Section \ref{ephemeris} to assign the phase to each photon observed within the ROI during the flaring state, and split the data over 6 equal phase bins. We perform the binned likelihood analysis (see Section \ref{fermi}) in each of the bins\footnote{We note that FLC09 performed an aperture analysis to determine the orbital modulation, which left an uncertainty about the background level, see their fig.\ 3B. The present method avoids this problem, and relies on the standard templates, see Section \ref{fermi}.}. The resulting energy fluxes as a function of the orbital phase are shown in Fig.\ \ref{orbital_g}. We have not found any statistically significant dependence on the energy range, which we looked for by using the photon energy ranges of 0.1--1, 1--10 and 10--100 GeV. We have also verified the consistency of our analysis by checking that the light curve of the nearby pulsar PSR J2032+4127 remains constant in all considered phase-bins.

\begin{figure}
\centerline{\includegraphics[width=\columnwidth]{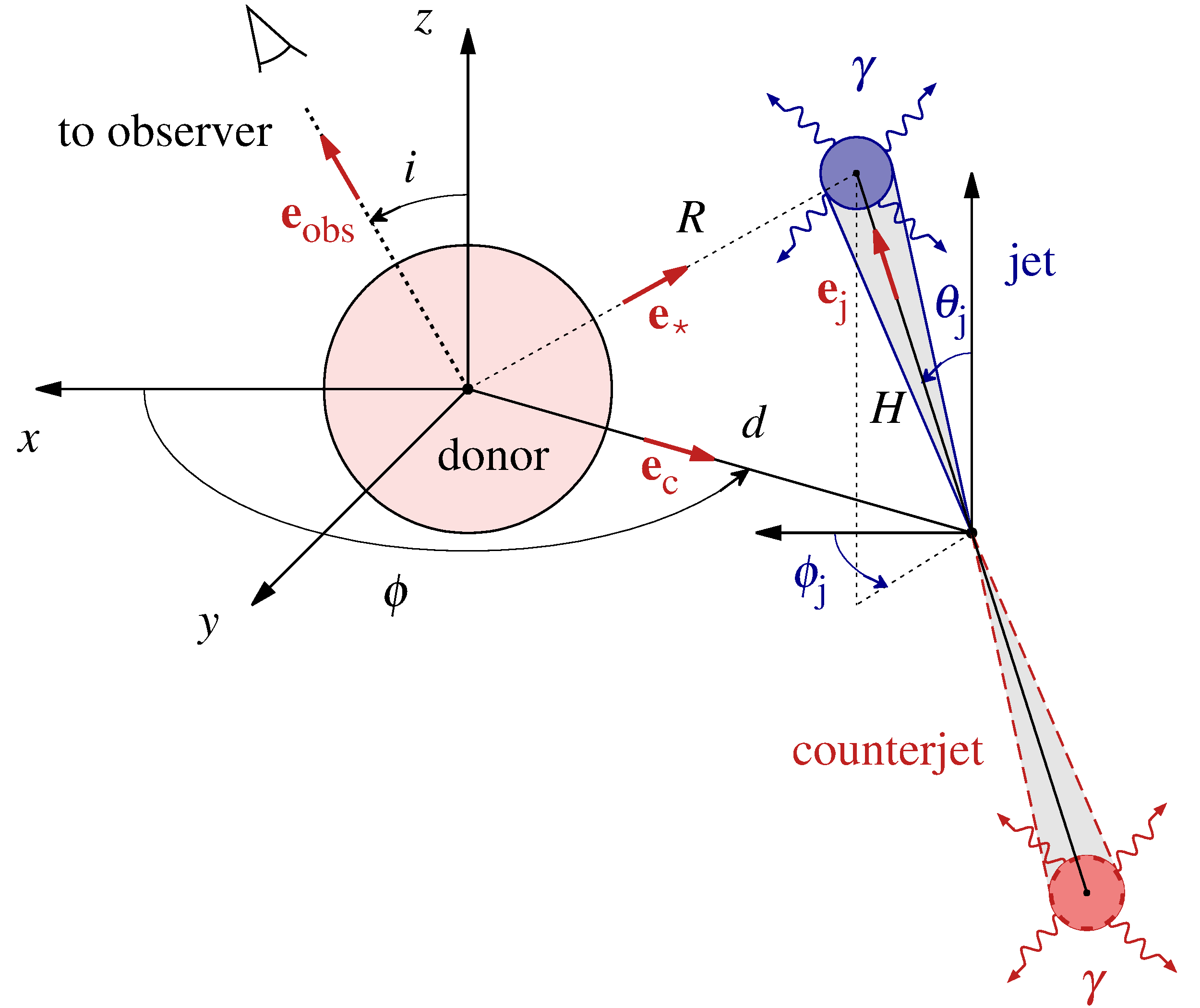}} 
\caption{The geometry of Compton scattering of the blackbody photons. The axes $x$ and $y$ are in the binary plane, and the $+z$ direction gives the binary axis. The $+x$ direction gives the projection of the direction toward the observer onto the binary plane. The observer is at an angle, $i$, with respect to the orbital axis, $\phi$ is the orbital phase, $\phi=0$ and $\upi$ correspond to the superior and inferior conjunction, respectively, $\theta_{\rm j}$ is the inclination of the jet with respect to the binary axis, $\phi_{\rm j}$ is the angle of the projection of the jet onto the binary plane with respect to $x$ axis, $d$ is the distance between the stars, $H$ is the distance of the \g-ray source from the centre of the compact object, the vectors \vec{$e$}$_{\rm obs}$, \vec{$e$}$_{\rm c}$ and \vec{$e$}$_*$ point from the donor toward the observer, the centre of the compact object, and the \g-ray source, respectively, and \vec{$e$}$_{\rm j}$ points from the centre of the compact object toward the \g-ray source.
} \label{geometry}
\end{figure}

\begin{figure}
\centerline{\includegraphics[width=5.8cm]{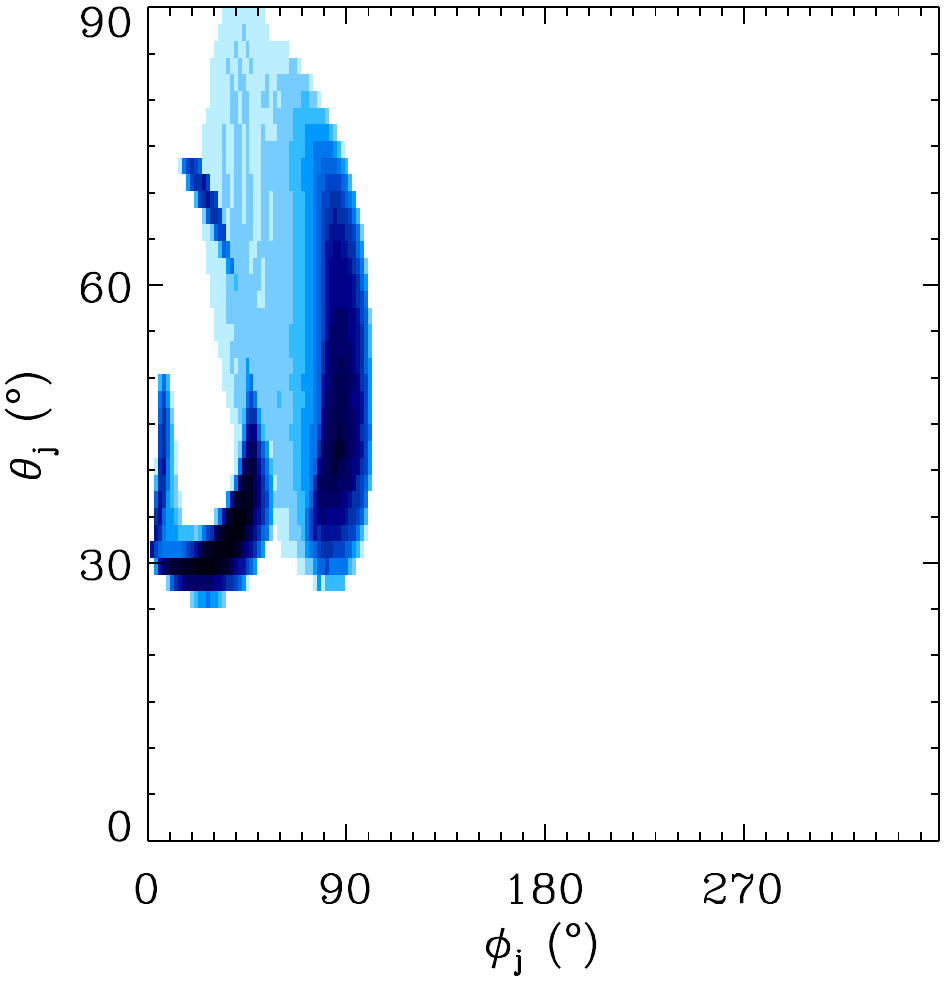}}
\centerline{\includegraphics[width=6.cm]{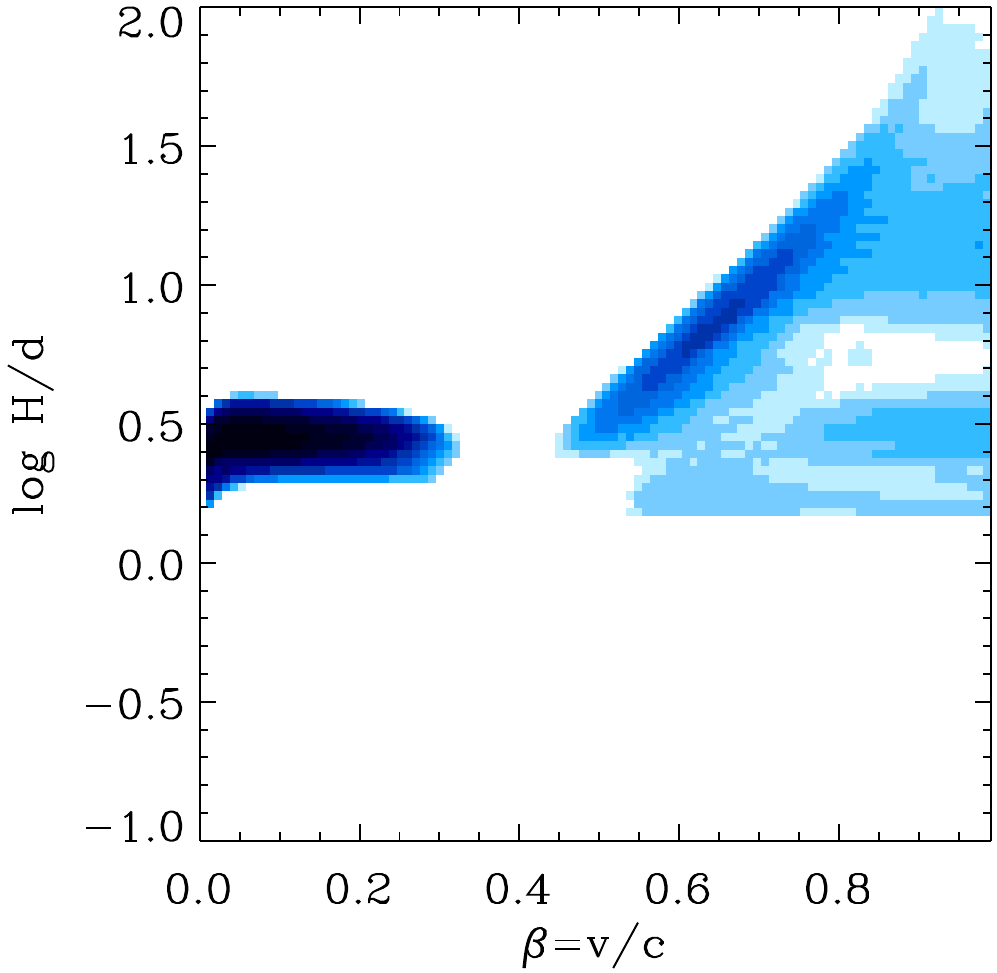}} 
\caption{The distributions of the mutual dependencies of the found acceptable parameters (within 68 per cent confidence region), $\phi_{\rm j}$ vs.\ $\theta_{\rm j}$ (top panel), and $H/d$ vs.\ $\beta$ (bottom panel), obtained with the Compton-anisotropy model applied to the observed \g-ray orbital modulation. The colour changes from white to dark blue $\propto\sqrt{n}$, where $n$ is the number of acceptable models per unit area.
} \label{orbital_dependencies}
\end{figure}

We then model the orbital light curve obtained by Compton anisotropy. This method utilizes two features of Compton scattering of stellar emission by relativistic electrons. First, the scattering probability is maximized for head-on collisions, i.e., for electrons moving towards the star. Second, a relativistic electron emits the scattered photon predominantly along its direction of motion. Thus, most of the Compton-scattered emission is towards the star and almost no emission is directed to an observer located along the line connecting the star centre and the \g-ray source. Therefore, the observed emission is maximized when the \g-ray source is behind the star. For a jet perpendicular to the orbital plane, this would be at the superior conjunction, and the modulation would be symmetric around it. Departures from that indicate that the jet is inclined with respect to the binary axis.

We use the method of \citet{dch10b}, including minor corrections given in \citet{z12b}, and use the coordinate system shown in Fig.\ \ref{geometry} (in which $\phi=0$ corresponds to the superior conjunction). We assume the blackbody photons to be emitted by a point source with the luminosity of $L_*=4\upi R_*^2\sigma_{\rm B} T_*^4$, where $R_*$ and $T_*$ are the stellar radius and temperature, respectively. We also assume the \g-ray source to be a point source, located at a distance, $H$, from the centre of the compact object. We assume the Thomson limit of Compton scattering, see equation (A9) of \citet{z12b}. We take into account the emission of both the jet and counterjet, and exclude fits to the observed modulation in which the counterjet is obscured by the star. We calculate the power injected into the non-thermal electrons in the jet+counterjet with a power-law distribution with the index of $p=4.1$ (corresponding to the fitted power-law index in the Thomson limit, $p=2\Gamma-1$) and the minimum Lorentz factor of $\gamma_{\rm min}=10^3$. We assume fast cooling, and thus that power equals the Compton-scattered luminosity emitted by the jet in all directions. In the calculations, we assume the donor mass of $M_*=14\msun$, $M_{\rm c}=4.5\msun$ (yielding the separation of $d=2.65\times 10^{11}$ cm) and the orbital inclination of $i=31\degr$, which correspond to the solution with the largest allowed masses in \citet{zmb13}. We also assume no eccentricity, $R_*=10^{11}$ cm, $T_*=10^5$ K and $D=7$ kpc. The assumed stellar radius is less than the Roche lobe radius, $\simeq 1.16\times 10^{11}$ cm (for the assumed masses; \citealt{eggleton83}).

Our best-fit solution, shown in Fig.\ \ref{orbital_g}, gives the jet velocity of $\beta\simeq 0.73$, the location of the \g-ray source along the jet at $H\simeq 6.1\times 10^{11}$ cm ($\simeq 2.3 d$), the jet inclination with respect to the orbital axis of $\theta_{\rm j}\simeq 37\degr$, with an azimuthal angle $\phi_{\rm j}\simeq 5\degr$ (Fig.~\ref{geometry}). Hence, the jet direction is $\simeq 6\degr$ off from the line-of-sight i.e. the jet is nearly pointed towards us. The total $\chi^2$ of the fit is 1.35 (with 6 orbital flux measurements and 5 fitted parameters, including normalization to the flux). The power injected into the non-thermal electrons (and/or e$^\pm$ pairs) is $\simeq 3.1\times 10^{36}$ erg s$^{-1}$, and the energy content of the electrons is $\simeq 1.1\times 10^{38}$ erg. In most of the acceptable solutions, the injected power is between $10^{36}$ and $10^{37}$ erg s$^{-1}$. Fig.\ \ref{orbital_dependencies} shows the mutual dependencies between $\theta_{\rm j}$ and $\phi_{\rm j}$, and $\beta_{\rm j}$ and $H/d$. We generally find the jet has to be inclined with respect to the binary axis by a relatively large angle, $\theta_{\rm j}\gtrsim 25\degr$, with the \g-ray emission zone located far from the compact object at $H\gtrsim d$. The acceptable ranges of our solutions are significantly narrower than those of \citet{dch10b}, but still consistent with them.

We see no evidence for precession within the epoch of the studied LAT observations, in either the power spectrum or by comparing the modulation shape at various epochs. It is likely that the inclined jet is aligned with the black-hole spin axis up to the location of the \g-ray emission. Occasional jet precession observed in radio \citep{m01,miller_jones04} occurs at much larger distances. If the \g-ray emitting jet precesses, the obtained parameters correspond to the average orbital modulation. Still, the observed large modulation amplitude, of $\sim$70--80 per cent, indicates the precession does not lead to its substantial reduction.

\subsection{Modulation of radio emission}
\label{orbital_radio}

\begin{figure}
\centerline{\includegraphics[width=6.cm]{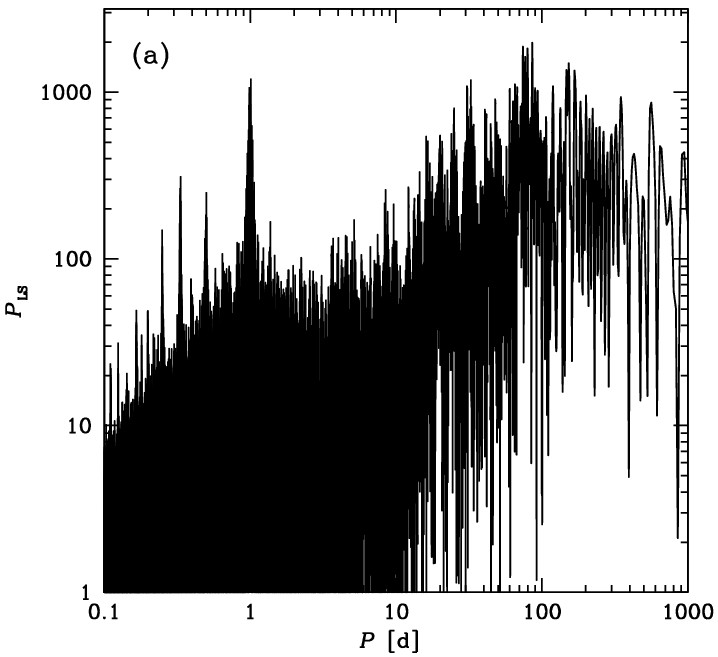}}
\centerline{\includegraphics[width=5.8cm]{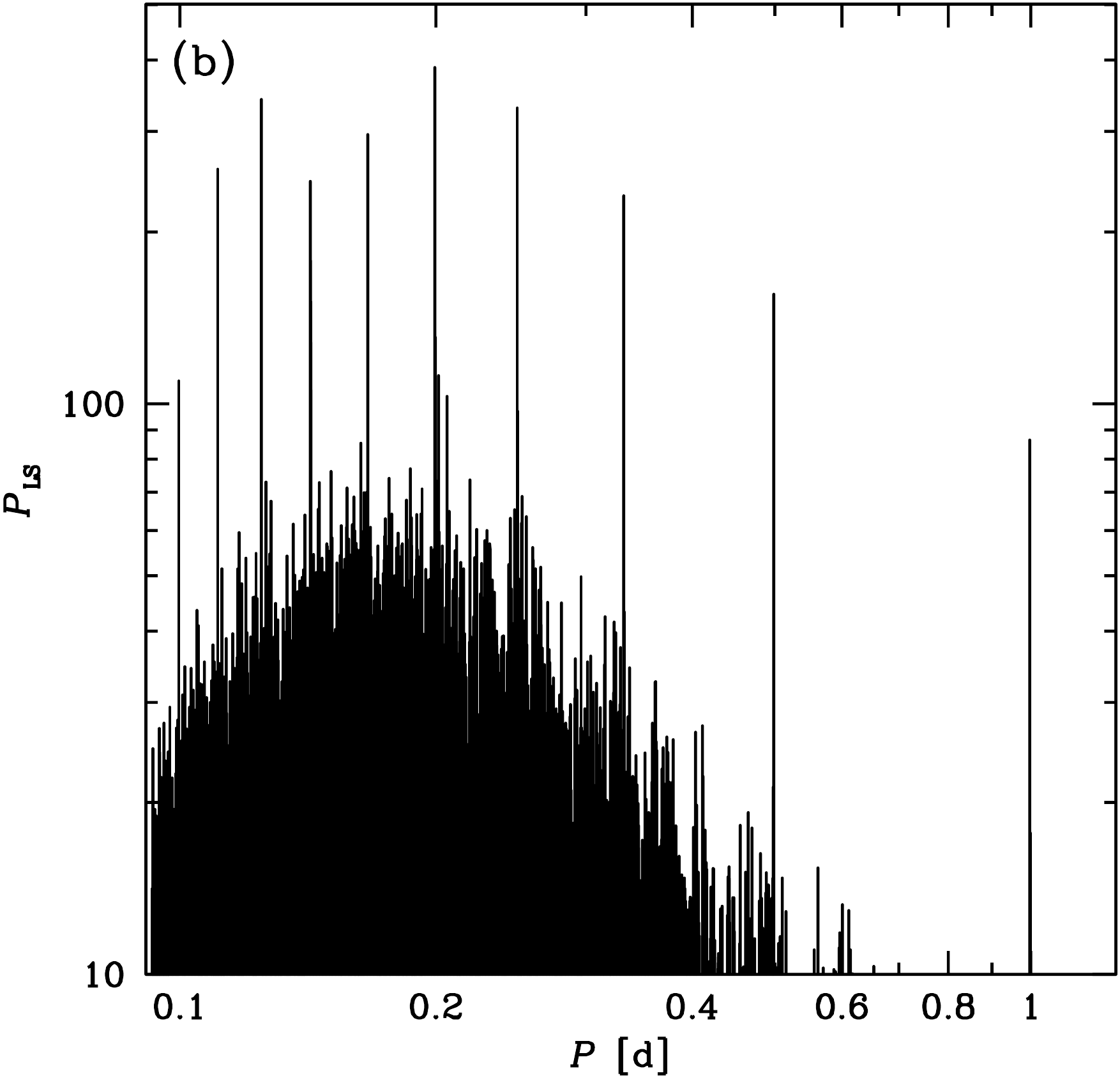}} 
\centerline{\includegraphics[width=5.8cm]{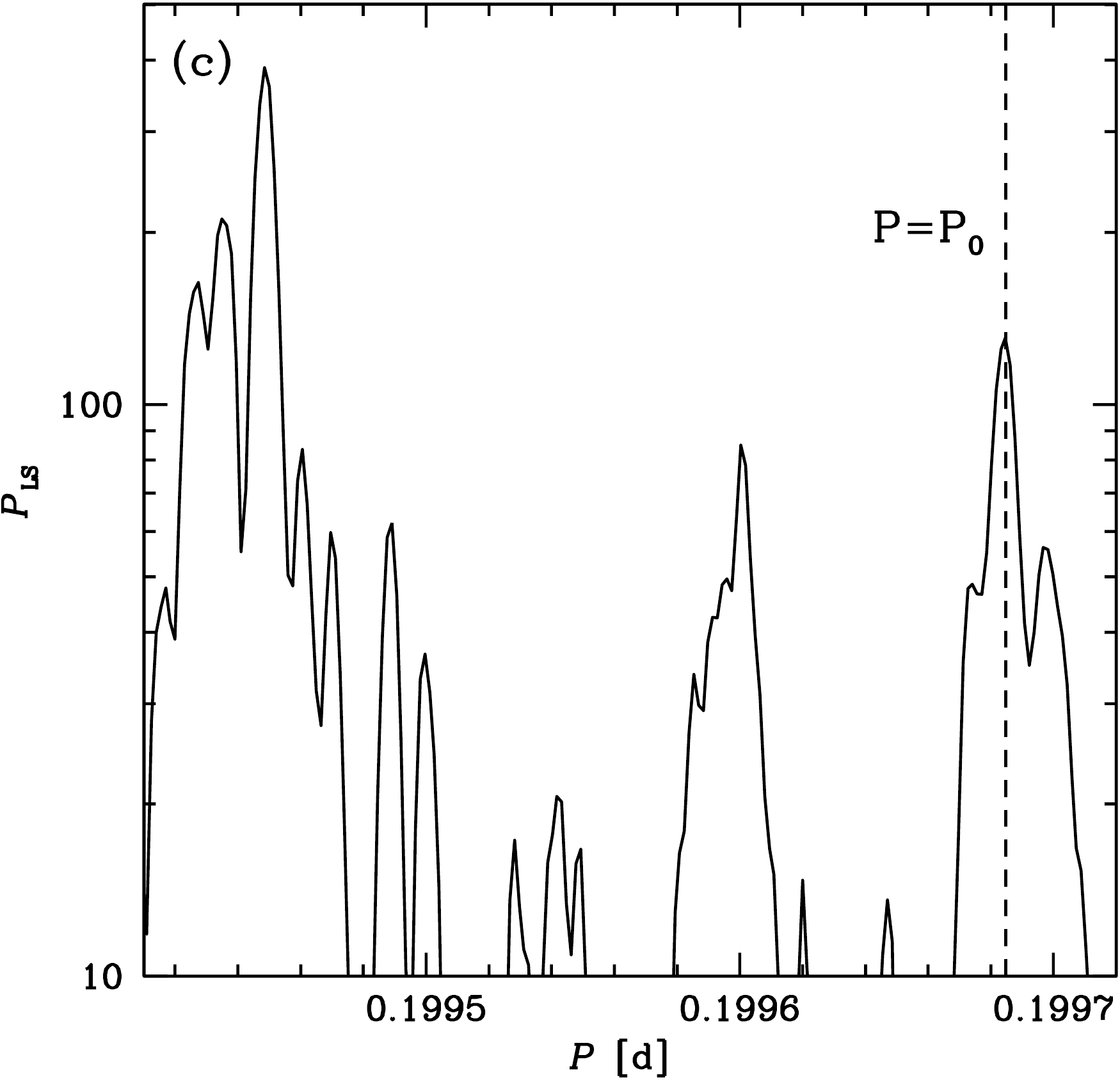}} 
\caption{(a) The Lomb-Scargle power spectrum (normalized as in \citealt{press92}) of the barycentre-corrected 15 GHz light curve from the Ryle and AMI telescopes obtained for $\ln F_\nu$. (b) The power spectrum after renormalizing $F_\nu$ to their running averages over $\pm 0.1$ d and transforming the time axis to that corresponding to the constant orbital period. (c) The same as in (b) but zoomed to the region containing both the orbital period, $P_0$ (marked by the dashed line), and the strongest peak.
} \label{radio_power_spectra}
\end{figure}

\begin{figure}
\centerline{\includegraphics[width=6.cm]{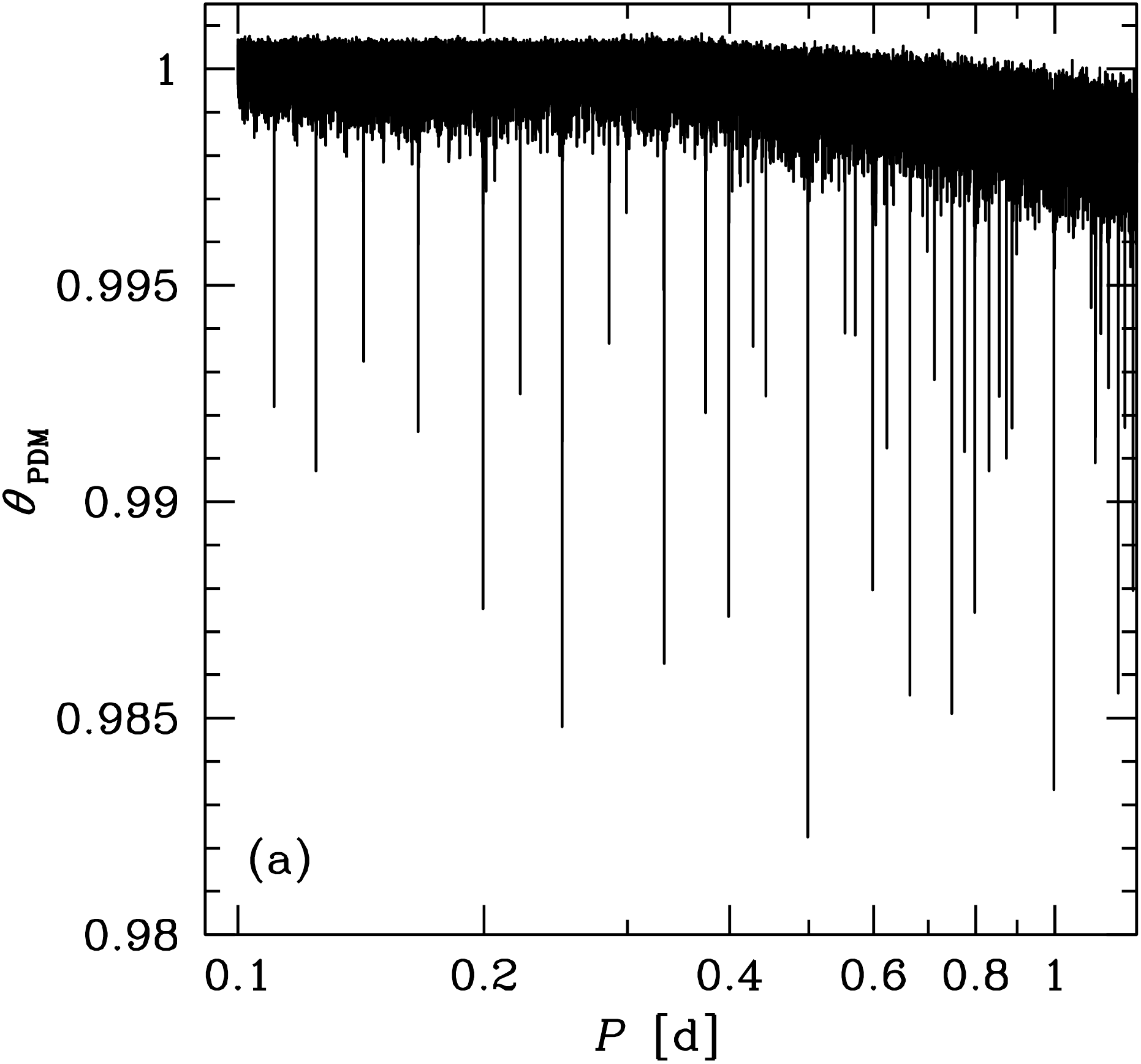}}
\centerline{\includegraphics[width=5.8cm]{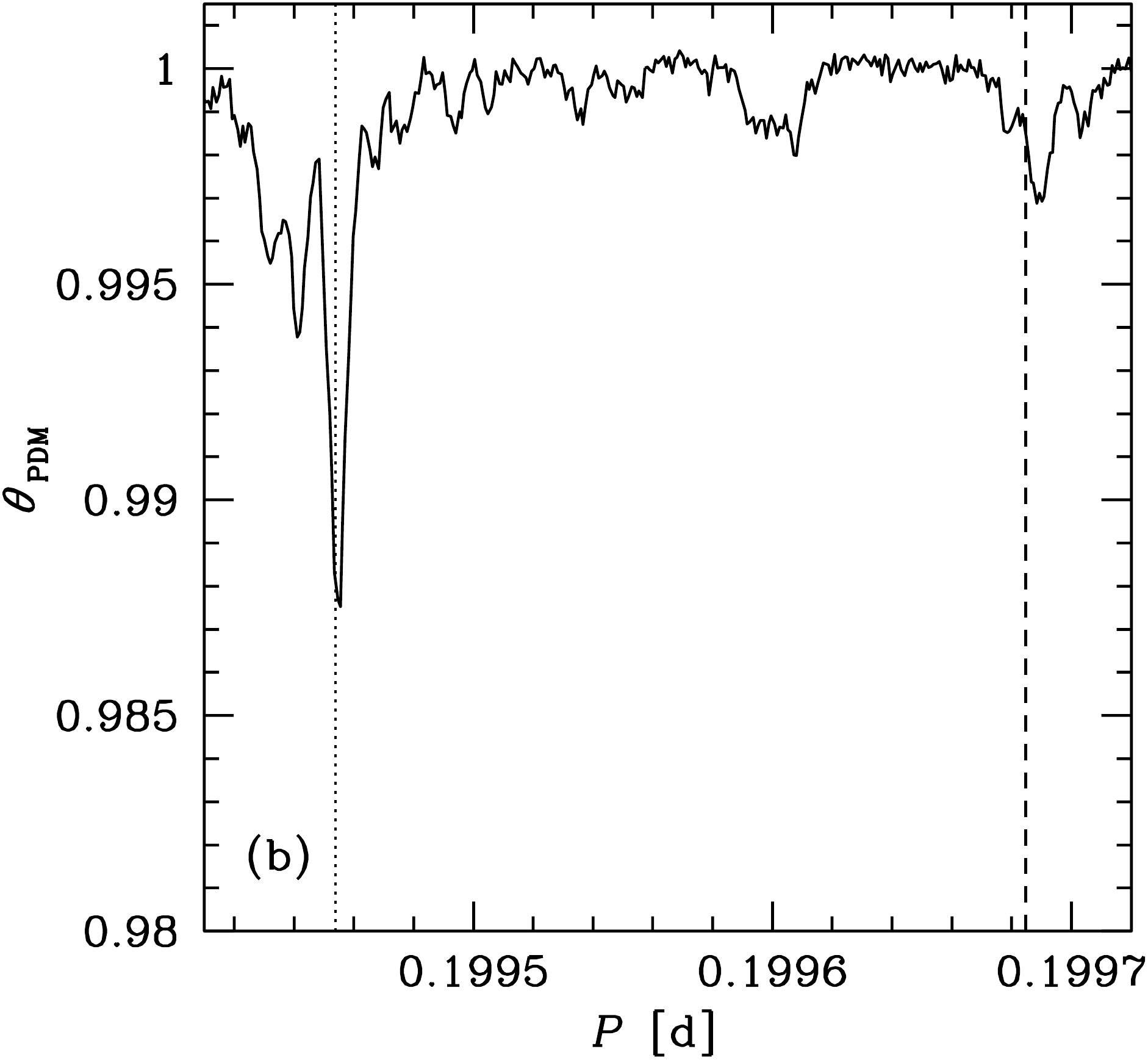}}  
\caption{(a) The results of the PDM analysis of the 15 GHz light curve without both the barycentric correction and the correction for the period increase, but after renormalizing $F_\nu$ to their running averages over $\pm 0.1$ d, shown for the range of trial periods of 0.1--1.25 d. (b) The same as in (a) but zoomed to the region containing both the orbital period, $P_0$, and the 1/5th of the sidereal day, $T_{\rm s}$, marked by the dashed and dotted line, respectively. The seen displacement of the former minimum from the exact value of $P_0$ is due to the the correction for $\dot P>0$ not being included in this calculation.
} \label{pdm}
\end{figure}

We expect to find some orbital modulation of the radio emission in Cyg X-3 caused by free-free absorption in the stellar wind. It is seen, e.g., in the BH HMXB Cyg X-1, where it is strong, with the total amplitude of $\simeq30\pm 1$ per cent at 15 GHz \citep{zdz12}. 

We note, however, that the radio observations of Cyg X-3 have been performed with the visibility window repeating each sidereal day, $T_{\rm s}=0.99726957$ d. Although the observations were scheduled at times determined by the current collection of other requests for observing, and their priorities, the presence of the visibility window results in a strong peak of the power spectrum around 1 d. In addition, harmonics appear, including the 5th one, which is very close to the orbital period. This has apparently prevented any detection of an orbital modulation of the radio emission in spite of many years of observations available. Indeed, we also do not find a significant peak at the orbital period in the power spectrum of the barycentre-corrected light curve. This is shown in Fig.\ \ref{radio_power_spectra}(a), where we see a strong broad peak around 1 d, and the peak around the orbital period is seen at a much lower power. The overall maximum power is at 85.95 d.

Thus, in order to see the orbital modulation in the power spectrum, we follow a technique used in \citet{z12a} for calculating folded light curves. In it, we normalize each flux density to its running linear average, see equation (4) in \citet{z12a}, determined in the present case by averaging the flux using the observations within $\pm \delta/2 = 0.1$--0.2 d of its time (i.e., within 1--2 orbital periods) and requiring at least 5 observations with the positive fluxes in each average. This reduces the number of usable observations by only 4 per cent, and the average number of observational points used for a renormalized flux is $26\pm 13$ for $\delta=0.2$ d and $37\pm 24$ for $\delta=0.4$ d. We note that this technique corresponds to imposing a high-pass filter in the frequency domain, i.e., it strongly reduces the variability on time scales longer than the orbital period (very significant in Cyg X-3), thus allowing us to detect the orbital modulation. We also convert the light curve to one corresponding to a constant orbital period in the same way as applied to the \g-ray light curve, see Section \ref{orbital_gamma} above. The power spectrum of the resulting light curve is shown in Fig.\ \ref{radio_power_spectra}(b). We see that now the strongest peak is around 0.2 d. We show a zoom of the periodogram to the vicinity of $P_0$ in Fig.\ \ref{radio_power_spectra}(c), in which we see that we clearly detect the orbital period of $P_0=0.19968476(3)$ of equation (\ref{bhargava}), with $P_{\rm LS}\simeq 130$. 

\begin{figure}
\centerline{\includegraphics[width=6.cm]{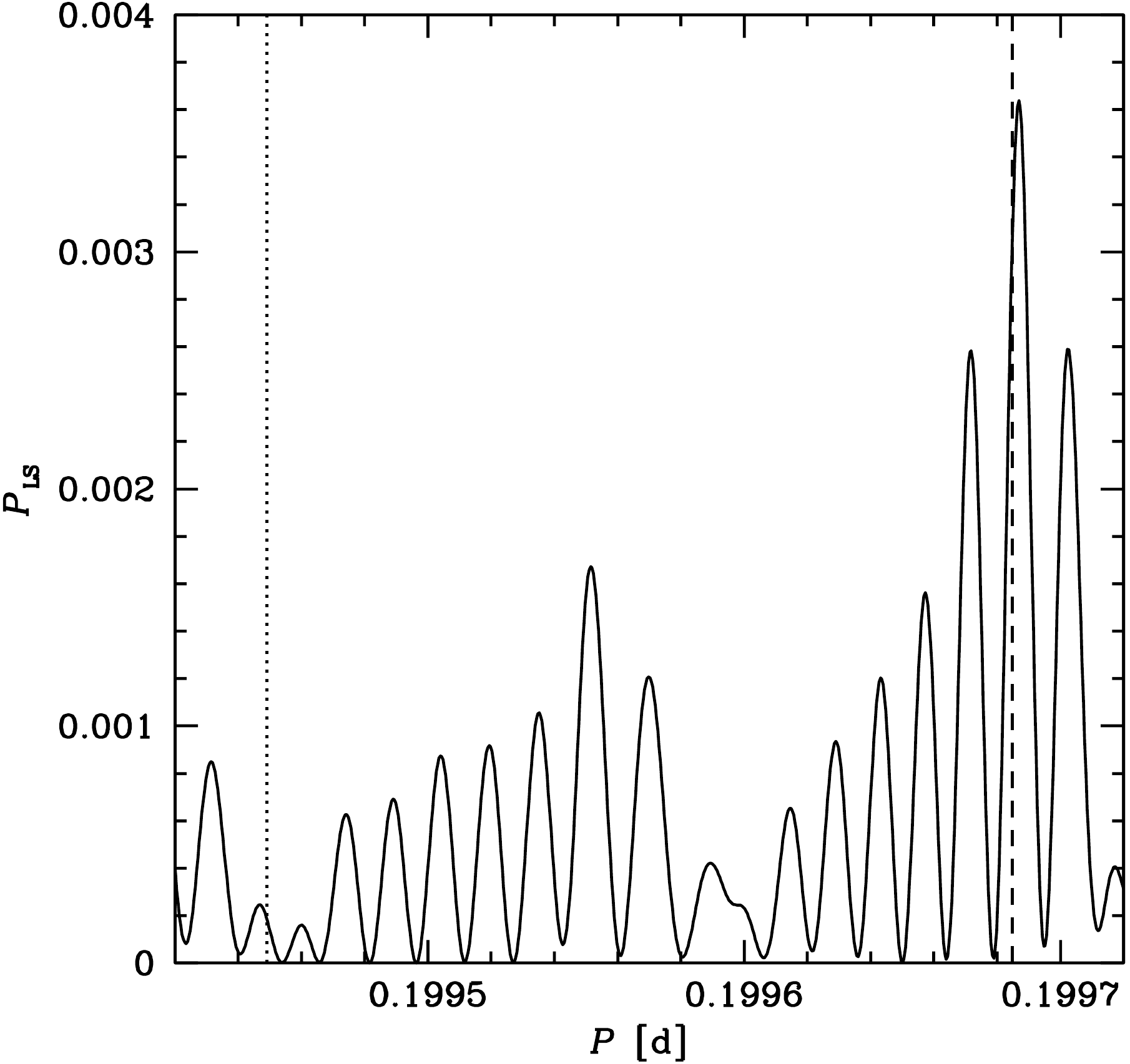}} 
\caption{The Lomb-Scargle periodogram for Cyg X-3 in the flaring state in the 0.1--100 GeV range (as in Fig.\ \ref{periodogram}, which takes into account the correction for $\dot P$) zoomed to the region containing the orbital period (shown by the dashed line) and the strongest peak in Fig.\ \ref{radio_power_spectra}(c) (shown by the dotted line).
} \label{power_gamma}
\end{figure}

On the other hand, we also see that the strongest peak of the periodogram is at $P_{\rm LS}\simeq 390$ at a period of $\simeq 0.19944848$ d, clearly different from $P_0$. We have searched for the origin of that peak, and found that it corresponds to the fifth harmonic of the sidereal day. In order to clearly see it, we have considered the 15 GHz light curve without the barycentric and $\dot P$ corrections. However, we still imposed our high-pass filter with $\pm \delta/2 = 0.1$, in order to see variability on time scales comparable to the orbital period. We have performed the timing analysis on this light curve using both the periodogram and the period-dispersion minimization (PDM; \citealt{stellingwerf78}) method. We show here the results only for the PDM analysis, with those from the periodogram being completely consistent with the PDM ones. Among others, we have found distinct minima of the PDM statistic, $\theta_{\rm PDM}$, at $T_{\rm s}$, $T_{\rm s}/2$, $T_{\rm s}/3$, $T_{\rm s}/4$ and $T_{\rm s}/5$, as shown in Fig.\ \ref{pdm}(a). A zoom to the region of $T_{\rm s}/5$ and $P_0$ is shown in Fig.\ \ref{pdm}(b). We see that the strongest peak seen in Fig.\ \ref{radio_power_spectra}(c) corresponds (after removing the time corrections) exactly to $T_{\rm s}/5$. We have also checked that the period observed in HE \g-rays is equal to $P_0$ and that no additional peaks appear in its vicinity, as shown in Fig.\ \ref{power_gamma}, which is consistent with the \g-ray observations being performed from space, thus not affected by the daily visibility window.

\begin{figure}
\centerline{\includegraphics[width=7.5cm]{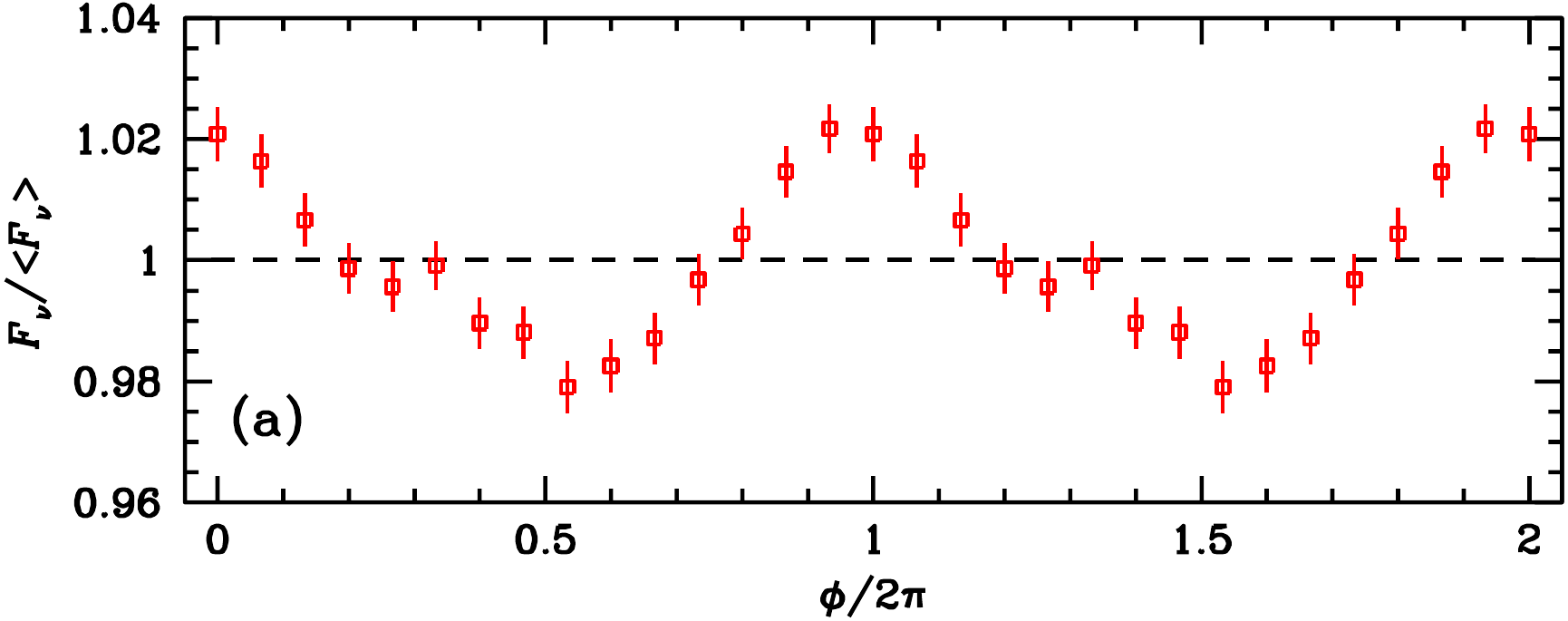}} 
\centerline{\includegraphics[width=7.5cm]{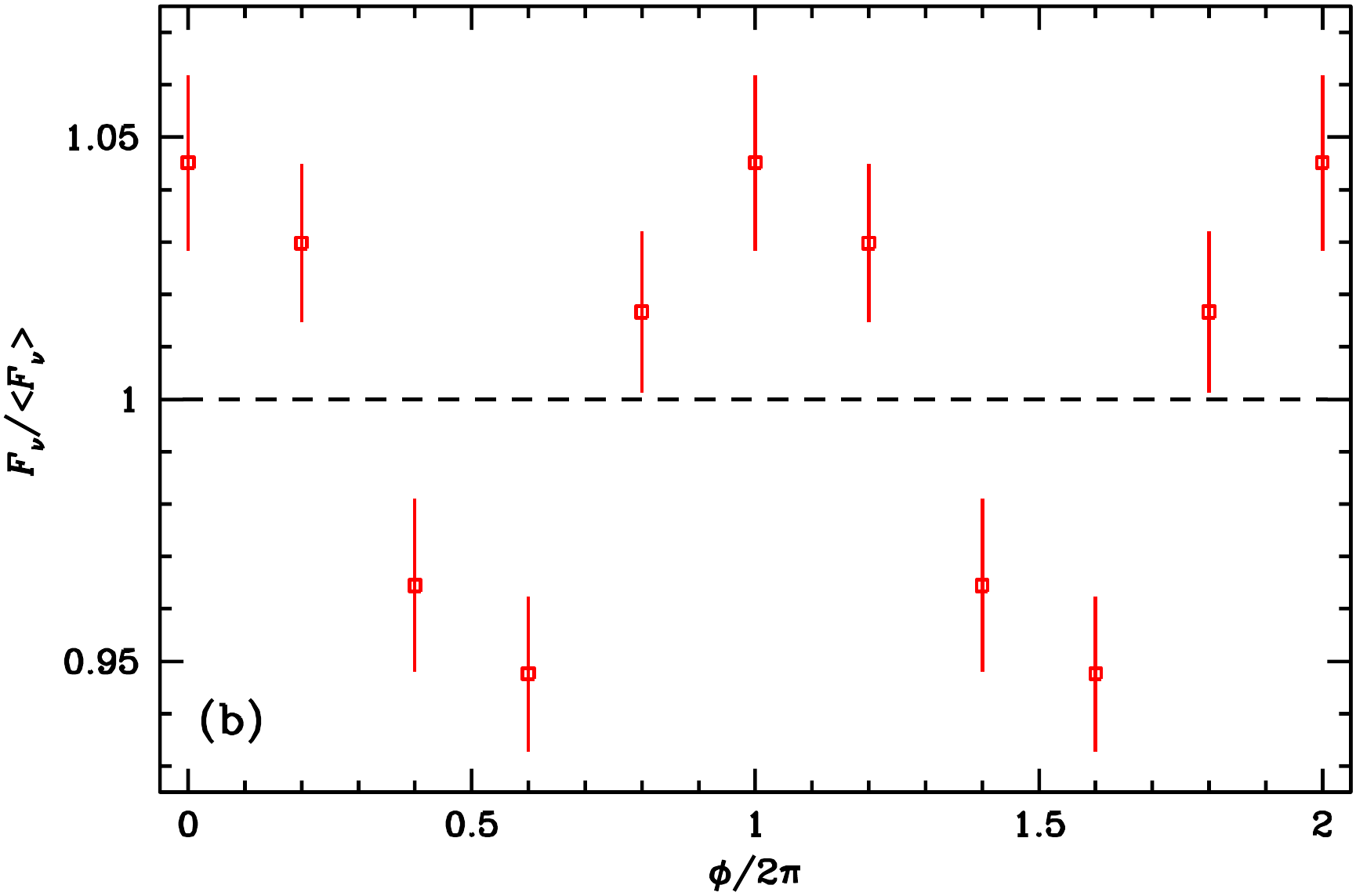}} 
\centerline{\includegraphics[width=7.5cm]{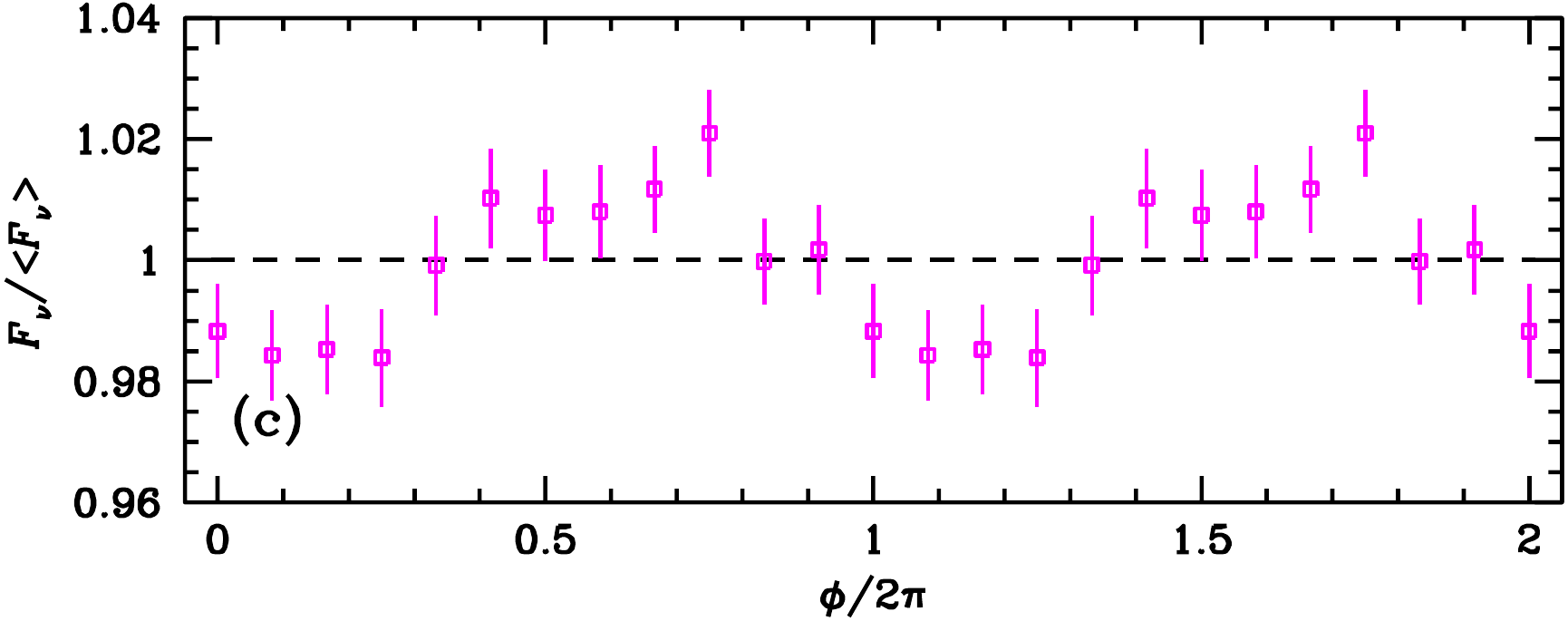}} 
\centerline{\includegraphics[width=7.5cm]{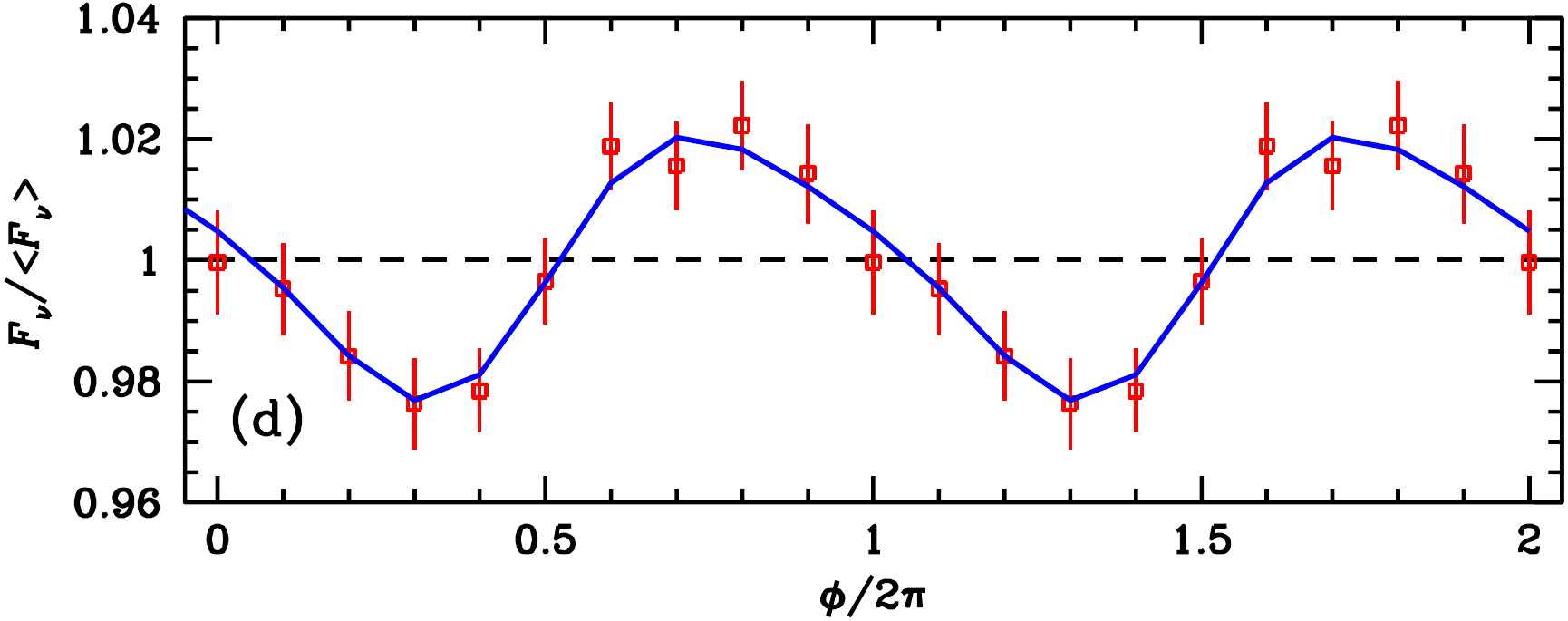}} 
\centerline{\includegraphics[width=7.5cm]{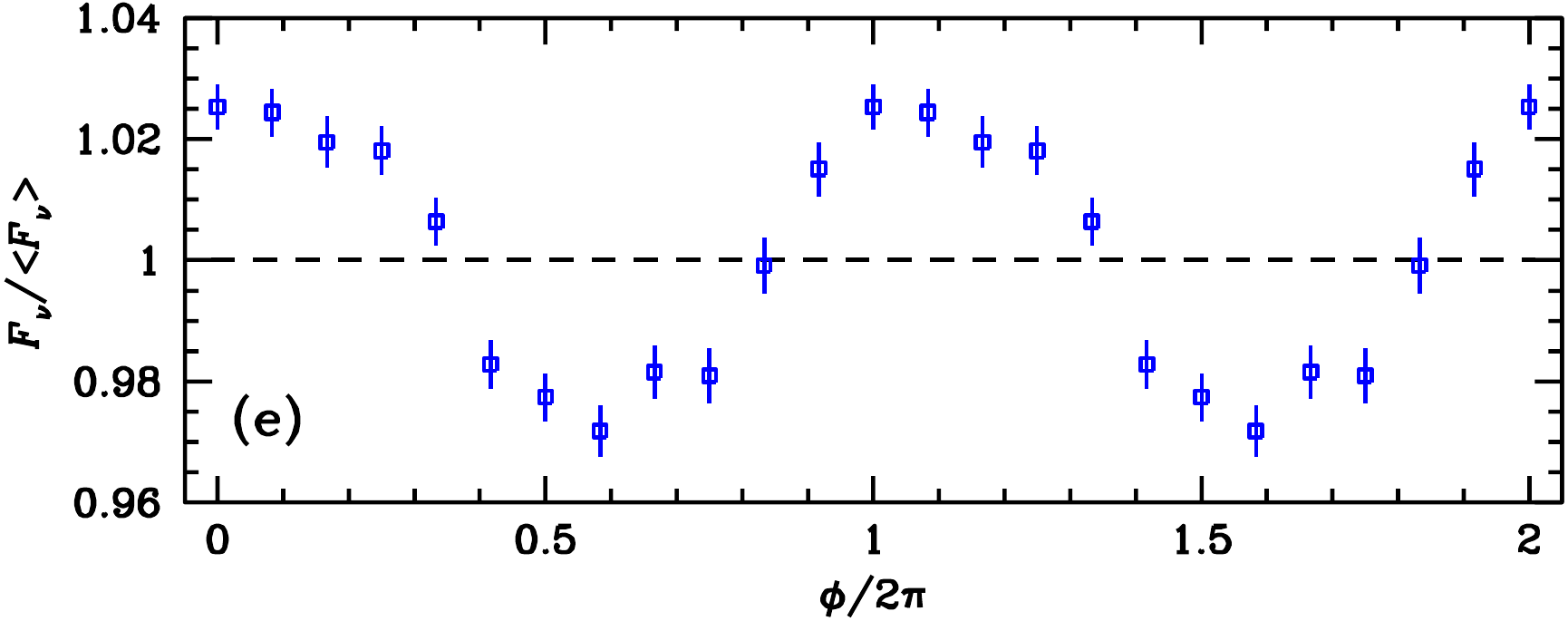}} 
\centerline{\includegraphics[width=7.5cm]{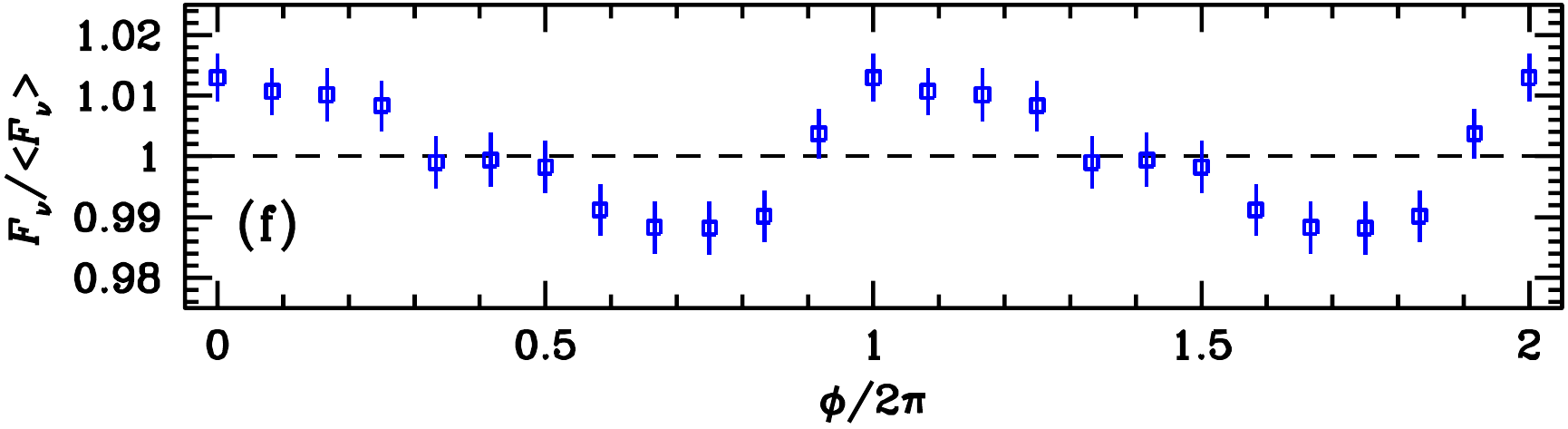}}
\caption{(a) Radio flux folded (using $\ln F_\nu$) on the orbital period for the entire 15 GHz data renormalized to the local running average (76129). Hereafter, the numbers in parentheses give the respective number of the measurements used. Below, we show the dependencies of the orbital modulation on the radio flux range in the soft/intermediate state (b--d) and the hard (e--f) state. We show the results for the data for (b) the lowest soft state ($F_\nu<30$ mJy; 5724); (c) the medium soft state ($30\,{\rm mJy}<F_\nu\leq 300\,{\rm mJy}$, $F$(3--5\,keV$)>0.5$ keV cm$^{-2}$ s$^{-1}$; 14607); (d) the highest soft state ($F_\nu>300\,{\rm mJy}$; 8487); (e) the lower hard state ($30\,{\rm mJy}<F_\nu\leq 100\,{\rm mJy}$, $F$(3--5\,keV$)<0.5$ keV cm$^{-2}$ s$^{-1}$; 8278); (f) the upper hard state ($100\,{\rm mJy}<F_\nu\leq 300\,{\rm mJy}$, $F$(3--5\,keV$)<0.5$ keV cm$^{-2}$ s$^{-1}$; 8606). The best fit with the wind-absorption model is shown by the solid line in panel (d). 
} \label{orbital_r}
\end{figure}

\begin{figure}
\centerline{\includegraphics[width=5.8cm]{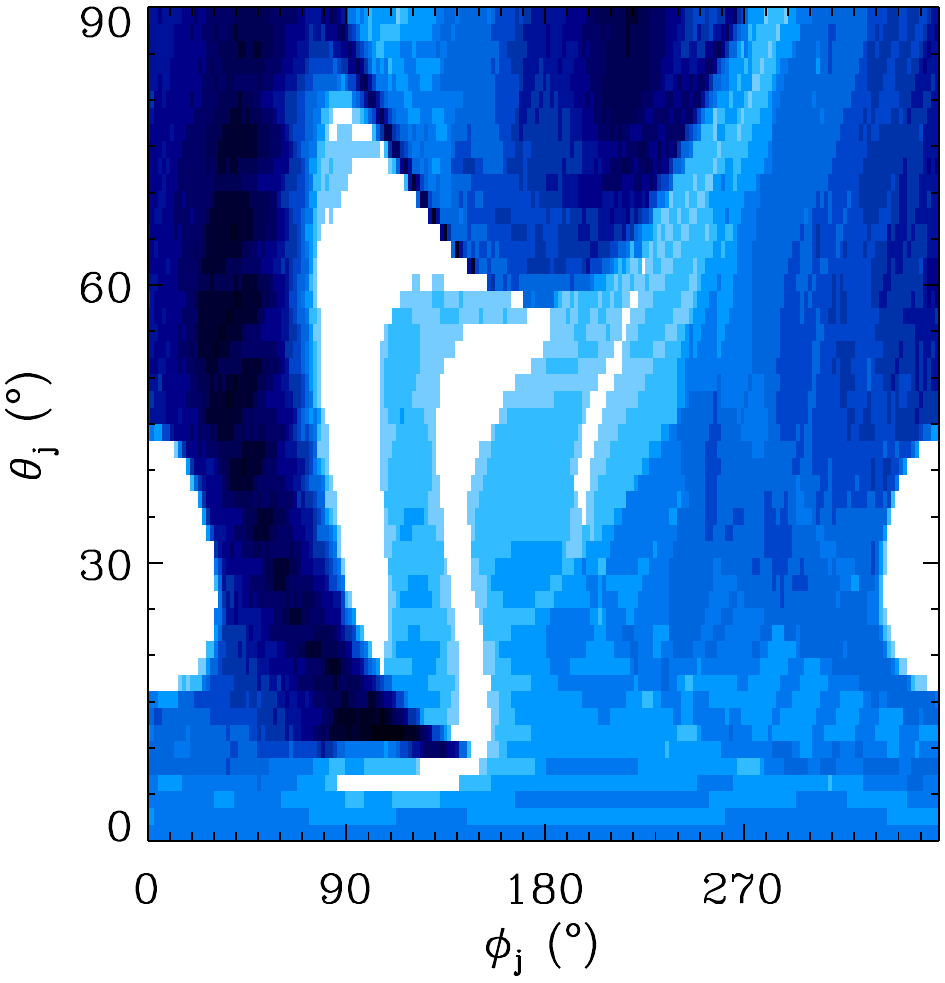}}
\centerline{\includegraphics[width=5.8cm]{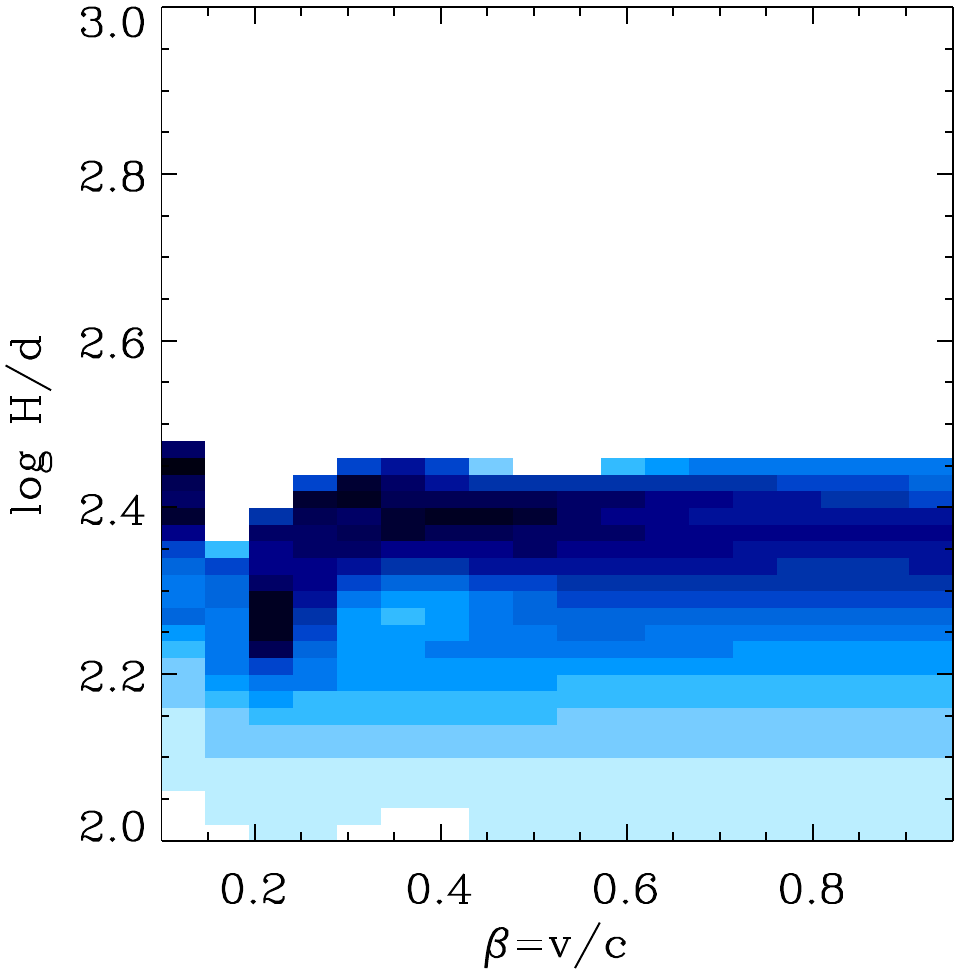}} 
\caption{The mutual dependencies of the acceptable parameters (within 68 per cent confidence region), $\phi_{\rm j}$ vs.\ $\theta_{\rm j}$, and $H/d$ vs.\ $\beta$, obtained with the isotropic wind model applied to the observed 15 GHz orbital modulation. The colour changes from white to dark blue $\propto\sqrt{n}$, as in Fig.\ \ref{orbital_dependencies}.
} \label{orbital_dependencies_radio}
\end{figure}

Therefore, we hereafter consider only the orbital period. We calculate the orbital modulation of the 15 GHz flux by using the light curve renormalized to the running average and take account of the $\dot P$, as described above. We calculate the average flux and its standard deviation within a given phase bin. We first present our results for the entire (renormalized) radio light curve, which corresponds to the modulation averaged over all flux and spectral states of Cyg X-3, see Fig.\ \ref{orbital_r}(a). We see a distinct modulation with a depth of $\simeq$4 per cent. 

We then separate the data into subsets based on the radio flux and the X-ray spectral state. Based on fig.\ 2(b) in ZSP16 and Figs.\ \ref{states}(b--c) above, we see that the hard state in Cyg X-3 is characterized by the radio flux density changing in the range of $30\,{\rm mJy}\lesssim F_\nu \lesssim 300\,{\rm mJy}$ and the 3--5 keV flux of $\lesssim 0.5$ keV cm$^{-2}$ s$^{-1}$. Thus, in order to select the hard state in our data, we use these two criteria for the radio data with available ASM coverage within a day only. Then, the combined soft/intermediate state has the 3--5 keV flux of $\gtrsim 0.5$ keV cm$^{-2}$ s$^{-1}$ and the values of the radio flux density anywhere between undetectable flux and 20 Jy, i.e., also including the 30--300 mJy range. We find that there are virtually no measurements with the 3--5 keV flux being $< 0.5$ keV cm$^{-2}$ s$^{-1}$ outside the 30--300 mJy range so we do not need to impose any condition on that flux. Thus, in order to select the soft/intermediate state in our data, we use the radio data with available ASM coverage within a day with the 3--5 keV flux $> 0.5$ keV cm$^{-2}$ s$^{-1}$ for the radio data within 30--300 mJy, and all the data with $F_\nu<30$ mJy and $F_\nu>300$ mJy. 

We show the results for the soft/intermediate state in Fig.\ \ref{orbital_r}(b--d). We see a significant dependence on $F_\nu$. The modulation amplitude is highest for the lowest radio fluxes, with the amplitude of $\simeq$10 per cent, and the modulation extrema are at the phases similar to those for the entire data. Then the intermediate and high fluxes show the amplitudes similar to those for the entire data, $\simeq$4 per cent. There is also a clear shift of the phase of the modulation minima, at $\phi/2\upi \simeq 0.6$, $\simeq$1.1--1.3, 1.3--1.4 for the lowest, intermediate and highest radio-flux range, respectively. Next, we consider the hard state. We also see a significant dependence on $F_\nu$, see Fig.\ \ref{orbital_r}(e--f). The modulation amplitude is higher for the lower range of the flux, 30--100 mJy, with the amplitude of $\simeq$6 per cent, than for the upper range, 100--300 mJy, with the amplitude of $\simeq$2.5 per cent.  We also see a shift of the phase of the modulation minima, at $\phi/2\upi \simeq 0.6$, $\simeq$0.65--0.85, for the lower and higher radio-flux range, respectively.

The amplitude decrease with the increasing flux within the hard state can be explained by an increase of the height along the jet to where the partially self-absorbed emission becomes optically thin. In the soft/intermediate state, the synchrotron emission is mostly optically thin, but still a larger fraction of the radio flux appears to be emitted close to the compact object for low radio fluxes.  Also, the phase of the modulation minimum increasing with the radio flux can be explained by the distance of the location of the bulk of radio emission increasing with the increasing flux and the jet being inclined with respect to the orbital axis. Similar effects are seen in Cyg X-1, see \citet{zdz12}.

The radio orbital modulation and its dependence on the flux can be fitted by the same method and geometry as for \g-rays (Section \ref{orbital_gamma}) except that now free-free absorption on the wind is included as the modulation process. We assume a constant wind velocity and temperature, $v_{\rm w}=1600$ km s$^{-1}$, $T_{\rm w}=10^5$ K, respectively, the mass-loss rate by the donor of $\dot M_{\rm w}=7.5\times 10^{-6}\msun$ yr$^{-1}$ (\citealt{zmb13} and references therein), the He composition ($X=0$) and $M_*$, $M_{\rm c}$, and $i$ as in Section \ref{orbital_gamma}. Since the radio emission zone is far from the system, we take into account the effect of the non-zero orbital velocity and finite jet velocity, which can cause a significant change in jet orientation at large distances and a phase delay in the radio modulation. These effects are negligible when fitting the \g-ray orbital modulation.

We fit the orbital modulation averaged over parts of the entire light curve corresponding to the brightest part of the soft state, $F>0.3$ Jy (Fig.\ \ref{orbital_r}d), since the strong \g-ray emission is observed predominantly in that state and we wish to compare to the jet parameters derived from the \g-ray modulation. We find the acceptable regions are wide and include values for the jet inclination, $\theta_{\rm j}$, and orientation, $\phi_{\rm j}$, that are compatible with those found for \g-rays, see Fig.\ \ref{orbital_dependencies_radio}  Only the location of the 15 GHz source is robustly constrained (by the amplitude of the modulation) to around $\sim\! 200 d$. The best-fit solution is plotted in Fig.\ \ref{orbital_r}(d). The region of dominant 15 GHz emission is located at the distance along the jet of $H\simeq 4.4\times 10^{13}$ cm, for a jet velocity of $\beta\simeq 0.45$, with $\theta_{\rm j}\simeq 64\degr$, $\phi_{\rm j}\simeq 140\degr$, and a total $\chi^2\simeq 1.9$ (10 orbital flux measurements, 5 fitted parameters). For the best fit, the relative contribution of the counterjet is quite large, in the range of 0.33--0.48. A problem with this solution is that it is heavily attenuated, with only a fraction $\simeq 5\times 10^{-5}$ of the intrinsic flux making it to the observer. We note that the average optical depth, $\langle \tau\rangle$, from the source to infinity is larger by a factor of the order of $\sim\! H/d$ than the difference in the optical depths between their maximum and minimum values, $\Delta\tau$ (roughly equal to the modulation amplitude), see equations (21--22) in \citet{zdz12} derived for a perpendicular jet. This explains the large attenuation of this solution.

If we impose the same jet velocity and orientation for both the \g-ray and radio modulation models, we obtain $\beta\simeq 0.55$, $\theta_{\rm j}=30\degr$, $\phi_{\rm j}=43\degr$, with a total $\chi^2\simeq 9.5$ (with contributions of 5.5 and 4.0 for the \g-ray and radio part, respectively). The locations of the sources are $H_\gamma\simeq 1.1\times 10^{12}$ cm and $H_{\rm r}\simeq 4.7\times 10^{13}$ cm. The average attenuation of this solution is more moderate, $5.4\times 10^{-2}$.

Finally, we mention that in our investigations we also considered the hypothesis that the strongest peak in Fig.\ \ref{radio_power_spectra}(c) is due to a beat with a precession of the jet. The difference of the period of the strongest peak and the orbital period of $\simeq$20 s corresponds to a precession period of $\simeq$170 d (which, interestingly, is about twice the period of the strongest peak in the periodogram of Fig.\ \ref{radio_power_spectra}a at $\simeq$86 d). We have then searched for a dependence of the orbital modulation on the precession phase, and, surprisingly, we have found a rather regular dependencies on it of both the precession amplitude and the orbital phases of the modulation extrema. Still, the exact coincidence of the found period with the fifth harmonic of the sidereal day convinced us of its origin as an artefact of the visibility window of the radio telescopes. The regular behaviour we found could thus be spurious.

\section{Cross-correlations}
\label{cross}

\begin{figure}
\centerline{\includegraphics[width=6.5cm]{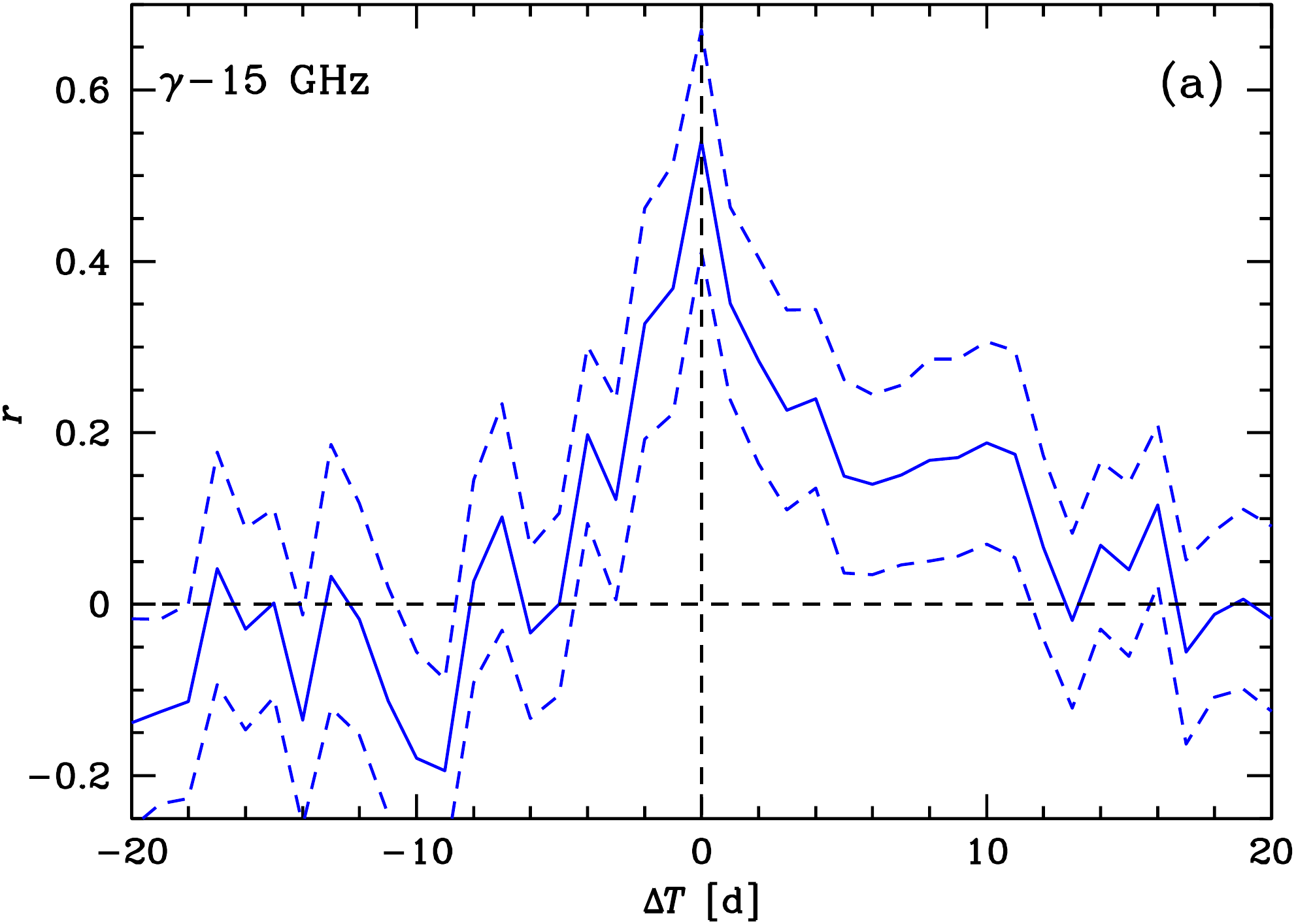}} 
\centerline{\includegraphics[width=6.5cm]{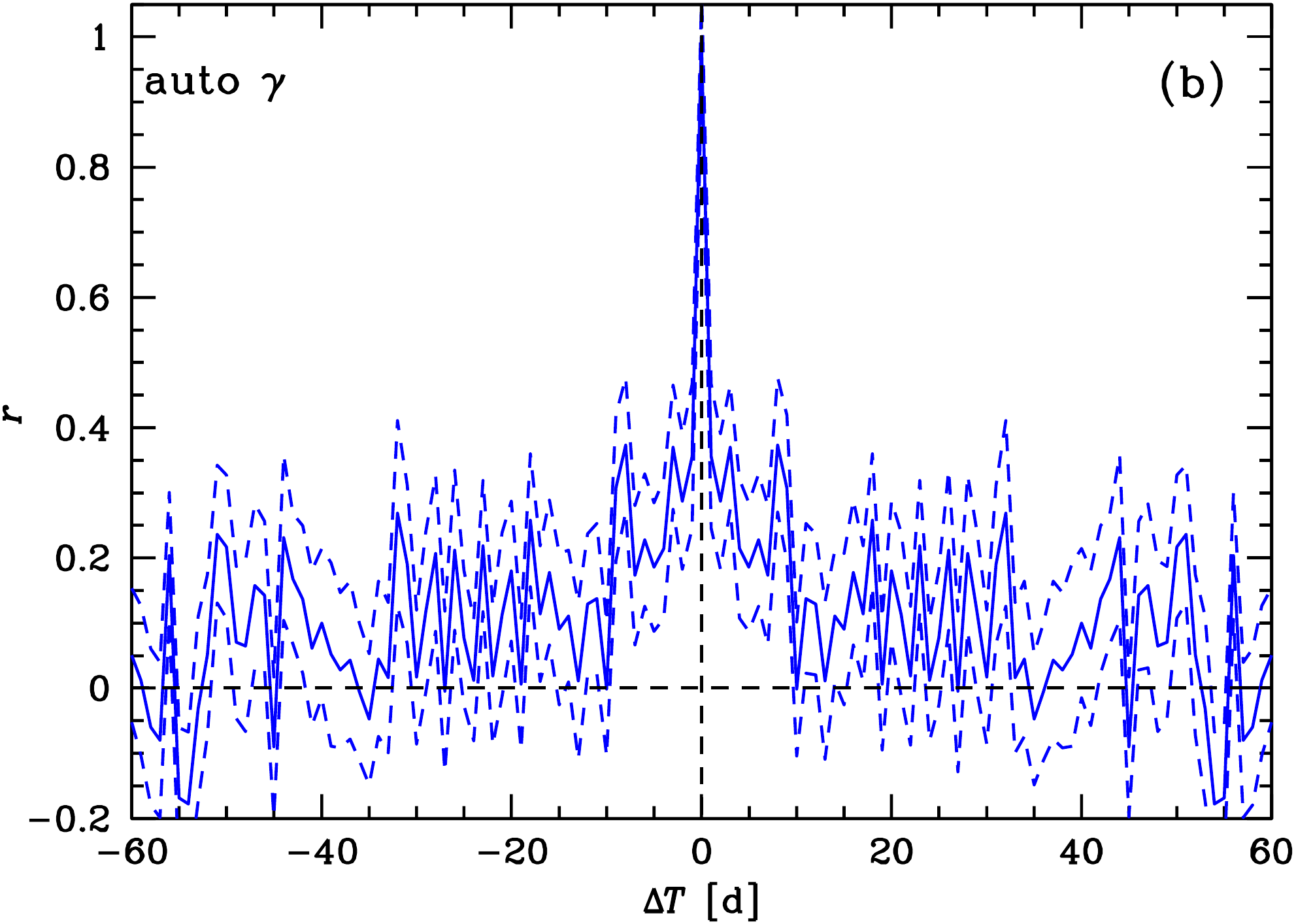}} 
\centerline{\includegraphics[width=6.5cm]{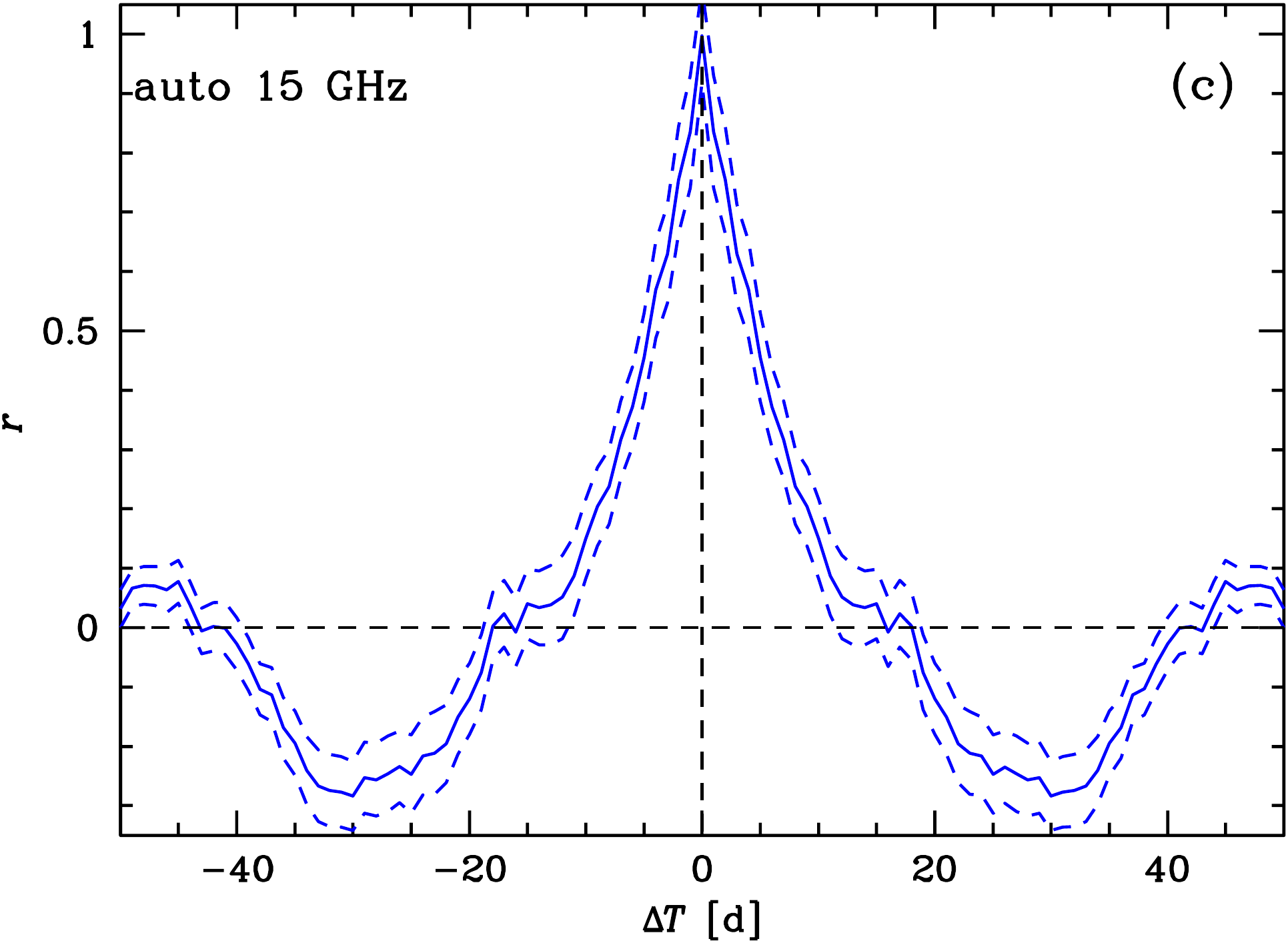}} 
\caption{(a) The cross-correlation (solid curve) between the one-day LAT detections with the fractional error $\leq$0.5 and the 15 GHz radio emission from the AMI. Hereafter $\Delta t>0$ corresponds to the signal in the second photon energy range given in the plot label lagging behing the signal in the first range (in the present case, 15 GHz flux lagging the \g-ray one), and dashed curves give the estimated uncertainty range of a correlation. $\Delta t>0$ corresponds to the radio emission lagging the \g-rays. We do not find a measurable lag of the radio emission. (b) The auto-correlation of the \g-ray detections (with the fractional error $<$1 within a day). (c) The auto-correlation of the 15-GHz emission during the same epoch as that of the \fermi\/ observations.
} \label{gamma_radio}
\end{figure}

\begin{figure*}
\centerline{\includegraphics[height=4.2cm]{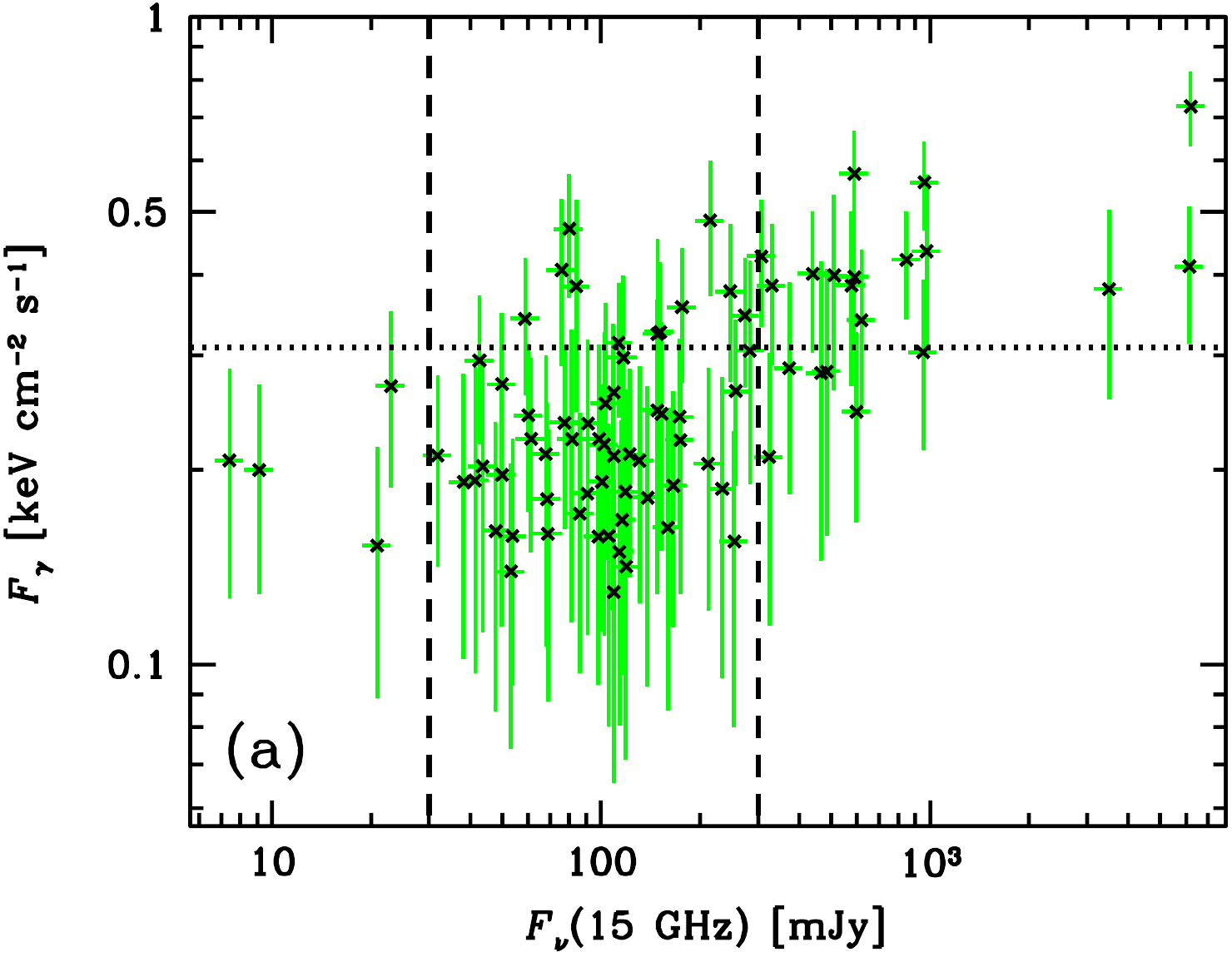}
\includegraphics[height=4.2cm]{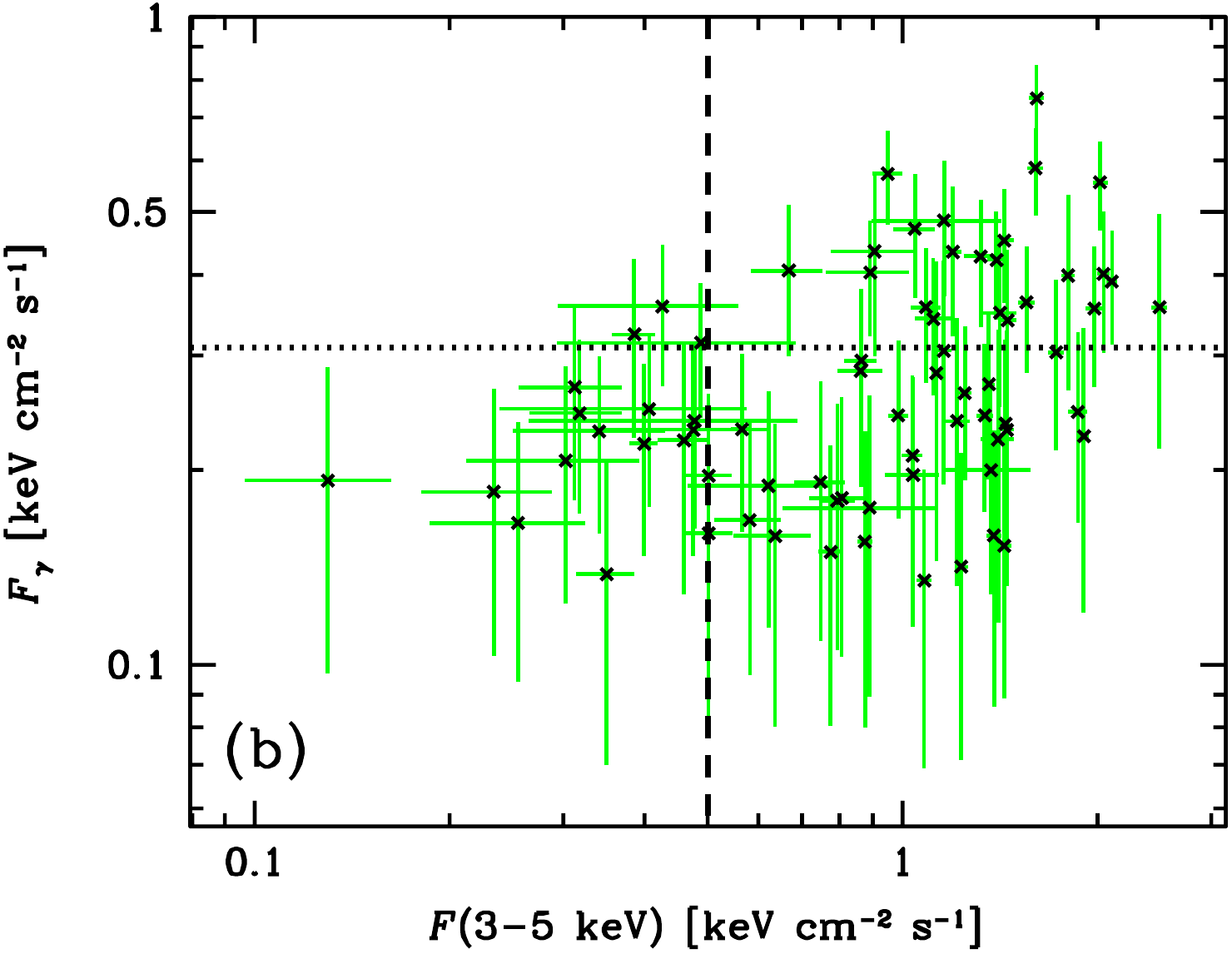}
\includegraphics[height=4.2cm]{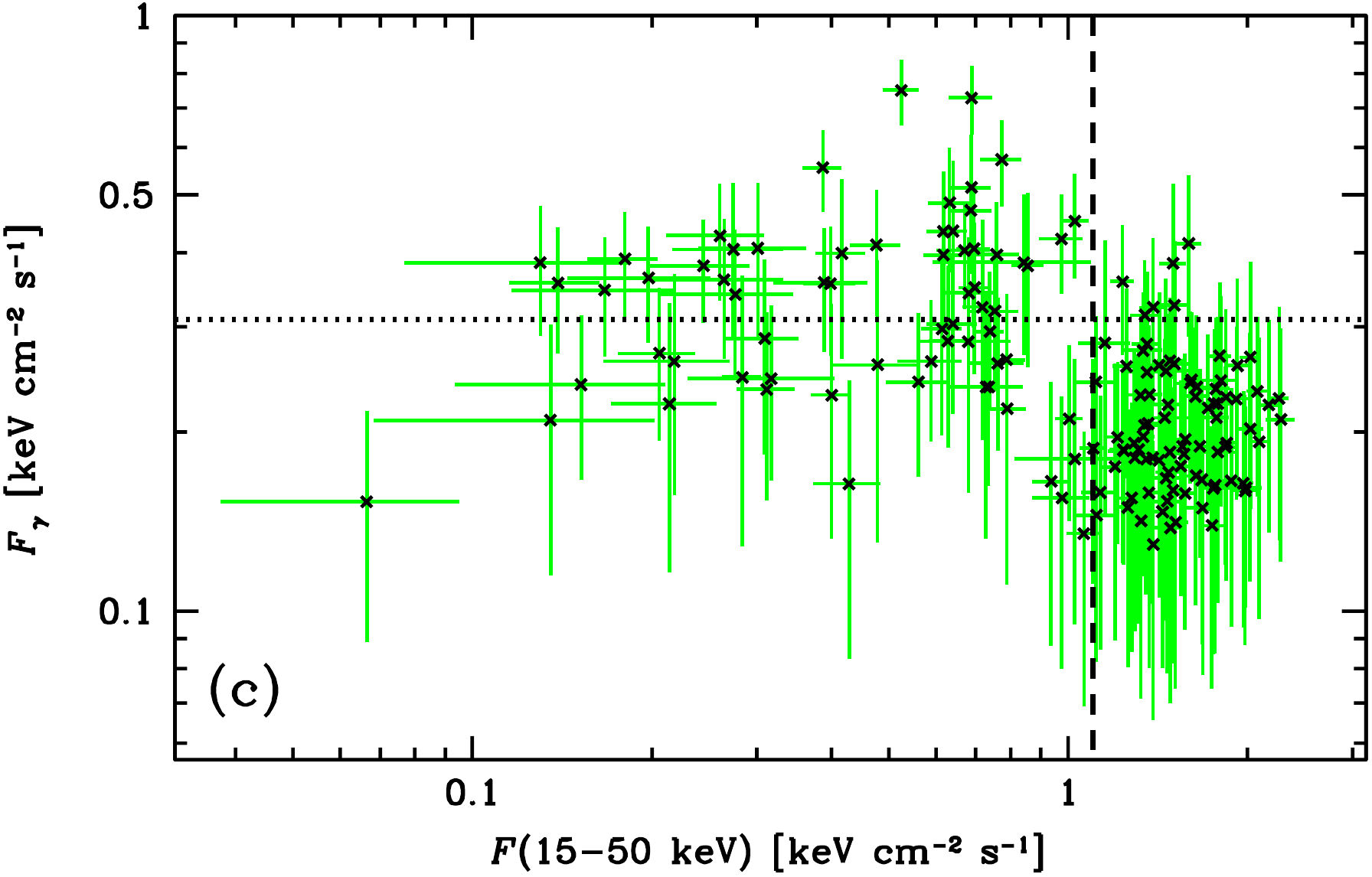}} 
\caption{The relationships between the daily 0.1--100 GeV energy fluxes with the fractional error $<$0.5 and the average of (a) the 15 GHz radio flux, (b) the ASM 3--5 keV flux, and (c) the BAT 15--50 keV flux, on the corresponding MJDs, respectively. The horizontal dotted line gives the boundary of the flaring \g-ray state, and the vertical dashed lines give the boundaries between the hard and soft/intermediate states. In (a), the hard state occurs only between the two dashed lines, but this range also corresponds to the soft/intermediate state. In (b) and (c), the soft state occurs to the right and left of the dashed line, respectively.
} \label{F_gamma_RX}
\end{figure*}

\begin{figure*}
\centerline{\includegraphics[width=5.7cm]{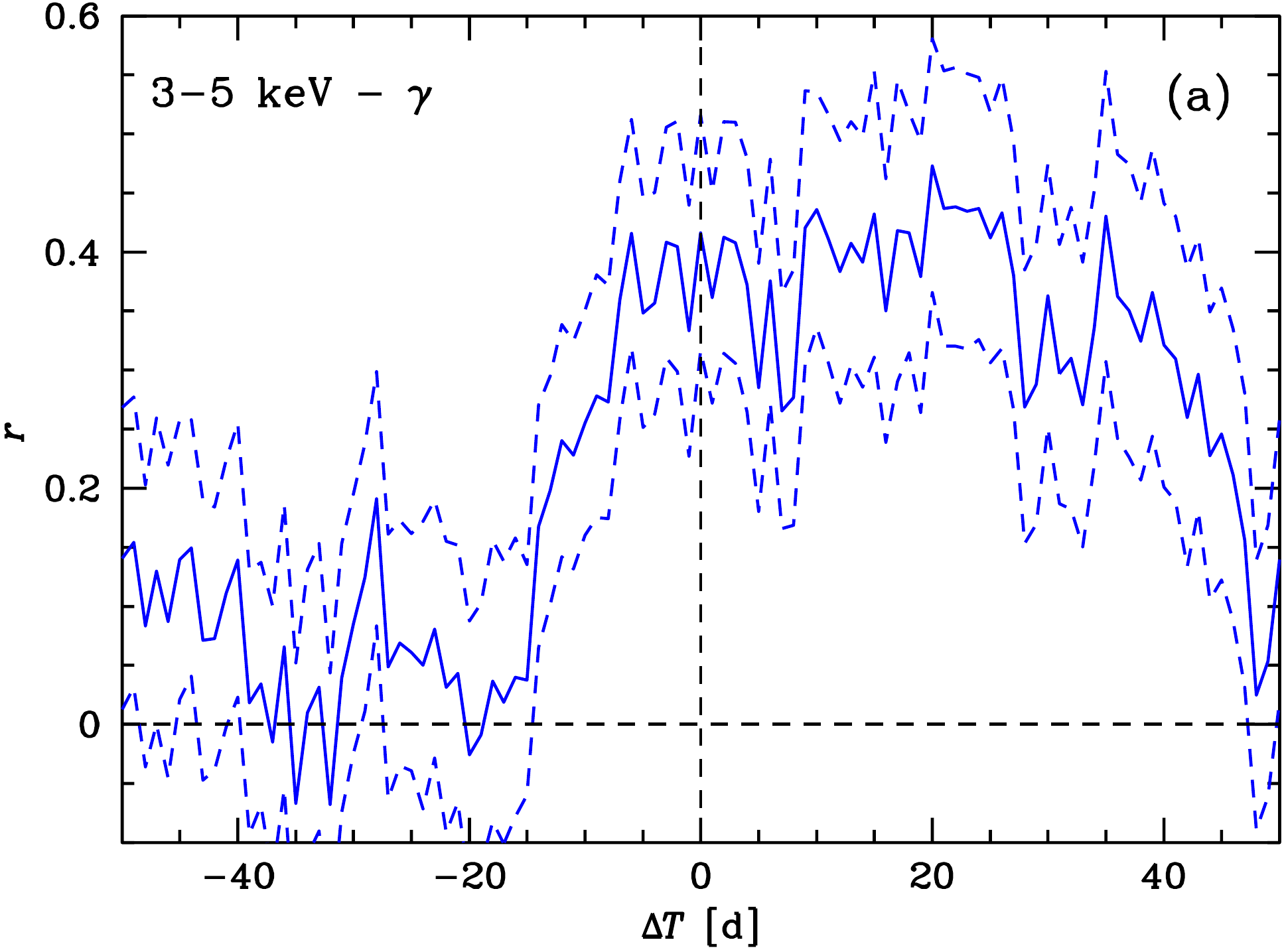} \includegraphics[width=5.7cm]{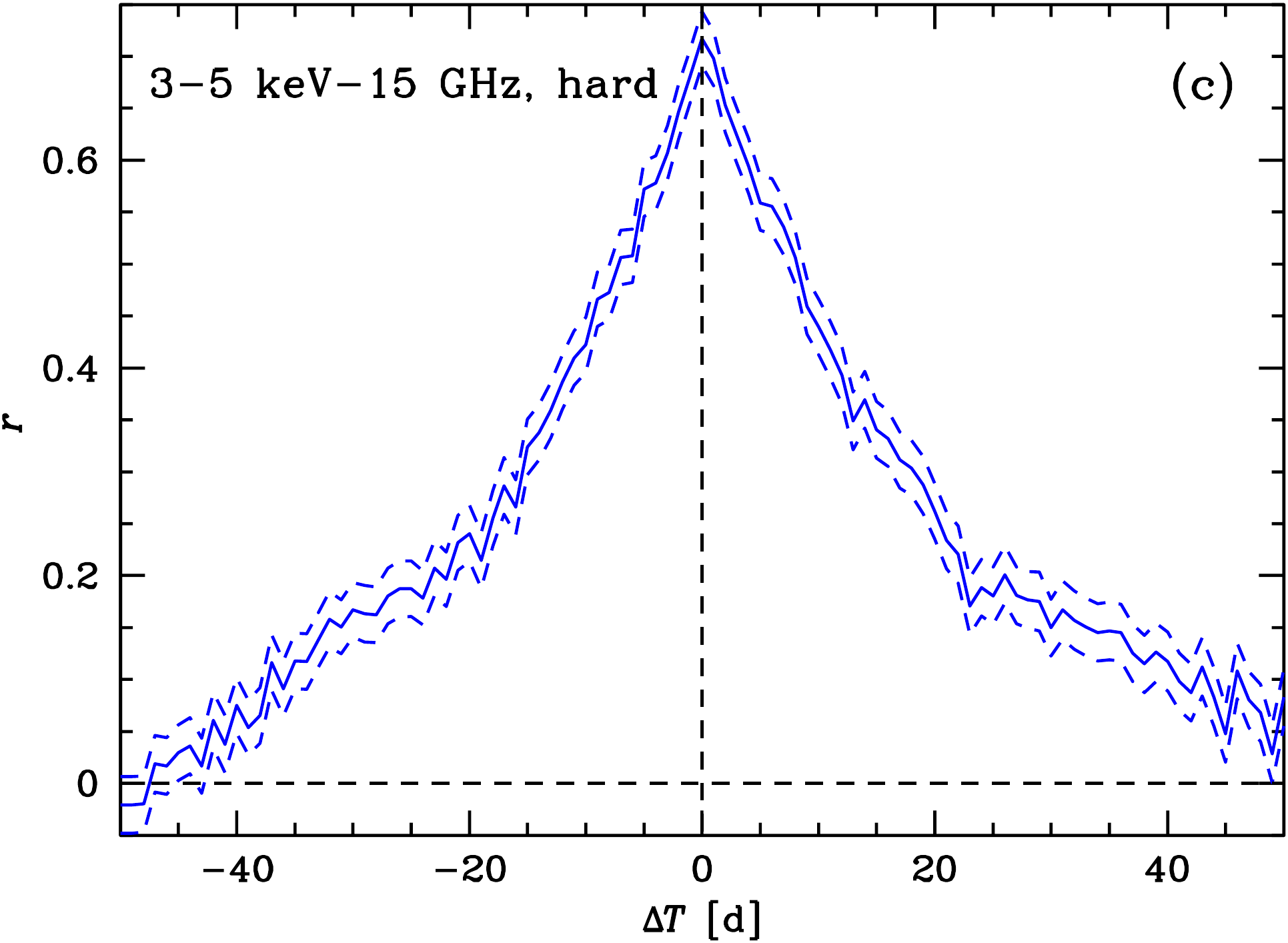} \includegraphics[width=5.7cm]{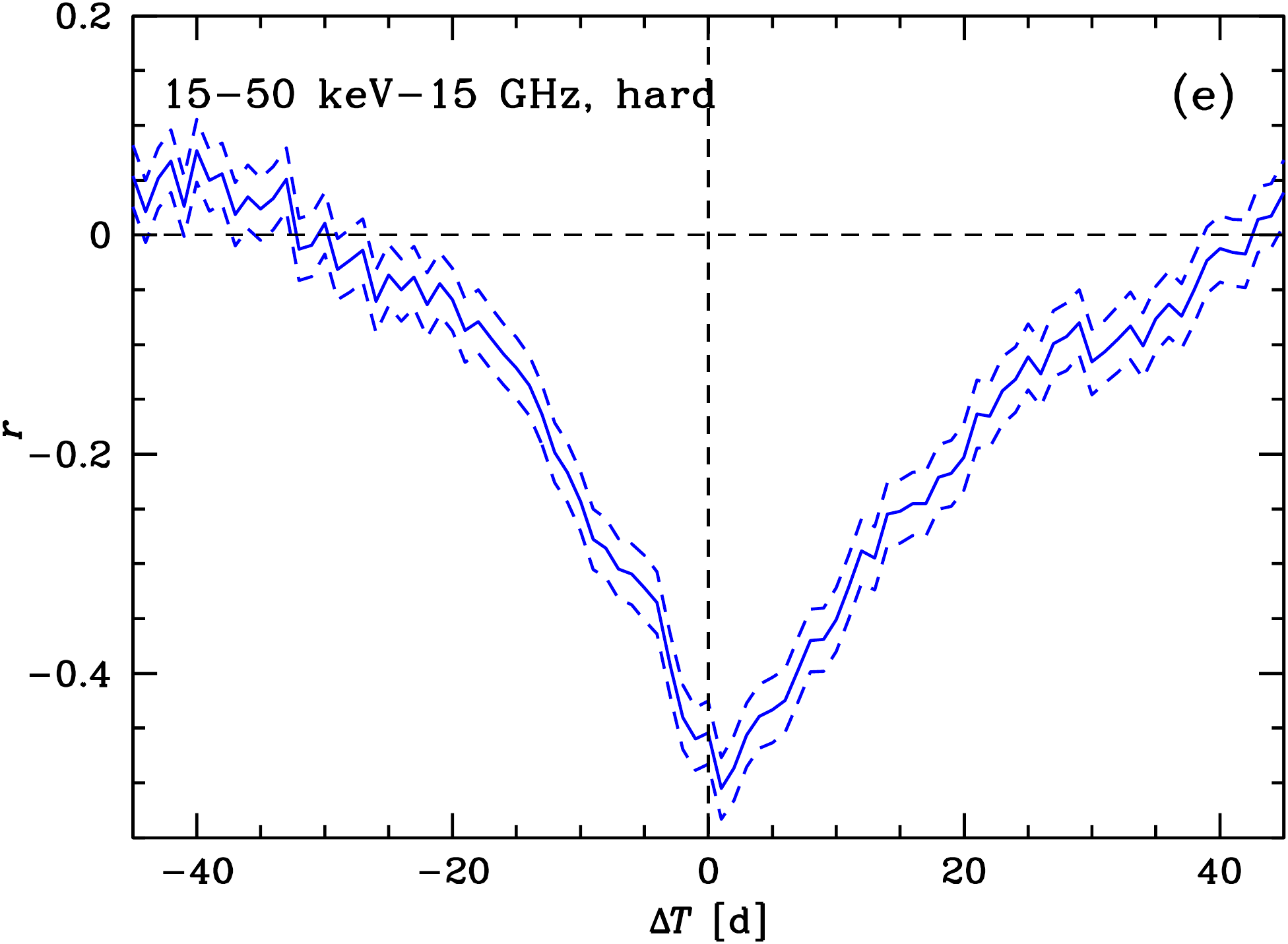}} 
\centerline{\includegraphics[width=5.7cm]{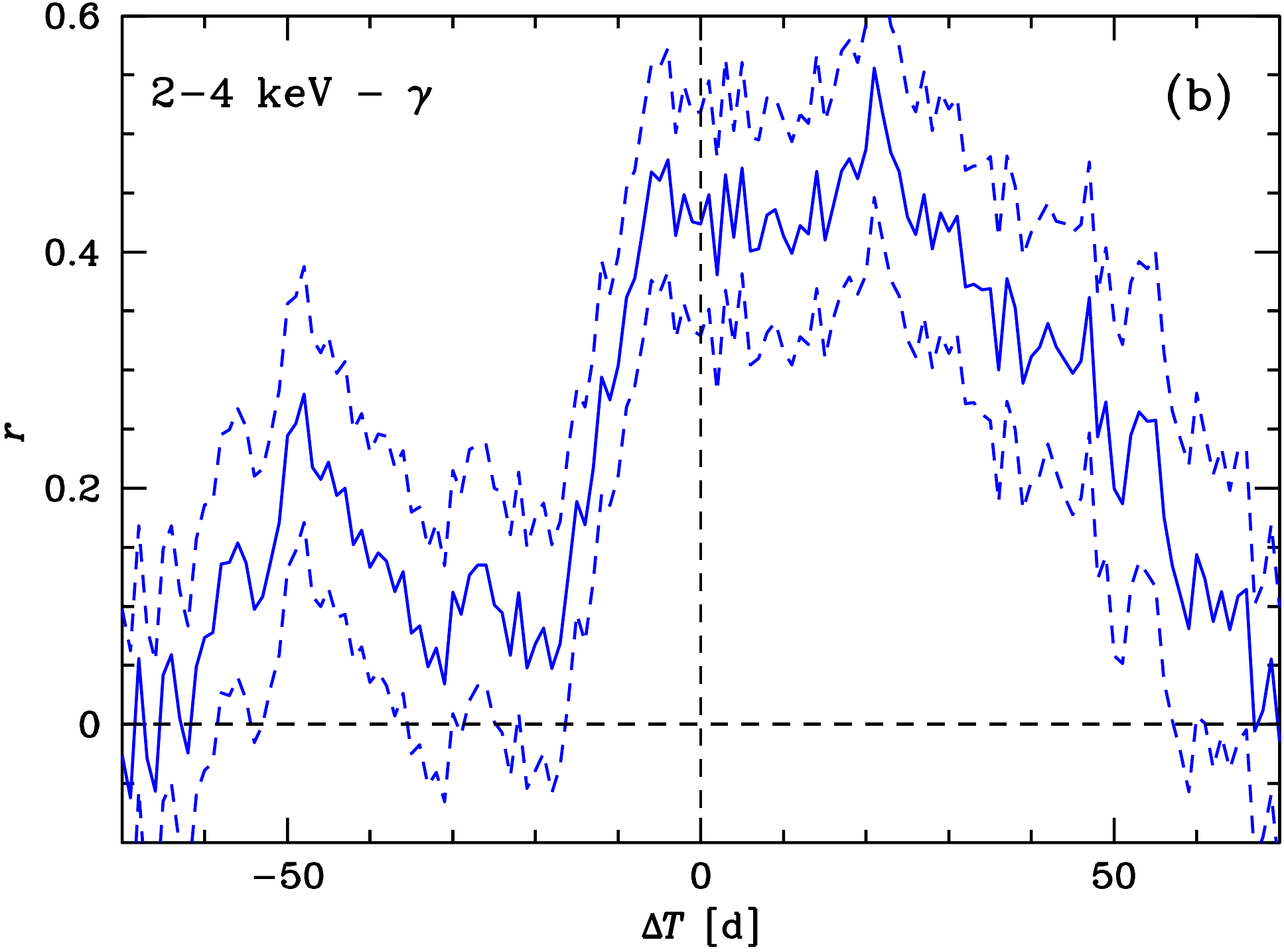} \includegraphics[width=5.7cm]{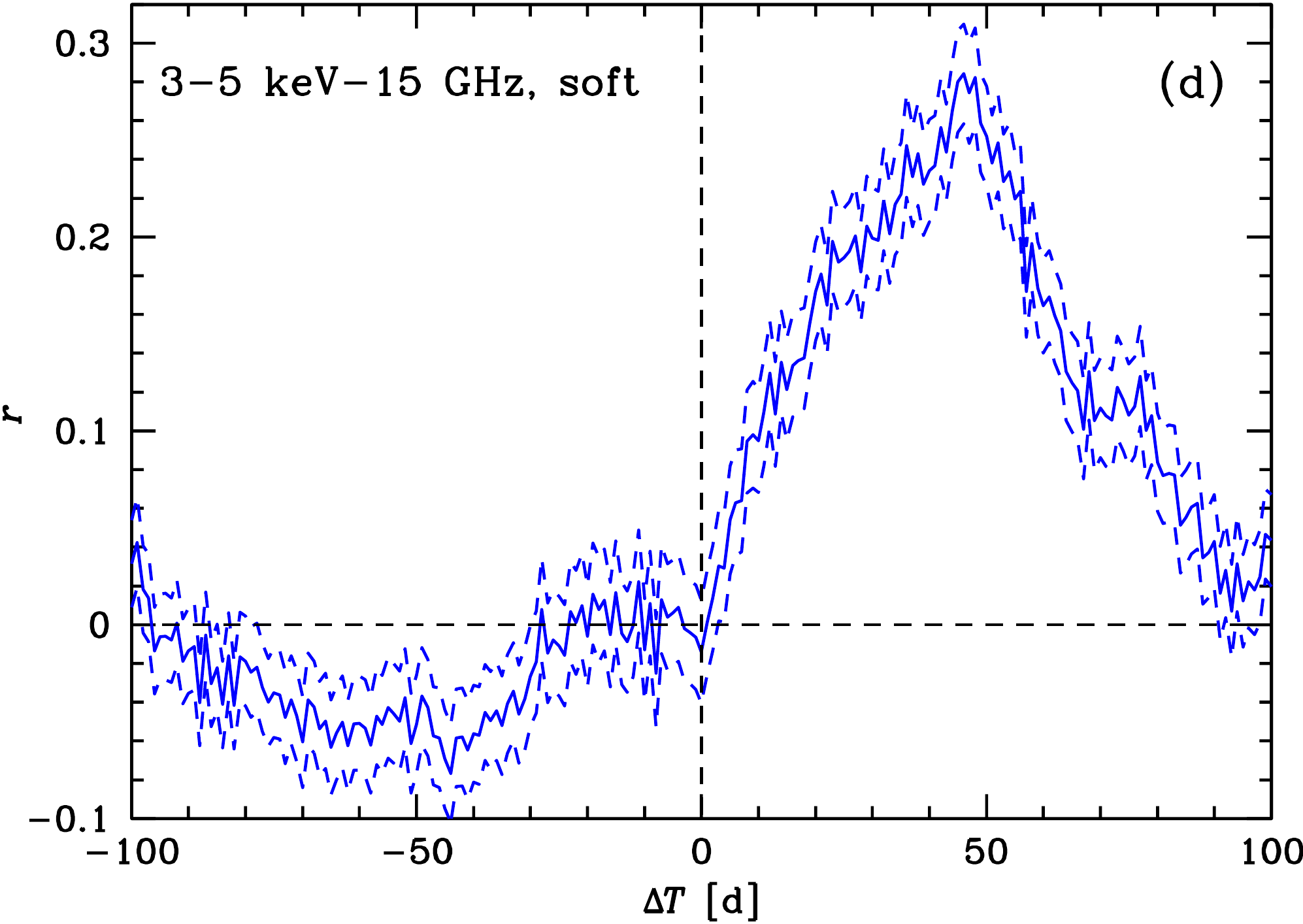}
\includegraphics[width=5.7cm]{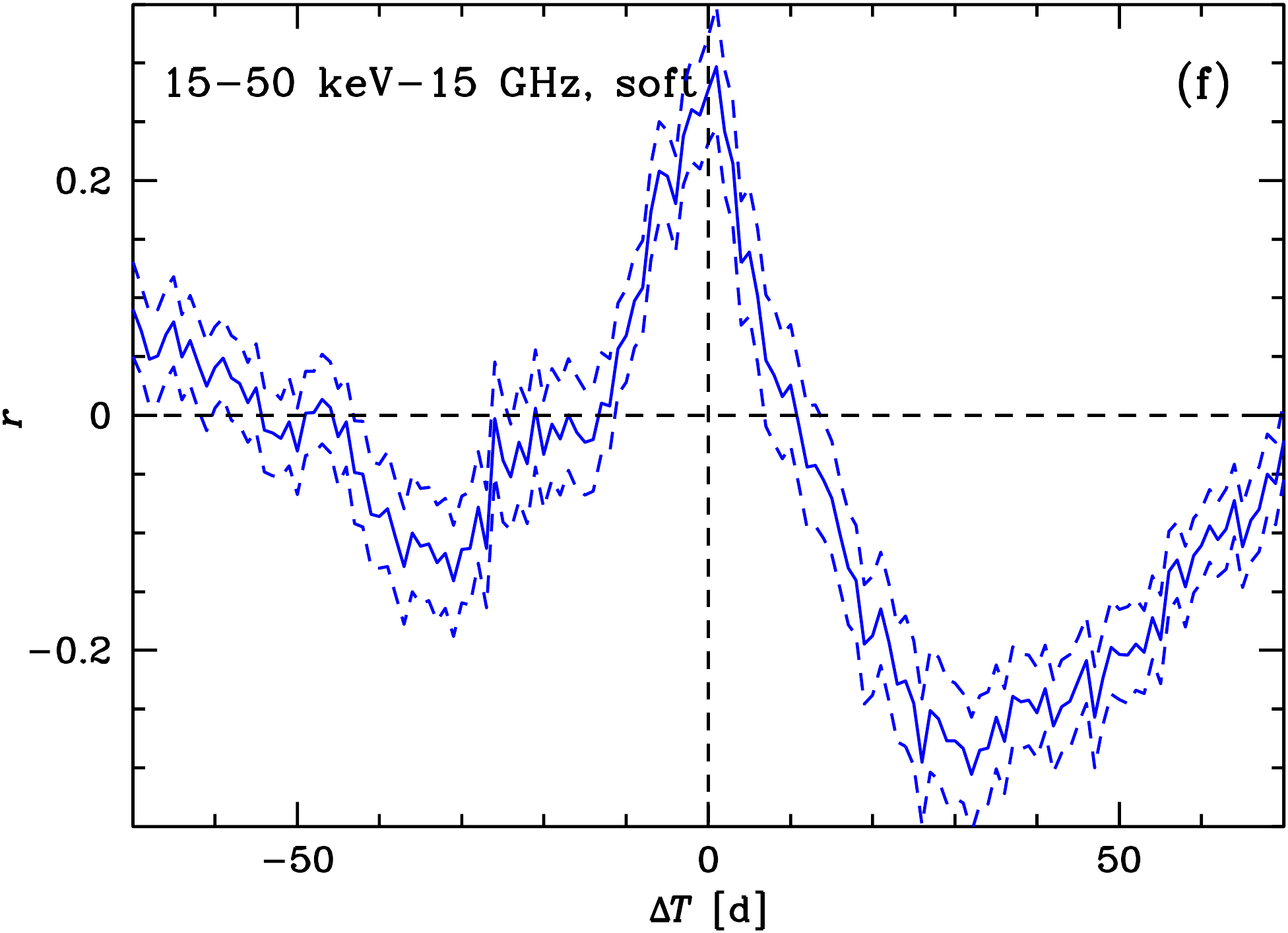}} 
\caption{Left panels: the cross-correlation between (a) the 3--5 keV ASM rate, and (b) 2--4 keV MAXI rate, vs.\ the LAT detections (positive lags correspond to \g-rays lagging the X-rays). Middle panels: the cross-correlation between the 3--5 keV ASM rate in (c) the hard state ($F_{\rm X}< 0.5$ keV cm$^{-2}$ s$^{-1}$, $30 < F_\nu< 300$ mJy), and in (d) the soft/intermediate state ($F_{\rm X}> 0.5$ keV cm$^{-2}$ s$^{-1}$). Right panels: the cross-correlation between the 15--50 keV BAT rate in (e) the hard state ($F_{\rm X}> 1.1$ keV cm$^{-2}$ s$^{-1}$, $30 < F_\nu< 300$ mJy), and in (f) the soft/intermediate state ($F_{\rm X}< 1.0$ keV cm$^{-2}$ s$^{-1}$). Positive lags correspond to radio lagging the X-rays. For all data sets, only data with the fractional error $\leq$0.5 were considered. 
} \label{X_gamma_radio}
\end{figure*}

\subsection{The method}
\label{method}

We calculate the Pearson's correlation coefficient between a discrete light curve, $x_i$, $i=1,...,I$, and another one, $y_j$, $j=1,...,J$, shifted in time by $\Delta t$,
\begin{equation}
r(\Delta t)=\frac{\sum [x_i-\bar{x}(\Delta t)][y_i-\bar{y}(\Delta t)]/K}{\sqrt{\sum_i [x_i-\bar{x}(\Delta t)]^2/I'}\sqrt{\sum_j [y_j-\bar{y}(\Delta t)]^2/J'}},
\label{pearson}
\end{equation}
where the summation in the numerator is over all pairs, $(i,j)$, satisfying 
\begin{equation}
\Delta t-\delta/2\leq t(y_j)-t(x_i)< \Delta t+\delta/2,
\label{delta}
\end{equation}
$\delta$ is the bin size of the time shift, $\Delta t$, $K$ is the number of such pairs, and the values of $\bar{x}$ and $\bar{y}$ and the sums in the denominator and in $\bar x$, $\bar y$, are over only $I'$, $J'$, values of the $x_i$ or $y_j$ satisfying equation (\ref{delta}), respectively. The standard deviation of $r$ is calculated using equation (5) of \citet{edelson88}.

This method differs slightly from that of \citet{edelson88}, who considered the values of the standard deviations and the averages for each light curve based on all of their respective points, while here we include only those entering a given bin, as proposed by \citet{lehar92}. Using global averages and standard deviations can lead to substantial inaccuracies if either a light curve has long-term trends and the cross-correlation is carried over a section of it, or it is strongly varying. In particular, we found that the value of the auto-correlation at zero lag is substantially greater than unity in a number of cases considered here. In fact, the value of $r$ is within the range of $[-1,1]$ only if $I'=J'=K$. Imposing that requires multiple counting (for each occurrence of the condition \ref{delta}) in the mean and standard deviation in cases in which a given $x_i$ satisfies the condition (\ref{delta}) for more than one $y_j$ (or vice versa). Similarly to the case of using global averages and standard deviations, the auto-correlation can exceed unity and the cross-correlation can be not correctly normalized if the multiple counting for the mean and standard deviation is not allowed.

Then, as an option, we average each light curve in the pair within its bin of the size $\delta$ before calculating $r$. This alleviates the above problem of the normalization of $r$, since then a given $x_i$ satisfies the condition (\ref{delta}) for (typically) only one value of $y_j$. We have found this to be especially important in the case of correlating the 15 GHz light curve from the AMI with the LAT \g-rays, in which case the LAT light curve has one point per day while the AMI one has typically several. We find then the correlation coefficient with a very noisy dependence on $\Delta t$, caused by variations of both the average values and the standard deviations. However, for each shown correlation, we have tested that using different options leads to similar overall shapes of the correlations. Also, we use logarithms of the fluxes, since the flux distributions in Cyg X-3 are much closer to log-normal than to normal (ZSP16), and the calculation of $r$ of equation (\ref{pearson}) assumes that the distributions of $x_i$ and $y_i$ are normal. Log-normal flux distributions have been found in other accreting systems \citep*{uttley05}.

\subsection{Cross and auto-correlations in Cyg X-3}

We first cross-correlate the \g-ray and radio light curves. We show the results for the LAT detections with the fractional error $\le$0.5 within a day (yielding 174 days) in Fig.\ \ref{gamma_radio}(a) (hereafter the dashed curves show the uncertainty ranges estimated as above). The highest \g-ray fluxes are found in the soft and intermediate states, though we also find a relatively large number of detections in the hard state. The cross-correlation peaks at zero lag, implying that the lag averaged over all frequencies of the variability is $<$1 d. This is a much more accurate result than the early one of FLC09, who obtained a peak lag of $5\pm 7$ d. However, the cross-correlation shows a significant asymmetry, indicating that some radio photons still lag the \g-ray emission by $\la$10 d. $F_\gamma\simprop F_{\rm R}^{1/3}$. When the required maximum fractional error is increased, the cross-correlation still peaks at zero lag, but becomes weaker, with a lower value of $r$. We have also considered the case with the radio data split into parts with $F_\nu>0.3$ and $<$0.3 Jy. For both ranges, the cross-correlations peak at zero lag. Fig.\ \ref{F_gamma_RX}(a) shows the relationship between the two daily-averaged energy fluxes on the same MJDs, where we see a positive correlation of the \g-ray flux with the radio one,

\begin{figure}
\centerline{\includegraphics[width=6.cm]{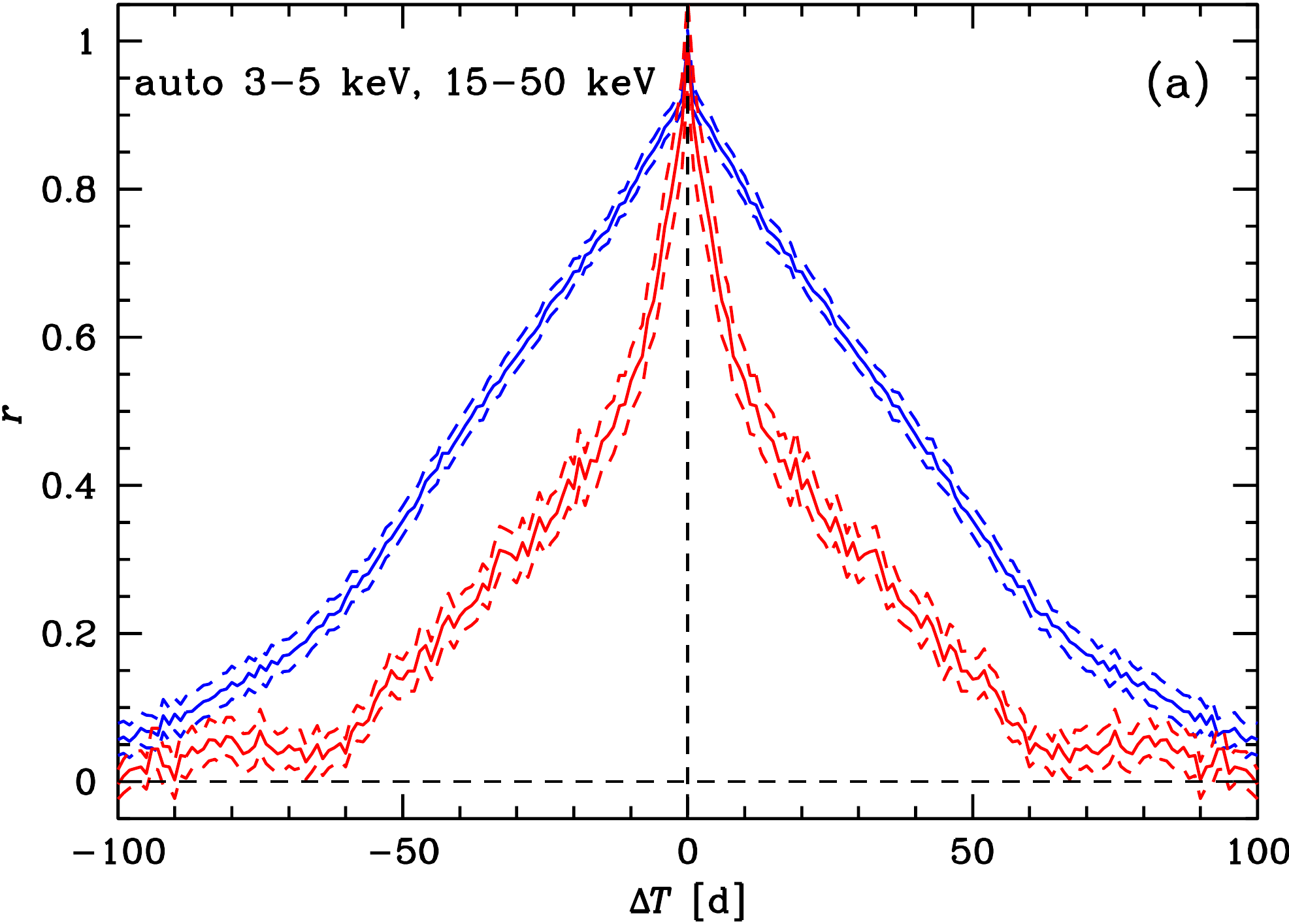}} 
\centerline{\includegraphics[width=6.cm]{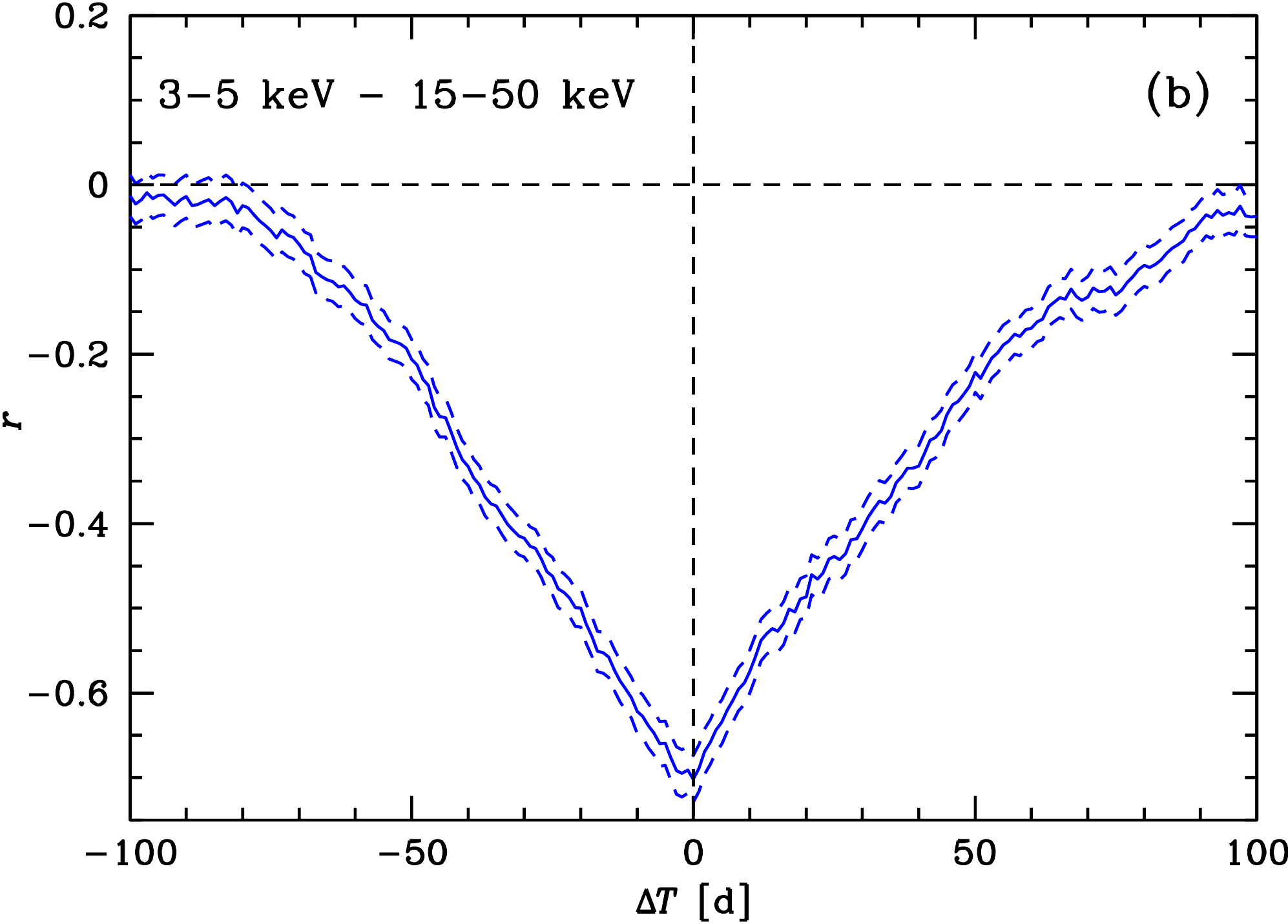}}
\caption{(a) The auto-correlation for the 3--5 keV ASM (upper blue curve) and the 15--50 keV BAT (lower red curve) rates. (b) The cross-correlation between the 3--5 keV and 15--50 keV rates in all states.
} \label{X_auto}
\end{figure}

The \g-ray emission (all detections with the fractional error $<$1; 486 days) has a narrow auto-correlation with the width of $<$1 d, see Fig.\ \ref{gamma_radio}(b). We also see a weak auto-correlation tail, dropping to null at $\gtrsim$10 d. On the other hand, the radio emission has a relatively wide auto-correlation, with the half-width at $r=0.5$ of 4.5 d, see Fig.\ \ref{gamma_radio}(c). We then see that the radio emission is anti-correlated with itself for $\Delta t\sim 30$ d. In order to investigate the origin of it we have split the data set into two parts, above and below 0.3 Jy. We find that the auto-correlation for the low fluxes is similar to the one for all the data, and the $\sim$30 d time scale appears to be related to the typical duration of a single occurrence of the hard state. On the other hand, the auto-correlation for the high fluxes is narrower and it becomes negative already at $\Delta t\simeq 8$ d, and reaches the global negative minimum at 20 d. This appears to be related to the radio flares both preceded and followed by states with weak radio emission (e.g., \citealt{szm08}). 

We then correlate the X-ray and \g-ray emission. Fig.\ \ref{F_gamma_RX}(b) shows the correlation with the soft X-rays on the same MJDs. We see that almost all \g-ray detections in the hard state have relatively low fluxes, below the boundary of the flaring state. Then, there is a positive correlation in the soft state. The cross-correlation with soft X-rays are shown in Figs.\ \ref{X_gamma_radio}(a, b), where the positive correlation at zero lag continues up to a few tens of days, indicating the \g-ray emission continues after an occurrence of a peak in soft X-rays. This corresponds to occurrences of \g-ray flares during transitions from the softest (hypersoft) X-ray states to harder ones, confirming \citet{koljonen10}. However, some \g-ray flares also take place during the opposite transtions, which may correspond to weaker peaks at negative lag in Figs.\ \ref{X_gamma_radio}(a, b),

The correlation with hard X-rays is more complicated, see Fig.\ \ref{F_gamma_RX}(c). While most of \g-ray detections with high fluxes occur for low hard X-ray flux, i.e., in the soft state, which corresponds to an overall anti-correlation, the detections within the soft state show a weak positive correlation. We also see a large number of detections in the hard state below the boundary of the flaring state. This results in a strong anti-correlation between the \g-ray emission and hard X-rays, with the minimum of the cross-correlation at a lag of \g-rays of $\sim$5 d (not shown here). However, that lag, given the measurement errors, may be not statistically significant.

We next consider cross-correlations between X-rays and radio in different states. We show the cross-correlations between the soft X-rays (3--5 keV) and radio in the hard and soft/intermediate states in Figs.\ \ref{X_gamma_radio}(c) and (d), respectively. We see the hard-state relationship gives a strong positive correlation peaking at zero lag (for 1-d bins). The cross-correlation is relatively wide, with a half-width of $\sim$15 d, and relatively symmetric, showing that some radio photons in the hard state lead and and some lag the soft X-rays. We see also some asymmetry at $|\Delta t|\ga 20$ d, indicating that the lag dominates at long $\Delta t$. 

On the other hand, while the soft/intermediate state shows almost no correlation at zero lag, it shows a strong positive peak at the radio band delayed by $\simeq$45--48 d, see Fig.\ \ref{X_gamma_radio}(d). To investigate it further, we have calculated the relationship between the soft X-ray and radio fluxes in the soft/intermediate state at 0 and 46 d lags. Those plots, not shown here, confirm the lack of a correlation at zero lag changing into a weak positive correlation at the 46-d lag. We note that the 45--48-d lag appears to correspond to the average time spent in the ultra/hypersoft X-ray state that directly precedes major radio flares. For instance, in 2011, the radio emission was quenched for a month while the source was in the ultra/hypersoft state, ending with a major 10-Jy radio flare \citep{corbel12}. The interpretation of the physical nature of this lag appears, however, unclear. It may correspond to a timescale linked to magnetic field re-arrangement in the disc so that a jet can be launched. It may also correspond to the propagation time scale from the centre to the dominant large-scale jet component of the radio emission (at tens of mas; \citealt{tudose10}). This interpretation, however, implies a rather low speed of the jet, $\sim\! 0.1 c$. The radio emission is also seen on arcsec scales, as found by \citet{marti01}, which may contribute to that long lag as well. We have also calculated the soft X-rays vs.\ 15 GHz cross-correlation without separating into the states, to be able to directly compare it to the 3--5 keV vs.\ \g-ray cross-correlation. It shows the shape relatively similar to that of Fig.\ \ref{X_gamma_radio}(b), with a positive $r$ at zero lag and a peak below 50 d.

Hard X-rays, 15--50 keV, are anti-correlated with the radio band in the hard state (i.e., for large X-ray fluxes), see Fig.\ \ref{X_gamma_radio}(e). There appears to be a 1-d lag of the radio emission here, but it is not statistically significant. On the other hand, the hard X-rays are positively correlated with the radio at zero lag in the soft/intermediate state, see Fig.\ \ref{X_gamma_radio}(f). However, they also show an anti-correlation with radio peaking at a lag of $\sim$30 d, see Fig.\ \ref{X_gamma_radio}(h). This behaviour is likely to be related to the 45--48 d lag at soft X-rays, Fig.\ \ref{X_gamma_radio}(d). 

We then show the auto-correlation functions for the 3--5 and 15--50 keV energy ranges, and their cross-correlations in Figs.\ \ref{X_auto}(a) and (b), respectively. Interestingly, the 15--50 keV auto-correlation is substantially narrower, with the half-width of $\simeq$20~d, than the 3--5 keV one, with the half-width of $\simeq$40 d. This difference may correspond to the hard X-ray emission region being closer to the compact object, and thus smaller than that for soft X-rays. On the other hand, it may also correspond to the 15--50 keV emission leading the 3--5 keV one. Interestingly, the widths of the X-ray auto-correlations are of the same order as the radio vs.\ X-ray lags in the soft state, Figs.\ \ref{X_gamma_radio}(d, f). 

We also note that \citet{tudose10} argued that the radio/X-ray correlation observed at zero lag (e.g., \citealt{szm08}) is not theoretically expected in bright radio states, where the bulk of the radio emission is in the jet rather than in the core. However, this does not seem to present a problem given the projected jet distance from the core of $\sim$1--2 light days and the widths of the auto-correlations of both radio and X-rays are $\gg$2 d. 

\section{Conclusions}
\label{conclusions}

We have obtained the following main results. 

Based on nine years of the \fermi\/ data, we have searched for occurrences of significant HE \g-ray emission from Cyg X-3. We have found a large number of days with significant LAT detections, and among them, 49 days with the 0.1--100 GeV energy fluxes above the flaring level, which we defined at $4.94\times 10^{-10}$ erg cm$^{-2}$ s$^{-1}$ based on the level $3\sigma$ above the average of non-flaring detections. Out of them, 43 days are in the soft spectral state. 

We have calculated the average \g-ray spectrum during strong flares (with positive detections in the 0.08--15 GeV energy range) and the spectrum averaged over all the occurrences of the soft spectral state (detected in the 0.5--15 GeV range). On the other hand, we have found only upper limits in the hard and intermediate states. The flaring-state spectrum is well modelled by Compton scattering of the blackbody photons from the donor by jet relativistic electrons with a power-law distribution with the spectral index of $\simeq$3.5 and the low-energy cutoff at the Lorentz factor of $\sim\! 10^3$. 

From the LAT data, we have also obtained the profile of the orbital modulation of \g-rays in the flaring state, which is significantly more accurate than the previous one of FLC09. The amplitude of the modulation is large, by a factor of $\simeq$5, and it is well modelled by Compton scattering of stellar blackbody, which agrees with the modelling of the spectrum. The modulation model implies the location of the \g-ray source at the distance along the jet similar to the separation between the binary components, a mildly relativistic jet velocity, and it requires that the jet is inclined at an angle $\gtrsim\! 25\degr$ with respect to the binary axis. 

We have then studied the 22 years of 15-GHz radio observations of Cyg X-3 by the Ryle and AMI telescopes. We have discovered pronounced modulation of the radio emission at the orbital period. The amplitude of the modulation depends on both the spectral state and the flux level. It changes from $\simeq$2.5 to $\simeq$10 per cent, and it is $\simeq$4 per cent when averaging over all the data. We model the observed modulation as free-free absorption in the stellar wind of the jet radio emission. We find the 15 GHz source to be located at a distance of $\sim\! 10^2$ times the binary separation. 

Finally, we have studied cross-correlations between the HE \g-ray and radio light curves, as well as between either of them and light curves from the ASM, MAXI and BAT X-ray monitors. We have found the correlation coefficient between the HE \g-ray and 15 GHz light curves peaks at zero lag. However, its asymmetry indicates that some radio photons lag the \g-rays by $\la$10 d. Then, the \g-rays lag soft X-rays by some tens of days, but without showing a clear peak of the correlation coefficient. This is consistent with occurrence of \g-ray flares mostly during the soft-to-hard transitions. Also, we have not found measurable lags between the X-ray and radio emission in the hard spectral state. On the other hand, we have found the lag peaking at 45--48 d of the 15 GHz emission with respect to 3--5 keV soft X-rays in the soft state. This lag is similar to the time spent in the X-ray ultra/hypersoft and radio-quenched state before a major flare occurs.

\section*{ACKNOWLEDGMENTS}

We thank the referee for valuable comments. This research has been supported in part by the Polish National Science Centre grants 2013/10/M/ST9/00729, 2015/18/A/ST9/00746 and 2016/21/P/ST9/04025, and the Carl-Zeiss Stiftung through the grant ``Hochsensitive Nachweistechnik zur Erforschung des unsichtbaren Universums" to the Kepler Zentrum f{\"u}r Astro- und Teilchenphysik at the University of T{\"u}bingen, and by the state of Baden-W\"urttemberg through bwHPC. The authors thank the SFI/HEA Irish Centre for High-End Computing (ICHEC) for the provision of computational facilities and support. The \textit{Fermi\/} LAT Collaboration acknowledges generous ongoing support from a number of agencies and institutes that have supported both the development and the operation of the LAT as well as scientific data analysis. These include the National Aeronautics and Space Administration and the Department of Energy in the United States, the Commissariat \`a l'Energie Atomique and the Centre National de la Recherche Scientifique / Institut National de Physique Nucl\'eaire et de Physique des Particules in France, the Agenzia Spaziale Italiana and the Istituto Nazionale di Fisica Nucleare in Italy, the Ministry of Education, Culture, Sports, Science and Technology (MEXT), High Energy Accelerator Research Organization (KEK) and Japan Aerospace Exploration Agency (JAXA) in Japan, and the K.~A.~Wallenberg Foundation, the Swedish Research Council and the Swedish National Space Board in Sweden. Additional support for science analysis during the operations phase is gratefully acknowledged from the Istituto Nazionale di Astrofisica in Italy and the Centre National d'\'Etudes Spatiales in France. This work performed in part under DOE Contract DE-AC02-76SF00515. The AMI arrays are supported by the University of Cambridge and by the UK STFC. This research has made use of the results provided by the ASM and BAT teams, and MAXI data provided by RIKEN, JAXA and the MAXI team.

\label{lastpage}
\end{document}